\documentclass[a4paper,11pt]{article}
\pdfoutput=1 

\usepackage{jheppub} 

\usepackage[T1]{fontenc} 

\usepackage{pdfsync}
\usepackage{mathptmx}
\usepackage[utf8]{inputenc}
\usepackage[T1]{fontenc}
\usepackage{slashed}
\usepackage{times}
\usepackage{amssymb,amsfonts,amsmath,amsthm}
\usepackage{dsfont,bbm}
\usepackage{dcolumn}
\usepackage{epsf}
\usepackage{graphicx}
\usepackage[caption=false]{subfig}
\usepackage{dsfont}
\usepackage{simplewick}
\usepackage{bm}
\usepackage{eucal}
\usepackage{slashed}
\usepackage[active]{srcltx}
\usepackage[usenames]{color}

\title{\boldmath Contour gauge: Compendium of Results in Theory and Applications}

\author[a, b]{I.~V.~Anikin}

\affiliation[a]{Bogoliubov Laboratory of Theoretical Physics JINR, 141980 Dubna, Russia}
\affiliation[b]{Institute of Modern Physics, Chinese Academy of Sciences, 730000 Lanzhou, P.R. China}

\emailAdd{anikin@theor.jinr.ru}

\abstract{In this review, we outline the main features of the non-local gauge, named the contour gauge.
The contour gauge belongs to the axial type of gauges and
extends the local gauge used in the most of approaches.
The geometry of gluon fields and the path-dependent formalism are the essential tools for the description of non-local gauges.
The principle feature of the contour gauge is that
there are no the residual gauges which are left in the finite domain of space.
In the review, we present the useful correspondence between the contour gauge conception and the
Hamiltonian (Lagrangian) formalism. The Hamiltonian formalism is turned out to be a very convenient framework
for the understanding of contour gauges.
The comprehensive comparison analysis
of the local and non-local gauges advocates the advantage of the contour gauge use.
As an example of practical worth,
we consider the Drell-Yan process and discuss the gauge invariance of the corresponding
hadron tensor. We show that the appropriate use the contour gauge leads to the existence
of extra diagram contributions. These additional contributions, first, restore the gauge invariance of the hadron tensor
and, second, give the important terms for the observable quantities.
We also demonstrate the significant role of the additional diagrams to form
the relevant contour in the Wilson path-ordered exponential.
Ultimately, it leads to the spurious singularity fixing.
Moreover, in the present review, we discuss in detail
the problem of spin and orbital angular momentum separation.
We show that in $SU(3)$ gauge theories the gluon decomposition on the physical and
pure gauge components has a strong mathematical evidence provided
the contour gauge conception has been used.
In addition, we prove that the contour gauge possesses the special kind of
residual gauge that manifests at the boundary of space.
Besides, the boundary field configurations can be associated with the pure
gauge fields.}

\begin{document}
\maketitle
\flushbottom

\section{Introduction}
\label{Intro}


As well-known, the gauge theories are identical to the systems with dynamical constraints
where the gauge conditions (or the additional conditions in the Hamiltonian formalism) play a very significant role for quantization.
Indeed, the gauge theory as the theory with the constraints
can be properly quantized if and only if all constraint conditions are uniquely resolved.
It should eliminate the unphysical degrees of freedom.
However,
the constraint conditions, which are expressed through the generalized momenta and coordinates,
have very nontrivial forms. Hence, to find unique solutions of these equations is not a simple task.
Even more, it sometimes becomes even impossible.

Fortunately,  the Faddeev-Popov method  \cite{Faddeev:1980be} allows us to avoid the direct solution of the gauge conditions.
Instead, the infinite group orbit volume leading to the quantization problems
can be factorized out to the insubstantial normalization factor.
It happens owing to the gauge invariance
of the giving Lagrangian (or Hamiltonian) of a theory.
It is, however, understood that in the spacial type of gauges the factorization of the infinite group orbit
is not enough thanks to the presence of the residual gauge freedom.

Nowadays, only the local type of gauges are traditionally
used for the practical applications.
At the same time, one of the most popular gauge which is the axial-type gauges
suffers from the residual gauge freedom.
It leads to the problem with the spurious singularity.
In contrast to the local axial gauge, the usage of contour gauge (we remind that the contour gauge forms
the class of non-local gauges) gives a possibility
to fix completely the gauge freedom in the finite region of the Minkowski (or Euclidian) space.
Moreover, the fixing of residual gauges can be implemented in
the most simple way and without an additional assumptions.
It is worth to notice that there is, however, the residual gauge freedom
at the infinite boundary of the space (see \cite{Anikin:2021oht}).

The principal preponderance of contour gauge is that they are constructed only with the
unique solution of the gauge condition from the very beginning.
Namely, within the frame of contour gauge conception
the gauge orbit representative
should, somehow, be first fixed and, then, a certain local gauge condition,
correlated with the given (gauge) orbit representative, has to be found.
It is the opposite logic of the gauge fixing if compared to the usual local gauges.

Among some theorists,
there is a prejudice that the study of any nonlocal gauge conditions, especially in QCD,
is not attractive because the contour gauge technique is rather a specific and
very complicated one.
In the present review, we attempt to break this superficial and wrong impression.

To this goal, we would need the knowledge of the basic differential geometry which does not, unfortunately,
receive a very wide spread in the phenomenology society.
In particular, the interpretation of gluons
as a connection on the principle fiber bundle are merely needed:
\begin{itemize}
 \item $(a)$ to demonstrate
that, within the contour gauge conception,
the gauge freedom can be uniquely fixed in the finite region of the space;
\item $(b)$ to prove that, in the local axial gauge (like $A^+=0$),
two representations of transverse gluon fields through the strength tensor $G_{\mu\nu}$,
related to the different paths of integrations integrations, $[+\infty;\, x]$ and $[-\infty;\, x]$,
are not equivalent each other.
\end{itemize}
We notice that stress that it is not possible
to distinguish the mentioned two different representations in the local gauge.
At the same time, the assumption of equivalence leads to many problems, for example, with the gauge invariance of
Drell-Yan-like hadron tensors.

Since the interest to the details of contour gauge applications
increases in the phenomenological community,
in the review, we explain the important subtleties based on the mathematical
technique adjusted to the physical language
which is almost missed in literature.
On the other hand, since in the recent literature
one can still find a wrong representation of the transverse gluon field through the strength tensor
considered in the local axial gauge,
we treat this fact as a strong motivation for the present review.

To demonstrate the practical profits of the contour gauge conception, in the review, we are focusing on
the investigation of nucleon (hadron) composite structure which is still the most important subjects of hadron physics.
In particular, we dwell on the calculation of several sorts of the single spin asymmetries (SSA).
From the experimental point of view, SSAa are the wide-spread and useful instruments for such studies and
they open the access to the three-dimensional nucleon structure.
It takes place thanks for
the non-trivial connection between the transverse spin and the parton transverse momentum dependence
(see, for example, \cite{Angeles-Martinez:2015sea, Boer:2011fh, Boer:2003cm, Kang:2011hk, Boer:2011fx})
In QCD, SSA related to the Drell-Yan (DY) process was first considered in the case of
the longitudinally polarized hadron \cite{P-R,Carlitz:1992fv}.
This SSA is important because the second hadron is a pion.
It is necessary to mention on the sensitivity
to the shape of pion distribution amplitude, being currently the object of major interest
\cite{Brandenburg:1995pk, Bakulev:2007ej, Radyushkin:2009zg,Polyakov:2009je}
(see also \cite{Mikhailov:2009sa} and the references therein).
In \cite{P-R,Carlitz:1992fv, Brandenburg:1995pk,Bakulev:2007ej},
it was shown that the imaginary phase of SSA,
which is associated with the longitudinally polarized nucleon,
appears due to either the hard perturbative gluon loops
or twist four contribution of the  pion distribution amplitude.

Previously, in the study of transverse SSA of DY process,
the imaginary part has been only extracted from the quark propagator in
the standard diagram (that is, the diagram without radiations from the
quark legs in the correlators) with quark-gluon twist three correlator
\cite{Boer:1997bw}.
In these approaches, the ambiguity in the boundary conditions for gluons provide the purely real
quark-gluon function $B^V(x_1,x_2)$ which parameterizes
$\langle\bar\psi\gamma^+A^\perp\psi\rangle$ matrix element.
On the other hand, the real $B^V(x_1,x_2)$-function kills the contribution from the
non-standard diagram (that is, the diagram with radiations from the
quark legs in the correlators)
which is, however, absolutely necessary
to ensure the QED gauge invariance of the DY hadron tensor.
To resolve this discrepancy, in the series of papers \cite{Anikin:2010wz, Anikin:2015xka}, with the help of the contour gauge conception,
it has been proven that
the twist three $B^V(x_1,x_2)$-function is, in fact, the complex function.
With this inference, the non-standard diagram does give the non-zero contribution and does
produce the imaginary phase required to have the SSA.
Besides, the additional contribution of the non-standard diagram
also leads to an extra factor of $2$ for SSA.

The other motivating point for the development of the contour gauge formalism
is related to the problem of the spurious singularity fixing (see, for example, \cite{Boer:2003cm, Belitsky:2002sm, Chirilli:2015fza}).
The light-cone axial gauge condition imposed on gluon fields, $A^+=0$,
naturally enables the parton number (probability) interpretation of parton density functions in the tree level
\cite{Collins:1981uw}.
However, any perturbative calculations beyond the tree level demand
the careful treatments of the spurious uncertainties in gluon propagators
\cite{Leibbrandt:1987qv, Slavnov:1987yh,Bassetto:1991ue, Bassetto:1984dq}.
It is worth to remind that the spurious singularities
arise as ill-defined pole singularities of the form  like $\sim (k^+)^{-1}$.
Basically, they are associated with the residual gauge freedom due to incomplete gauge fixing
by $A^+=0$.
For this reason, the direct calculations in the axial light-cone gauge in higher perturbative orders
are cumbersome and sometimes even contradictory \cite{Collins:1989gx, Ivanov:1990vy}.
From one hand, one would try
to overcome this difficulty by working in the well-defined general
covariant gauge setting the gauge parameter to
$\xi = - 3 + 0 (\alpha_s)$,
which is known to effectively `imitate' non-covariant gauges \cite{Stefanis:1983ke, Ivanov:1990vy}.
Or, from the other hand, we would follow
another approach to keep working in the light-cone gauge and to get rid of the residual gauge freedom
by an appropriate extra gauge-fixing condition.
The latter can be obtained in terms of the various boundary conditions for the gluon fields and/or
their spatial derivatives \cite{Belitsky:2002sm, Boer:2003cm, Chirilli:2015fza}.

In the present review, we give the detail description of an alternative approach to formulation of the more general
gauge-fixing condition from the very beginning. It is supposed to entail the ``right'' pole
prescriptions for the gluon propagator.
With the help of the contour gauge conception, we
demonstrate how the spurious uncertainties in the gluon propagator can ultimately be fixed
in the nonlocal axial gauges.
Within the framework of collinear factorization, we specially
emphasize the substantial role of the nonstandard diagram
to get the relevant contour in the Wilson
path-ordered exponential needed to fix ultimately the spurious singularity in the gluon propagator.


The other demonstrative example of the practical profits presented in the review is given by
the problem of separation of the parton spin and orbital angular momentum (AM).
It is one of interesting subjects of modern disputes in both the theoretical and experimental communities \cite{Ji:2020ena}.
There are currently two parallel decompositions known as the Jaffe-Manohar decomposition (JM-decomposition)
\cite{Jaffe:1989jz} and Ji's decomposition (J-decomposition) \cite{Ji:1996ek} which
have been widely discussed.
The JM-decomposition refers to the complete expansion of the nucleon spin into
spin and orbital parts of quarks and gluons separately.
While, the J-decomposition is
gauge-invariant by construction and this decomposition does not lead
to the separated quark and gluon contributions of spin and orbital AM to the total spin of the nucleon.

In \cite{Chen:2008ag,Chen:2009mr}, the analogue of JM-decomposition that possesses 
the gauge invariance has been discussed. Working with the Coulomb gauge, 
the gluon field is promoted to be formally presented in the form of
$A_\mu= A_\mu^{\text{phys}}+A_\mu^{\text{pure}}$.
It is worth to notice that this decomposition has being assumed as the fist-step ansatz in all existing discussions on
the gauge-invariant separation of spin AM from the orbital AM
(see, for example,  \cite{Ji:2009fu,Wakamatsu:2014zza,Wakamatsu:2014toa,Wakamatsu:2017isl,Lorce:2012ce,Lorce:2013gxa,Leader:2013jra,Wakamatsu:2011mb,Wakamatsu:2010cb,
Wakamatsu:2010qj,Zhang:2011rn,Bashinsky:1998if}).

In the Abelian $U(1)$ gauge theory the physical components $A_\mu^{\text{phys}}$
correspond to the transverse components $A_\mu^{\perp}$ which are gauge invariant in contrast to the
longitudinal components $A_\mu^{L}$ associated with $A_\mu^{\text{pure}}$ which are gauge-transforming and
they should be eliminated by the gauge condition used in the Lagrangian approach.
As a result, in the Abelian theory, the mentioned decomposition is absolutely natural and
there is no doubt of its validity at all.

In the non-Abelian $SU(3)$ gauge theory, both the transverse and longitudinal components are gauge-transforming.
Hence, the mentioned decomposition is actually questioned regarding the definition of the physical components.
In particular, the use of covariant-type gauge conditions should inevitably lead to the inability to
separate the spin and orbital AMs in the gauge-invariant manner because the coordinate dependence of gluon configurations
cannot be independently determined for every of components, see for example  \cite{Belitsky:2005qn}.
Meanwhile, the decomposition plays a role of keystone in many discussions devoted to the
gauge-invariant separation of spin AM from orbital AM.

In the review, we propose to consider the gluon decomposition as a statement which must be proven, if it is possible,
within the non-Abelian theory.
It turns out that the proof can be implemented and elucidated only with the help of the contour gauge conception
(cf. \cite{ContourG1,ContourG2}).
It is necessary to remind that
the contour gauge can be readily understood in the frame of the Hamiltonian formalism
where the contour gauge condition defines the manifold surface crossing over a group orbit of the fiber uniquely
(see \cite{An-Sym} for further details).
It is also shown that even in the contour gauge one deals with a special kind of residual gauge freedom.
However, this residual gauge is positioned on
the non-trivial boundary formed by pure gauge configurations at infinity.

Note that any physical quantities are gauge-independent, as they should.
The axial gauge is always, as a rule, associated with a certain fixed direction in space.
In this case, the gauge independence should be considered as the independence respecting the choice of direction.
which is provided by additional requirements \cite{Anikin:2009bf}.

\section{The contour gauge in use: the general principles}
\label{Sec:CG-I}

\subsection{The Lagrangian and Hamiltonian systems with the dynamical constraints}
\label{Sec:II}

The local axial gauge, $A^+=0$, suffers from the residual gauge freedom.
Hence, it demands the additional requirements to fix the remained gauge freedom.
In the most of cases, the formulation of additional requirements
is not a trivial task within the Lagrangian system.
Indeed, if one demands simultaneously $A^+=A^-=0$, the maximal gauge fixing is
available only in a classical theory.
In a quantum theory, the simultaneous conditions $A^+=0$ and $A^-=0$
as delta-function arguments in the corresponding
functional integration (which lead to the effective Lagrangian with
$1/\xi_1 (n\cdot A)^2$ and $1/\xi_2 (n^*\cdot A)^2$)
result in the absence of the well-defined gauge propagator.
It happens because the corresponding kinematical operator cannot be inverted.
In this connection, it is necessary to develop
an alternative method of gauge fixing compared to the “classical” approaches with the effective Lagrangian
including both $(n\cdot A)^2$ and $(n^*\cdot A)^2$.
The contour gauge conception gives us such an alternative and effective method.

In order to understand all subtleties of the contour gauge,
we remind the main stages of the Lagrangian (L-system) and Hamiltonian (H-system) approaches to the quantization of gauge fields
(see, for example, \cite{Konopleva:1981ew}).
In this section,  our efforts are concentrated on the demonstration that the H-system is the most adequate approach
for our purposes.
At the same time, due to the special role of a time-space
in H-system, the L-system is more suitable for a practical use.
In other words, we need the H-system as a convenient intermediate instrument to see how
the contour gauge fixes uniquely the total gauge freedom.
However, the main computation procedure has been formulated in terms of L-system related, of course, with the
corresponding H-system.


Let us consider the H-system, defined by $H(p_i, q_i)$, where the phase space $\Gamma$ is formed by
the generalized momenta $p_i$ and coordinates $q_i$. In addition, we have
$2m$ constraints on $p_i$ and $q_i$ which have been imposed on the system.
Traditionally, these $2m$ constraints are denoted by $\varphi_a(p_i, q_i)$ and $\chi_a(p_i, q_i)$.

We suppose that the H-system has an equivalence orbit which is nothing but the gauge group orbit in the gauge theory.
Since the phase space $\Gamma$ is overfilled by unphysical degrees, in the ideal case, we have to resolve all kinds of constraints.
The additional constraints $\chi_a(p_i, q_i)$ are necessary to fix uniquely the orbit representative.
It is needed to quantize the H-system.
From the pure theoretical viewpoint, 
after all restrictions are resolved, we are dealing with a quantized H-system
where the physical phase space $\Gamma^*$ of dimension $2(n-m)$
is a subspace of the original space $\Gamma$ of dimension $2n$. 
This is the groundwork of the H-system where physical configurations exist only,
$H^*(p_i^*, q^*_i)$.

In the functional method of quantization, the amplitude between the initial $| q_1^{in}, ..., q^{in}_n; t^{in} \rangle$ and the
final $\langle q_1^f, ..., q^f_n; t^f |$ states takes the form of (modulo the unimportant normalization factors)
\cite{Konopleva:1981ew,Faddeev:1980be}
\begin{eqnarray}
\label{Amp-1}
&&\langle q_1^f, ..., q^f_n; t^f | q_1^{in}, ..., q^{in}_n; t^{in} \rangle =
\nonumber\\
&& N\,
\int {\cal D} p_i(t)  {\cal D} q_i(t) \, \delta(\varphi_a) \delta(\chi_a)\, \text{det} \big\{ \varphi_a, \chi_a \big\} \times
\nonumber\\
&&\text{exp} \Big\{
i\int_{t_{in}}^{t_{f}} dt \big[ p_i \partial_0 q_i  - H(p_i, q_i) \big]
\Big\},
\end{eqnarray}
where $\{ ..\, , \, ..\}$ denotes the Poisson brackets.

The delta-function $\delta(\varphi_a)$ of Eqn.~(\ref{Amp-1}) can be presented through
the integration over the Lagrange factor $\lambda_a$ as
\begin{eqnarray}
\label{Lag-1}
\delta(\varphi_a)= \int (d\lambda_a) \,e^{i \lambda_a \, \varphi_a(p_i, q_i)}.
\end{eqnarray}
It gives the generalized Hamiltonian of system which reads
\begin{eqnarray}
\label{H-prime}
H^\prime (p_i, q_i) = H(p_i, q_i) + \sum_a \, \lambda_a \, \varphi_a(p_i, q_i).
\end{eqnarray}

If we suppose that the constraint conditions (see the delta-function arguments)
have somehow been resolved, the amplitude is given by
\begin{eqnarray}
\label{Amp-1-2}
&&\langle q_1^f, ..., q^f_n; t^f | q_1^{in}, ..., q^{in}_n; t^{in} \rangle =
 N\,
\int {\cal D} p^*_i(t)  {\cal D} q^*_i(t) \times
\nonumber\\
&&
\text{exp} \Big\{
i\int_{t_{in}}^{t_{f}} dt \big[ p^*_i \partial_0 q^*_i  - H^*(p^*_i, q^*_i) \big]
\Big\},
\end{eqnarray}
where only the physical generalized momenta and coordinates are forming the integration measure together with the
Hamiltonian. The generating functional of (\ref{Amp-1-2}) corresponds to the H-system
with the dynamical constraints which have been resolved. Therefore,
there is no a (gauge) freedom associated with the arbitrary Lagrange factor $\lambda_a$.
It would be an ideal situation which cannot be realized practically in the most of cases.
However, the contour gauge, as a class of nonlocal gauges, gives a possibility to realize
the mentioned ideal situation. It is so because the contour gauge condition has a unique solution by construction, see below.

\begin{figure}[t]
\centerline{\includegraphics[width=0.6\textwidth]{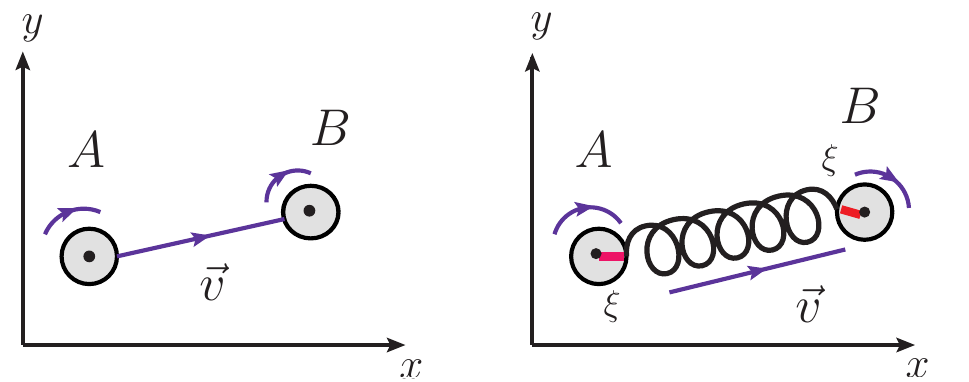}}
\caption{The H-system with unresolved and resolved constrains. The left panel represents
the system with the inner (gauge) symmetry; the right panel corresponds to
the system with the fixed (preferable) site $\xi$ on the ball surface.}
\label{Fig-S-1}
\end{figure}

It is instructive to illustrate the difference between the H-system with unresolved and resolved constraint
conditions. It can be done with the help of a trivial mechanical example, see Fig.~\ref{Fig-S-1}.
Consider the homogenous flat ``ball'' which moves from the point $A$ to the point $B$.
The ball has a spherical symmetry under the rotation around the inertia center.
For the sake of simplicity, we focus on the ball rotation in a two-dimensional plane.
Since the flat ball surface is homogenous, we do not have a chance to observe
the rotation of the moving ball unless we mark some point on the surface.
The invisible ball rotation around its center of inertia corresponds to the inner (gauge)
transformations of the H-system. It does not affect much the trajectory provided
the angle velocity is constant, see the left panel of Fig.~\ref{Fig-S-1}.
In this case, the inertia center plays a role of ``physical'' configurations of the H-system, while
the moving of different sites on the flat ball surface is invisible and relates to the ``unphysical'' configurations.

If we mark the site on the flat ball surface by a dash, we break down the
rotation symmetry. It means that we choose, so to say, the preferable site of the ball surface
and the ball rotation becomes visible.
In this case, we can describe the moving dash together with the inertia center
as ``physical'' configurations of the H-system when the ball position varies from $A$ to $B$,
see the right panel of Fig.~\ref{Fig-S-1}.

In the context of the contour gauge conception, the marked dash on the flat ball corresponds to the
certain gauge function $\theta$ of the gauge transformations which has been fixed by the contour gauge condition.
If we focus on the above mentioned ideal case of resolved constraints,
we would merely deal with the gauge fields considered as a massless vector fields
(transforming as a spinor of $2$-rank under the Lorentz group) that are described by the Hamiltonian
without the gauge transforms.

In both the Abelian and non-Abelian gauge theory,
the correspondences between
the canonical variables and the dynamical constraints (cf. Eqn.~(\ref{Amp-1}))
can be expressed as
\begin{eqnarray}
\label{Rep-1}
\big( p_i;\, q_i  \,; t \big) &\Longrightarrow& \big( (E_0, E_i) ; \, (A_0, A_i) \,; x \big),
\nonumber\\
\delta(\varphi_a) &\Longrightarrow& \delta\left(  \partial_i E_i \right) \, \delta\left( E_0\right),
\nonumber\\
\delta(\chi_a) &\Longrightarrow& \delta\left(  \partial_iA_i \right) \, \delta\left( A_0\right),
\nonumber\\
\text{det} \big\{ \varphi_a, \chi_a \big\}  &\Longrightarrow& \Phi(A).
\end{eqnarray}

%
%
%


In the typical gauge theories, resolving the additional (or gauge) conditions is not a simple task.
The gauge conditions, leading to the equation system for the gauge function $\theta(x)$, would have an unique
solution $\theta_0$. In fact, it might be practically impossible.

Within the frame of L-system, Faddeev and Popov have proposed
the method (FP-method)
to avoid the needs for finding the unique solution $\theta_0$.
In FT-method, the infinite group orbit volume can be factorized out to the insubstantial normalization factor.
It becomes thanks to the gauge invariance
of the corresponding Lagrangian (or Hamiltonian) of theory  \cite{Faddeev:1980be}.

For the simplicity, we focus on $U(1)$ gauge theory.
The H-system can be traced from the Lagrangian of first-order formalism. The first-order Lagrangian is determined by
\begin{eqnarray}
\label{Lag-1ord}
{\cal L}_{(1)}= -\frac{1}{2} \Big[
\partial_\mu A_\nu(x) - \partial_\nu A_\mu - \frac{1}{2}F_{\mu\nu}
\Big] F_{\mu\nu},
\end{eqnarray}
where $A_\mu$ and $F_{\mu\nu}$ are supposed to be independent field configurations. The Lagrangian of Eqn.~(\ref{Lag-1ord})
can be written in $3$-dimensional form. We have
\begin{eqnarray}
\label{Lag-1ord-2}
{\cal L}_{(1)}= E_i \,\partial_0A_i + A_0 \,\partial_i E_i + \frac{1}{2} E^2
+ \frac{1}{4} B_i \, \epsilon_{i k \ell }\partial_k A_\ell + \frac{1}{2} B^2,
 \end{eqnarray}
where  $A_i$ and $E_i=F_{i 0}$ are coordinates and the generalized momenta, respectively.
$B_i=1/2\,\epsilon_{i j k} F_{jk}$ are the variables in the integration measure, see below.
Using the Euler-Lagrange equations, one gets the equations which are not including $\partial_0$.
These equations define the following constraint conditions: 
\begin{eqnarray}
\label{Const-Cond}
&&E_0=0, \quad \partial_i E_i =0 
\end{eqnarray}
giving the primary and secondary conditions, respectively.

We remind that the gauge theory is nothing but
the theory with the (dynamical) constraints applied on the fields.
In addition to the constraint conditions of Eqn.~(\ref{Const-Cond}), we have to introduce by the supplementary (or gauge) conditions.
They read
\begin{eqnarray}
\label{Gauge-Cond}
A_0=0 \,\,\, \text{and}\,\,\,   \partial_i A_i =0.
\end{eqnarray}
The full set of conditions defined by Eqns,~(\ref{Const-Cond}) and (\ref{Gauge-Cond})
should eliminate all unphysical degrees of freedom for the correct quantization of H- (or L-) system.

Using the FP-method, we begin with the functional integration written for the L-system \cite{Konopleva:1981ew}.
We have
\begin{eqnarray}
\label{FI-L-1}
\int {\cal D}\theta(x) \, \int {\cal D}A_\mu \,\delta\left( F[A] \right) \Phi(A)\, e^{i S[A]},
\end{eqnarray}
where the infinite group orbit volume given by ${\cal D}\theta(x)$
has been factorized out in the integration measure
due to the gauge invariant action $S[A]$ and functional $\Phi(A)$.

The exact magnitude of the group volume prefactor, {\it i.e.}
\begin{eqnarray}
\label{G-orbit}
v_g\equiv \int {\cal D}\theta(x) =\int \prod_x [d\theta(x)]= \big\{ \infty, \,0\big\},
\end{eqnarray}
is irrelevant because
the prefactor should be cancelled by the corresponding normalization of Green functions.
In Eqn.~(\ref{G-orbit}), the infinite group volume corresponds to the standard case of unresolved gauge conditions.
The zero group volume appears in the case of resolved gauge conditions.
Indeed, $[d\theta(x)]$ is defined on a group manifold as an invariant measure. That is, we have 
\begin{eqnarray}
\label{G-inv-m}
\int [d\theta] f(\theta) \sim \sum_{\theta\in G} f(\theta),
\end{eqnarray}
where $\prod_x [d\theta(x)]$ is the product of the invariant measures
on the structure group $G$ of fiber corresponding to every point of the Minkowski space.
If we suppose that the gauge function $\theta$ is unfixed, we obtain the infinite integration over
the group invariant measure.
Otherwise, if the gauge function $\theta$ is somehow fixed, the integration is equal to zero for
every point of the Minkowski space. The value of fixed $\theta_i$ can, however, vary from one point to another.

Returning to Eqn.~(\ref{FI-L-1}), it can be identically rewritten as
\begin{eqnarray}
\label{FI-L-2}
\int {\cal D}A_\mu {\cal D} F_{\mu\nu} \,\delta\left( F[A] \right) \Phi(A)\, e^{i S[A, F]},
\end{eqnarray}
where the functional of action $S[A, F]$ is given by the Lagrangian of Eqn.~(\ref{Lag-1ord}).
In the three-dimensional forms, we obtain that
\begin{eqnarray}
\label{FI-L-3}
\int {\cal D}A_i  {\cal D}A_0 \,
 {\cal D} B_i   {\cal D} E_i \,\delta\left( F[A] \right) \Phi(A)\, e^{i S[A, E, B]},
\end{eqnarray}
where $S[A, E, B]$ is now defined by Eqn.~(\ref{Lag-1ord-2}).
Integrations over $B_i$ and $A_0$ in Eqn.~(\ref{FI-L-3}) lead to the functional integral which is given by
\begin{eqnarray}
\label{FI-L-4}
&&
\int {\cal D}A_i \,
{\cal D} E_i \, \delta\left( F[A] \right) \Phi(A) \delta\left(  \partial_i E_i \right)
\times
\nonumber\\
&&
\text{exp}\Big\{
i \int dz \Big[
E_i \,\partial_0A_i + H(E_i, A_i)
\Big]
\Big\},
\end{eqnarray}
where the Hamiltonian is
\begin{eqnarray}
\label{H-1}
H(E_i, A_i) = \frac{1}{2} E^2
- \frac{1}{2} \left( \epsilon_{i j k} \partial_j A_k\right)^2.
\end{eqnarray}
In Eqn.~(\ref{FI-L-4}),
the gauge condition is chosen to be $F(A)= \partial_i A_i$.

On the other hand, Eqn.~(\ref{FI-L-4}) can be presented in the equivalent form. It reads
\begin{eqnarray}
\label{FI-L-5}
&&
\int {\cal D}A_i  {\cal D}A_0\,
{\cal D} E_i {\cal D} E_0\, \delta\left(  \partial_iA_i \right) \Phi(A) \delta\left(  \partial_i E_i \right)
\delta\left( A_0\right)\delta\left( E_0\right)
\times
\nonumber\\
&&
\text{exp}\Big\{
i \int dz \Big[
E_i \,\partial_0 A_i + E_0\partial_0 A_0 + H(E_i, A_i; E_0, A_0)
\Big]
\Big\},
\end{eqnarray}
where the primary and secondary constraints together with the gauge conditions of Eqn.~(\ref{Gauge-Cond})
have been explicitly shown in the functional integrand.
This representation of H-system resembles the functional integration presented by Eqn.~(\ref{Amp-1}).

Notice that, in H-system, there are no problems to write all constraints through the delta functions.
It is true because, in contrast to L-system which is forming the Feynman rules,
we have no needs to invert the kinematical-like operator.
At the same time, the H-system approach is not convenient for the practical computation in QFT.

This section is basically written in the textbook style.
The basic reason for the usage of this style is that
the section is preparing a reader for the main features of contour gauge uses.
Indeed, we have reminded the differences between the H-systems with and without resolved additional
dynamical constraints presenting the mechanical illustration in Fig.~(\ref{Fig-S-1}).
In the framework of the FP-method for both L- and H-systems, we do not need to find a unique solution
of the gauge condition respecting the gauge function $\theta(x)$.
Owing to the gauge invariance of Lagrangian (Hamiltonian) and $\Phi(A)$,
the volume of infinite group orbit (defined by $[d\theta]$) has been separated out in the 
functional integration measure. It gives a possibility to quantize the gauge theory
unless the residual gauge problems present.
If the gauge condition is resolved with respect to $\theta$, the group representative corresponding to each orbits
should be uniquely fixed. Hence, the separated (or, in other words, factorized) group integral should be equal to zero
in a sense of the Riemann summations.
Generally speaking, the group volume has to be cancelled by the suitable normalization and, therefore,
it does not lead to the nullification of the functional integration of Eqn.~(\ref{FI-L-1}) at all.
This just-mentioned case takes place in the approach where the contour gauge conception has been used.

\subsection{The basic idea of contour gauge approach}
\label{Sec:III}

We now describe the basic idea of the contour gauge conception.
Within the contour gauge approach, 
the gauge condition is assumed to be resolved, at least, in the formal manner and
the unique gauge orbit representative is found (see, for example,
\cite{ContourG1, ContourG2,Shevchenko:1998uw}).

We claim that the gauge function can be completely fix (in the H-system, see Eqn.~(\ref{Amp-1})) and
the unphysical gluons can be eliminated (in the L-system, see Eqn.~(\ref{FI-L-1}))
provided by the following condition:
\begin{eqnarray}
\label{CG-1}
{\bf g}(x | A)\Big |_{P(x_0,x)}
\equiv\mathbb{P}\text{exp} \Big\{ ig\int_{P(x_0,x)} d\omega_\mu A_\mu(\omega)\Big\}={\bf 1}
\end{eqnarray}
where the path $P(x_0,x)$ between the points $x_0$ and $x$ is fixed.
That is, we demand that the path dependent functional (or, in other words, the Wilson path functional)
should be equal to unity.

In QCD, the axial light-cone gauge, $A^+=0$, is a particular case of the reduced nonlocal contour gauge, see below,
determined by Eqn.~(\ref{CG-1}) if the fixed path is given by the straightforward line
connecting $\pm\infty$ with $x$.
By construction, the contour gauge does not possess the residual gauge freedom in the finite space.
In \cite{Anikin:2021oht}, it is shown that the possible residual gauge
can be located at the corresponding boundary only.
In what follows the boundary gluon configurations have assumed to be equal to zero.
So, from the technical point of view, the contour gauge provides the simplest way to fix the gauge function completely.

The contour gauge conception is inspired by the path group formalism \cite{Mensky:2012iy,Mensky-book}.
Also, it can be traced from the Mandelstam approach \cite{Mandelstam:1962mi}.
For better understanding, it is worth to give a short introduction to the
geometry of gluons where the gluon field has been treated as a connection on the
principle fiber bundle ${\cal P}({\cal X}, \pi \,|\, G)$.

The principle fiber bundle ${\cal P}({\cal X}, \pi \,|\, G)$ is one of main ingredients that forms a
differential geometry basic. By definition, it is a combination of two sets, ${\cal P}$ and ${\cal X}$ together with
the given transformations $\pi$ between them. Moreover, the group structure $G$ is determined on the set ${\cal P}$
(strictly speaking,  on each of fibers). The set ${\cal X}$ is named by a base of the fiber bundle.
The base usually coincides with the Minkowski space.
 In the principle fiber bundle, we are able to define two directions. One direction is determined in the base ${\cal X}$
as the tangent vector of a curve going through the point $x\in {\cal X}$.
The other direction is defined in the fiber and can be uniquely determined as the
tangent subspace related to the parallel transport \cite{Mensky:2012iy,Mensky-book}.
These two directions allow us to introduce the horizontal vector defined by
\begin{eqnarray}
\label{H-vector}
H_\mu = \frac{\partial}{\partial x^\mu} -ig A^a_\mu(x) \, D^{a},
\quad D^{a}= {\bf D}^a \cdot {\bf g} \frac{\partial}{\partial {\bf g}},
\end{eqnarray}
where $D^a$ denotes the corresponding shift generator along the group fiber written in the differential form.
The vector coefficients (connection of the principle fiber bundle), $A^a_\mu(x)$, defines the algebraic vertical (tangent) vector field on the fiber \cite{Mensky:2012iy,Mensky-book}.
The horizontal vector $H_\mu$ is invariant under the structure group $G$ acting on the given representation
of the fiber by construction.

In ${\cal P}({\cal X}, \pi \,|\, G)$, the functional
${\bf g}(x | A)$ of Eqn.~(\ref{CG-1})
is a solution of
the parallel transport equation given by
\begin{eqnarray}
\label{H-vector-2}
\frac{d x_{\mu}(s)}{d s} H_\mu(A) {\bf g}(x(s)| A) = 0,
\end{eqnarray}
where the fiber point  $p(s)=\left( x(s), {\bf g}(x(s)) \right)$
with the curve $x(s)\in {\cal X}$ parametrized by $s$.
Eqn.~(\ref{H-vector-2}) being a differential equation takes place even if  ${\bf g}$ is fixed on the group.
it is true because, in this case, $D^a {\bf g}=0$ while $A^a_\mu(x)\not = 0$.
In this connection, the condition presented by Eqn.~(\ref{CG-1}) implies that the full curve-linear integration
goes to zero rather than the integrand itself.

\begin{figure}[t]
\centerline{\includegraphics[width=0.6\textwidth]{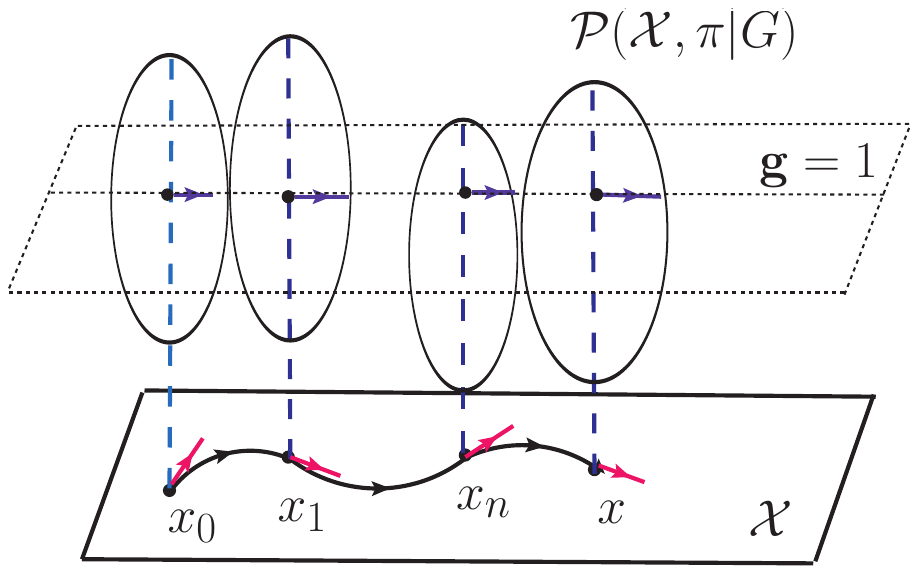}}
\vspace{-6.5cm}
\caption{The contour gauge: the plane of ${\bf g}=1$ in the principle fiber bundle ${\cal P}({\cal X}, \pi | G)$. }
\label{Fig-S-2}
\end{figure}

It can be shown \cite{Mensky:2012iy, Mensky-book} that every of points belonging to the fiber bundle, ${\cal P}({\cal X}, \pi \,|\, G)$,
has one and only one horizontal vector corresponding to the given tangent vector at $x\in {\cal X}$.
We remind that the tangent vector at the point $x$ is uniquely determined by the
given path passing through $x$.
In the frame of H-system based on the geometry of gluons
the condition of Eqn.~(\ref{CG-1}) corresponds to the determining of the surface
on ${\cal P}({\cal X}, \pi \,|\, G)$ that is parallel to the base plane with the path.
Moreover, it singles out the identity element, ${\bf g}=1$, in every fibers of ${\cal P}({\cal X}, \pi \,|\, G)$,
see Fig.~\ref{Fig-S-2}.
This choice can be traced to the Lagrange factor $\lambda_a$ which is formally
fixed in H-system, see Sec.~\ref{Sec:II}.
Roughly speaking, once the group (any) element of fiber is fixed, we deal with the Lagrange factor $\lambda_a$
of H-system which is also uniquely fixed.
On the other hand, if we fix the group element of fiber we fix the function theta of gauge transforms as well.
In this sense, we do not have the local gauge transforms (or the gauge freedom) anymore.

The path-dependent functional, see the {\it l.h.s.} of (\ref{CG-1}), can be also gauge-transformed. It reads
\begin{eqnarray}
\label{G-tr-wl}
&&\mathbb{P}\text{exp} \Big\{ ig\int_{P(x_0,x)} d\omega_\mu A^\theta_\mu(\omega)\Big\}=
\nonumber\\
&&
\omega(x)
\mathbb{P}\text{exp} \Big\{ ig\int_{P(x_0,x)} d\omega_\mu A_\mu(\omega)\Big\}
\omega^{-1}(x_0),
\end{eqnarray}
where the local gauge function is
\begin{eqnarray}
\label{U}
\omega(x)=e^{i\theta(x)}\equiv e^{iT^a \theta^a(x)}
\end{eqnarray}
with the corresponding generator $T^a$. From this, one can see if the Minkowski space has been realized as a loop space,
the path dependent functional becomes invariant under the local gauge transforms.
In the general case of arbitrary paths, imposing the condition (\ref{CG-1}) on the {\it r.h.s.} of (\ref{G-tr-wl}),
we are able to get the different contour gauge ${\bf g}(x | A)=C$
with $C=\omega(x)\omega^{-1}(x_0)$, see (\ref{U}). From the theoretical point of view, this new contour gauge
has the same status as the gauge of (\ref{CG-1}).
It would correspond to the different plane ${\bf g}=C$ in Fig.~\ref{Fig-S-2} which also transects the principle fiber bundle
and, therefore, it generates the different contour gauge condition which, generally speaking, is related to the previous contour gauge by the
local transform. However, from the practical point of view, the contour gauge
given by ${\bf g}(x | A)=C$ is not convenient to use for calculations because the representation of transverse gluon field
through the strength tensor has a more complicated form compared to, for example, (\ref{Ag1}).
Of course, the physical quantities are independent on the contour gauge choice.

Notice that since the gauge condition given by Eqn.~(\ref{CG-1}) selects only the
identity element of group $G$ on each fiber, it means that the gauge transforms have been reduced to the ``global''
gauge transforms. That is, the gauge function $\theta(x)$ becomes the coordinate independent and is fixed $\theta_0$.
This situation is identical to that one can see in Fig.~\ref{Fig-S-1} of Sec.~\ref{Sec:II}.
Namely, the red dash on the ball surface corresponds to
one particular choice of the group element fixed by the given contour gauge. However, if we mark the line on the ball surface
by the other dash, it would mean that we choose the other group element fixed by the different contour gauge.
In both cases, we deal with the same description of H-system.

Since the functional ${\bf g}(x | A)$ depends on the whole path in ${\cal X}$,
the contour gauge refers to the non-local class of gauges and
generalises naturally the familiar local axial-type of gauges.
It is also worth
to notice that two different contour gauges can correspond to the same
local (axial) gauge with the different fixing of residual gauges \cite{Anikin:2010wz, Anikin:2015xka}.
If we would consider the local gauge without the connection with the non-local gauge,
the residual gauge freedom would require the extra conditions to fix the given freedom.
This statement reflects the fact that, in contrast to the local axial gauge,
the contour gauge does not possess the residual gauge freedom in the finite region of a space
where the boundary gluon fields are absent \cite{Anikin:2021oht}.
%
%
%
\subsection{Comparison of local and nonlocal gauge}
\label{Sec:IV}

The contour gauge as a nonlocal kind of gauges generalizes or extends the standard local gauge of axial type.
Therefore, it is worth to discuss shortly the correspondence between the local and nonlocal gauge transforms.

Let us begin with the axial local gauge defined by $A^+(x)=0$.
The nonlocal gauge is given by Eqn.~(\ref{CG-1}) and
the local gauge can be obtained from  Eqn.~(\ref{CG-1}) if, as above-mentioned, the starting point is $x_0=-\infty$
and the path is fixed to be a straightforward line, $P=[-\infty, x]$.
The differences between the local and nonlocal gauges can symbolically be demonstrated by the
following trivial example.
Consider two different vectors $A$ and $B$. We stress that they are different by construction.
We now assume that these vectors
have the same projection on the certain direction given by another vector $N$.
It is clear that if the vectors have the same projections on the vector $N$,
it does not mean the equivalence of vectors $A$ and $B$, {\it i.e.}
\begin{eqnarray}
\Big\{ \text{if}\,\,\, (A\cdot N) = (B\cdot N)\Big\} \, \,\,\nRightarrow \Big\{ A=B \Big\}
\end{eqnarray}
In this example, the different vectors $A$ and $B$ can be associated with two different contour nonlocal gauges,
\begin{eqnarray}
\label{A-B-1}
A\Leftrightarrow  \Big\{ {\bf g}(x | A)\Big |_{P=[-\infty, x]}={\bf 1}\Big\}
\quad \text{and} \quad
B\Leftrightarrow \Big\{ {\bf g}(x | A)\Big |_{P=[+\infty, x]}={\bf 1}\Big\} .
\end{eqnarray}
While the local axial gauge plays a role of the projections on $N$,
\begin{eqnarray}
\label{A-B-2}
\Big\{ (A\cdot N) = (B\cdot N)\Big\} \Leftrightarrow \Big\{ A^+(x)=0 \Big\}.
\end{eqnarray}
On the other hand, focusing on only the $N$ vector projection, there is no additional information
to see from which vectors $A$ or $B$ the given projection has been performed.

We are coming back to the local axial gauge.
The local axial gauge suffers from the residual gauge transformations
which can lead to the
spurious pole uncertainties.
However, the preponderance of nonlocal (contour) gauge is that it fixes all the gauge freedom in the finite space
(see for instance \cite{Anikin:2021oht}).
To demonstrate it in the simplest way, we consider the local axial gauge as an equation on the gauge function $\theta(x)$.
We have
\begin{eqnarray}
\label{LGT-1}
A^{ +\, \theta}(x)= \omega(x)A^+ \omega^{-1}(x) + \frac{i}{g} \omega(x)\partial^+ \omega^{-1}(x)=0,
\end{eqnarray}
where $x=(x^-, \tilde x)$ with $\tilde x= (x^+, {\bf x}_\perp)$.
We can readily find a solution of this equation which takes the form of
the undetermined integration given by
\begin{eqnarray}
\label{Sol-gauge-0}
&&\omega_0(x)=C \, \mathbb{P}\text{exp}
\Big\{  - ig\int dx^- A^+(x)\Big\}.
\end{eqnarray}
At the same time, the solution $\omega_0(x)$ can be rewritten via the determined integration, it reads
\begin{eqnarray}
\label{Sol-gauge}
&&\omega_0(x^-, \tilde x)=C(\tilde x) \, \overline{\omega}(x^-, \tilde x),
\nonumber\\
&&
\overline{\omega}(x^-, \tilde x)= \mathbb{P}\text{exp}
\Big\{  - ig\int_{x_0^-}^{x^-} dz^- A^+(z^-, \tilde x)\Big\}.
\end{eqnarray}
Here, $x^-_0$ is fixed and $C(\tilde x)$ is an arbitrary function which does not depend on
$x^-$.  $C(\tilde x)$ is given by
\begin{eqnarray}
\label{C-norm}
\omega_0(x^-_0, \tilde x)=C(\tilde x).
\end{eqnarray}
The arbitrariness of $C$-function also reflects the fact that we deal with an arbitrary fixed starting point $x_0$.

Let us study the residual gauge freedom requiring both $A^{+,\, \theta}(x)=0$ and $A^{+}(x)=0$,
we then have
\begin{eqnarray}
\label{Sol-gauge-2}
&&\omega_0(x^-, \tilde x)\Big|_{\overline{\omega}=1}\equiv \omega^{\text{res}}(\tilde x)=
C(\tilde x)\equiv
e^{i\,\tilde\theta(\tilde x)}.
\end{eqnarray}
One can see that the function $C$ determines the residual gauge transforms.
This situation takes place in the local axial gauge defined by the only condition $A^+=0$
applied for (\ref{Sol-gauge}).

We go over to the nonlocal gauge which actually gives more information
on the gauge fixing.
The nonlocal (contour) gauge extends the local axial-type gauge
and it demands that the full (curve-linear) integral in the exponential of Eqn.~(\ref{CG-1})
has to go to zero. We stress that  in contrast to the local gauge where the corresponding exponential disappears thanks to
that the integrand $A^+$ goes to zero. One can demonstrate it on the example of solutions (\ref{Sol-gauge-0}) or
(\ref{Sol-gauge}).
Indeed, in the contour gauge
the residual gauge function $\tilde\theta(\tilde x)$ can be related to the path dependent functionals with  $A^-$ and $A^i_\perp$
which are also disappeared eliminating the gauge freedom and
giving the physical gluon representation in the form of (\ref{A-cg}) (see \cite{Anikin:2016bor,Belitsky:2002sm} for details).
That is, if we restore the full path in the path dependent functional for a given process, we can get that
\begin{eqnarray}
\label{C-full}
&&C(\tilde x)=\tilde C(x^+_0, x^-_0, {\bf x}^\perp_0)
\nonumber\\
&&\times
\mathbb{P}\text{exp}
\Big\{ ig\int_{{\bf x}_0^\perp}^{{\bf x}^\perp} d \omega_\perp^i A^i_\perp(x^+_0, x^-_0, \omega_\perp)\Big\}
\nonumber\\
&&\times
\mathbb{P}\text{exp}
\Big\{ -ig\int_{x_0^+}^{ x^+} d \omega^+ A^-(\omega^+, x^-_0, {\bf x}_\perp)\Big\}.
\end{eqnarray}
Then, requiring the conditions as
\begin{eqnarray}
\label{Con-cg-1}
&&A^-(\omega^+, x^-_0, {\bf x}_\perp)=0,
\\
&&
\label{Con-cg-2}
\int_{{\bf x}_0^\perp}^{{\bf x}^\perp} d \omega_\perp^i A^i_\perp(x^+_0, x^-_0, \omega_\perp)=0,
\end{eqnarray}
 we get that
\begin{eqnarray}
\label{C-full-2}
C(\tilde x)\Big|_{\text{c.g.}}=\tilde C(x^+_0, x^-_0, {\bf x}^\perp_0).
\end{eqnarray}
In the contour gauge, Eqn.~(\ref{C-full-2}) means that no the gauge freedom has left at all.
We emphasize that the condition of (\ref{Con-cg-1}) demands that the integrand is zero,
 while the condition of (\ref{Con-cg-2}) is imposed on the integration which leads to
 the corresponding representation for the transverse gluon field, see (\ref{Ag1}),  (\ref{Ag2}) and
 (\ref{A-cg}). Besides,
the exact value of the fixed starting point $x_0$ depends on the process under
our consideration \cite{Belitsky:2002sm}.

The path dependent functional also defines the path dependent gauge transformation
in the form of
\begin{eqnarray}
\label{p-g}
&&A^{p.g.}_\mu(x) =
{\bf g}^{-1}(x | A)\Big |_{P(x_0,x)}
A_\mu(x) {\bf g}(x | A)\Big |_{P(x_0,x)} +
\nonumber\\
&&
\frac{i}{g}
{\bf g}^{-1}(x | A)\Big |_{P(x_0,x)} \partial_\mu {\bf g}(x | A)\Big |_{P(x_0,x)},
\end{eqnarray}
where the starting point $x_0$ is now fixed.
Hence, having calculated the derivative of ${\bf g}$ in (\ref{p-g}), we get that
(here, for the simplicity, $A_\mu(-\infty)=0$)
\begin{eqnarray}
\label{p-g-1}
A^{p.g.}_\mu(x)=
\int_{-\infty}^x dz_\alpha \frac{\partial z_\beta}{\partial x_\mu}
{\bf g}^{-1}(z | A)\Big |_{P(-\infty,z)}
\, G_{\alpha\beta} (z | A)\,
{\bf g}^{-1}(z | A)\Big |_{P(-\infty,z)},
\end{eqnarray}
where $x_0=-\infty$.
If we are now focusing on the case of $x=(0^+, x^-, {\bf 0}_\perp)$, we readily obtain that
the gluon representation reads
\begin{eqnarray}
\label{p-g-2}
A^{p.g.}_\mu(x^-)=
\int_{-\infty^-}^{x^-} dz^-
[z^-; -\infty^-]_{A^+}^{-1}
G^{+\mu} (z^- | A)\, [z^-; -\infty^-]_{A^+},
\end{eqnarray}
where 
\begin{eqnarray}
\label{WL-def}
[z_2 ;\, z_1]_{A}= {\bf g} (z_2 | A) \Big|_{P(z_1, z_2)}.
\end{eqnarray}
The representation of Eqn.~ (\ref{p-g-2}) is a very important result for our further considerations.

\subsection{The advantages of the contour gauge use}
\label{Sec:V}

In the preceding subsection, 
it is shown that the non-local contour gauge is able to
illuminate the unphysical gluons. Meanwhile, the local axial gauge fixes the
gauge only partially and the residual gauge freedom is still presented.
In this subsection, we concentrate 
on the certain examples which, first, relate to the practical use of the contour gauge
and, second, demonstrate the preponderance of the nonlocal gauges compared to the local gauges.

As discussed in \cite{Anikin:2010wz,Anikin:2015xka,Anikin:2016bor}, the Drell-Yan-like processes
with the polarized hadrons give the unique example where
the contour gauge use shows the definite advantage from the practical point of view.
In particular, the contour gauge use allows to find the new contributions
to the Drell-Yan-like hadron tensor
which restore and ensure the gauge invariance of the corresponding hadron tensors \cite{Anikin:2010wz,Anikin:2015xka,Anikin:2016bor}.
It is important, however, to stress that the mentioned new contributions are invisible if we would work within the
frame in the local gauge.

In the similar manner, due to the contour gauge conception
the $\xi$-process of DVCS-amplitude which clarifies the gauge invariance
of the non-forward processes takes the closed form again  \cite{Anikin:2020ipg}.
From the practical point of view, it is instructive to consider the appearance of
standard and non-standard diagrams contributing to the well-known deeply virtual Compton scattering (DVCS) amplitude
in the frame of the factorization procedure.
The gluons radiated from the internal quark of the hard subprocess
generate the standard diagrams, while the non-standard diagrams are formed by
the gluon radiations from the external quark of the hard subprocess (see \cite{Anikin:2020ipg} for details).

In the most cases, it is sufficient to exponentiate only the longitudinal components of gluon field, $A^-$ and $A^+$,
which are related to the unphysical degrees.
Indeed,
the standard diagram contributions give
the gauge invariant quark string operator which reads
\begin{eqnarray}
\label{OP1}
\bar\Psi^{(st)}(0^+,0^-,{\bf 0}_\perp | A^+)\,
\big\{ \gamma \big\}\,
\Psi^{(st)}(0^+,z^-,{\bf 0}_\perp | A^+),
\end{eqnarray}
where
\begin{eqnarray}
\label{Psi}
&&\bar\Psi^{(st)}(0^+,0^-,{\bf 0}_\perp | A^+)=
\nonumber\\
&&\bar\psi(0^+,0^-,{\bf 0}_\perp) [0^+,0^-,{\bf 0}_\perp; \, 0^+,+\infty^-,{\bf 0}_\perp]_{A^+},
\\
&&\Psi^{(st)}(0^+,z^-,{\bf 0}_\perp | A^+)=
\nonumber\\
&&
[0^+,+\infty^-,{\bf 0}_\perp; \, 0^+,z^-,{\bf 0}_\perp]_{A^+} \psi(0^+,z^-,{\bf 0}_\perp)
\end{eqnarray}
and $\big\{ \gamma \big\}$ stands for $\gamma$-matrices the exact form of which is now irrelevant.

The non-standard diagram contributions result in the string operator defined as
\begin{eqnarray}
\label{OP2}
\bar\Psi^{(non-st)}(0^+,0^-,{\bf 0}_\perp | A^-)
\, \big\{\gamma\big\}\,
\Psi^{(non-st)}(0^+,z^-,{\bf 0}_\perp | A^-),
\end{eqnarray}
where
\begin{eqnarray}
\label{Psi-non-st}
&&\bar\Psi^{(non-st)}(0^+,0^-,{\bf 0}_\perp | A^-)=
\nonumber\\
&&\bar\psi(0^+,0^-,{\bf 0}_\perp) [-\infty^+,0^-,{\bf 0}_\perp; \, 0^+,0^-,{\bf 0}_\perp]_{A^-},
\\
&&\Psi^{(non-st)}(0^+,z^-,{\bf 0}_\perp | A^-)=
\nonumber\\
&&[0^+,z^-,{\bf 0}_\perp; \, -\infty^+,z^-,{\bf 0}_\perp]_{A^-} \psi(0^+,z^-,{\bf 0}_\perp).
\end{eqnarray}
Hence, using the contour gauge conception,
we eliminate all the Wilson lines with the longitudinal gluon fields $A^+$ and $A^-$
demanding that
\begin{eqnarray}
\label{C-G-2}
&&[0^+,0^-,{\bf 0}_\perp; \, 0^+,+\infty^-,{\bf 0}_\perp]_{A^+} = {\bf 1},
\nonumber\\
&&[0^+,+\infty^-,{\bf 0}_\perp; \, 0^+,z^-,{\bf 0}_\perp]_{A^+} = {\bf 1}
\end{eqnarray}
and
\begin{eqnarray}
\label{C-G-3}
&&[-\infty^+,0^-,{\bf 0}_\perp; \, 0^+,0^-,{\bf 0}_\perp]_{A^-} = {\bf 1},
\nonumber\\
&&[0^+,z^-,{\bf 0}_\perp; \, -\infty^+,z^-,{\bf 0}_\perp]_{A^-} = {\bf 1}.
\end{eqnarray}
Eqns.~(\ref{C-G-2}) and (\ref{C-G-3}) give rise to the local gauge conditions given by $A^+=0$
and $A^-=0$.

With respect to the Wilson line with the transverse gluons $A^i_\perp$, we remind
that we work here within
the factorization procedure applied for DVCS-amplitude.
In this case, the Wilson lines with the transverse gluon fields
are considered in the form of an expansion due to the fixed twist-order,
and the transverse gluons correspond to the physical configurations of L-system.

Thus, the DVCS process gives us the example how the unphysical gluon degrees can be
illuminated from the consideration with the help of contour gauge.

We are now going over to the Drell-Yan (DY) process with one transversely polarized hadron \cite{Anikin:2010wz}
\begin{eqnarray}
\label{DY}
N^{(\uparrow\downarrow)}(p_1) + N(p_2)\Rightarrow
\ell(l_1) + \bar\ell(l_2) + X(P_X)\,,
\end{eqnarray}
where the virtual photon producing the lepton pair ($l_1+l_2=q$) has a large offshelness, {\it i.e.}
$q^2=Q^2\to \infty$,
while all the transverse momenta are small and integrated out in the corresponding cross-sections $d\sigma$.
Here, the contour gauge use results in the gauge invariant hadron tensor and provides the
new contributions to single spin asymmetries.
Having considered this hadron tensor in
the asymptotical regime associated with the very large $Q^2$,
the factorization theorem can be applied for the given hadron tensor as well as for the DVCS process.
As a result, the DY hadron tensor takes a form of convolution as
\begin{eqnarray}
\label{Fac-DY}
\text{Hadron tensor} =
\{ \text{Hard part (pQCD)} \}
\otimes
\{ \text{Soft part (npQCD)}\}\,,
\end{eqnarray}
where both the hard and soft parts should be independent of each other and are in agreement with
the ultraviolet and infrared renormalizations.
Moreover, the relevant single spin asymmetries (SSAs), which is a subject of experimental studies,
can be presented as
\begin{eqnarray}
d\sigma(N^{(\uparrow)}) - d\sigma(N^{(\downarrow)})
\sim {\cal L}_{\mu\nu}\, H_{\mu\nu}\, ,
\end{eqnarray}
where ${\cal L}_{\mu\nu}$ and $H_{\mu\nu}$ are the lepton and
hadron tensors, respectively.
The hadron tensor $H_{\mu\nu}$ includes the
the polarized hadron matrix element which takes a form of
\begin{eqnarray}
\label{parVecDY}
&&\langle p_1, S^\perp | \bar\psi(\lambda_1 \tilde n)\, \gamma^+ \,
g A^{\alpha}_\perp(\lambda_2\tilde n) \,\psi(0)
|p_1, S^\perp \rangle
\stackrel{{\cal F}}{=}
\nonumber\\
&&
i\varepsilon^{\alpha + S^\perp -} (p_1p_2)
\, B(x_1,x_2),
\end{eqnarray}
where $\stackrel{{\cal F}}{=}$ stands for the Fourier transform between
the coordinate space, formed by positions $\lambda_i \tilde n$, and
the momentum space, realized by fractions $x_i$;
the light-cone vector $\tilde n$  is a dimensionful analog of $n$.
In Eqn.~(\ref{parVecDY}), the parametrizing function $B$ describes the corresponding parton distribution.


In the studies, see for example \cite{Boer:2003cm,BQ,Teryaev,Boer}, where the local light-cone  gauge $A^+=0$
has been used,
$B$-function of Eqn.~(\ref{parVecDY}) is given by a purely real function. That is, we have
\begin{eqnarray}
\label{g-pole-B}
B(x_1,x_2)= \frac{{\cal P}}{x_1-x_2} T(x_1,x_2)\,
\end{eqnarray}
where the function $T(x_1,x_2)\in\Re\text{e}$
parametrizes the corresponding projection of $\langle \bar\psi\, G_{\alpha\beta}\,\psi \rangle$ and
obeys $T(x,x)\not = 0$.

In Eqn.~(\ref{g-pole-B}), the pole at $x_1=x_2$ is treated as a principle value and it obviously leads to $B(x_1,x_2)\in\Re\text{e}$.
Indeed, within the local gauge $A^+=0$,
the statement on that $B$ is the real function stems actually from
the ambiguity in the solutions of the trivial differential equation, which is equivalent to the definition of $G_{\mu\nu}$,
\begin{eqnarray}
\label{DiffEqnsG}
\partial^+\, A^\alpha_\perp=G^{+\,\alpha}_\perp\,.
\end{eqnarray}
The formal resolving of Eqn.~(\ref{DiffEqnsG}) leads to two representations written as
\begin{eqnarray}
\label{Ag1}
&&A^\mu_{(1)}(z) =
\int_{-\infty}^{z} d\omega^-
G^{+\mu} (\omega^-)
+ A^\mu(-\infty),
\\
\label{Ag2}
&&A^\mu_{(2)}(z) =
- \int_{z}^{\infty} d\omega^-
G^{+\mu} (\omega^-)
+ A^\mu(\infty).
\end{eqnarray}

We stress that within the approaches backed on the local axial gauge use, there are no evidences to think that
Eqns.~(\ref{Ag1}) and (\ref{Ag2}) are not equivalent each other.
That is, the local gauge $A^+=0$ inevitably leads to the following logical scheme (see \cite{Anikin:2015xka} for details)
\begin{eqnarray}
\label{Loc-g-equiv}
\boxed{A^\mu_{(1)} \stackrel{loc.g.}{=}A^\mu_{(2)}} \rightarrow
\boxed{B(x_1,x_2)={\cal P}\frac{T(x_1,x_2)}{x_1-x_2} \in\Re\text{e}}.
\end{eqnarray}
This equation demonstrates that the equivalence of Eqns.~(\ref{Ag1}) and (\ref{Ag2})
causes the representation of $B$-function as in Eqn.~(\ref{g-pole-B}).
The discussion on the boundary configurations can be found in \cite{Anikin:2015xka}.

Regarding the DY process, the physical consequences of the use of $B$ presented by Eqn.~(\ref{g-pole-B}) are
the problem with the photon gauge invariance of DY-like hadron tensors and the losing of significant contributions
to SSAs \cite{Anikin:2010wz,Anikin:2015xka}.
Besides, based on the local gauge $A^+=0$, the representation of gluon field as a linear
combination of Eqns.~(\ref{Ag1}) and (\ref{Ag2}) has been used in the different studies, see
\cite{Belitsky:2005qn,Hatta:2011zs,Lorce:2013gxa}.


In contrast to the local gauge $A^+=0$, as discussed in Sec.~\ref{Sec:III}, we can infer that
the path dependent non-local gauge (see Eqn.~(\ref{CG-1}))
fixes unambiguously the representation of gluon field which is given by either Eqn.~(\ref{Ag1}) or Eqn.~(\ref{Ag2}).
Indeed, fixing the path $P(x_0,x)$, a solution of Eqn.~(\ref{CG-1}) takes the form of
(see \cite{Anikin:2010wz,Anikin:2015xka,Durand:1979sw} for details)
\begin{eqnarray}
\label{A-cg}
A^{c.g.}_\mu(x)=
\int_{P(x_0,x)} dz_\alpha \frac{\partial z_\beta}{\partial x_\mu}\,
G_{\alpha\beta}(z| A),
\end{eqnarray}
where the boundary configuration $A^{c.g.}_\mu(x_0)$ has assumed to be zero.
By direct calculation, we can show that the non-local gauge ${\bf g} (P(- \infty, x) | A)=1$ leads to the gluon filed
representation of Eqn.~(\ref{Ag1}),
while the non-local gauge ${\bf g} (P(x, + \infty ) | A)=1$ corresponds to the gluon field representation of Eqn.~(\ref{Ag2}).
Moreover, we can readily check that \cite{Anikin:2010wz,Anikin:2015xka}
\begin{eqnarray}
\label{2-cg}
\text{Eq.~(\ref{Ag1})}\rightarrow \boxed{B_{+}(x_1,x_2)} \not=
\boxed{ B_{-}(x_1,x_2)} \leftarrow \text{Eq.~(\ref{Ag2})},
 \end{eqnarray}
where
\begin{eqnarray}
\label{B-plus-minus}
&&
B_{+}(x_1,x_2)=\frac{T(x_1,x_2)}{x_1-x_2 + i\epsilon}\in \mathbb{C},
\\
&&
B_{-}(x_1,x_2)=
\frac{T(x_1,x_2)}{x_1-x_2 - i\epsilon}\in \mathbb{C}.
\end{eqnarray}

Notice that both the non-local contour gauges, {\it i.e.} ${\bf g} (P(- \infty, x) | A)={\bf 1}$ and
${\bf g} (P(x, + \infty ) | A)={\bf 1}$,
can be projected into
the same local gauge $A^+=0$. As mentioned, the projection given by $A^+=0$ does not give a possibility
to understand which of the non-local gauges generates the local gauge.

To conclude, we can state that, considering DY-like processes,
the corresponding non-local gauge gives rise to the correct representation
of $B$-functions, see Eqn.~(\ref{B-plus-minus}), which has the non-zero imaginary part.
This enables us to find the new significant contributions to the hadron tensors that ensure ultimately
the gauge invariance  \cite{Anikin:2010wz,Anikin:2015xka}. In a similar manner, with the help of
Eqn.~(\ref{B-plus-minus}) we can fix the prescriptions for
the spurious singularities in the gluon propagators \cite{Anikin:2016bor}.

\section{Drell-Yan hadron tensor: contour gauge and gluon propagator}
\label{CG-GP-1}

\subsection{Kinematics}

We begin with the kinematics of Drell-Yan process.
We study the Drell-Yan process with the transversely polarized hadron:
\begin{eqnarray}
N^{(\uparrow\downarrow)}(p_1) + N(p_2) &\to& \gamma^*(q) + X(P_X)
\nonumber\\
&\to&\ell(l_1) + \bar\ell(l_2) + X(P_X),
\end{eqnarray}
where the virtual photon producing the lepton pair ($l_1+l_2=q$) has a large mass squared
($q^2=Q^2$) while the transverse momenta are small and integrated out.
This kinematics (anticipating the collinear factorization procedure) suggests a convenient frame with
fixed dominant light-cone directions:
\begin{eqnarray}
\label{kin-DY}
&&p_1\approx \frac{Q}{x_B \sqrt{2}}\, n^*\, , \quad p_2\approx \frac{Q}{y_B \sqrt{2}}\, n,
\\
&&n^{*\,\mu}=\big(\frac{1}{\sqrt{2}},\,\vec{\bf 0}_\perp,\,\frac{1}{\sqrt{2}}\big)=
\big(1^+,0^-,\vec{\bf 0}_\perp \big),
\nonumber\\
&&n^{\mu}=\big(\frac{1}{\sqrt{2}},\,\vec{\bf 0}_\perp,\,\frac{-1}{\sqrt{2}}\big)=
\big(0^+,1^-,\vec{\bf 0}_\perp \big),
\nonumber\\
&&n^*\cdot n=n^{*\,+}n^{-}=1
\nonumber
\end{eqnarray}
It is also instructive to introduce the dimensionful analogs of $n, n^*$ as
\begin{eqnarray}
\label{ntilde}
\tilde n^-=\frac{p_2^-}{p_1p_2},\quad \breve n^+=\frac{p_1^+}{p_1p_2}.
\end{eqnarray}
With the above vectors as a basis, an arbitrary vector can be (Sudakov) decomposed as
\begin{eqnarray}
\label{Sud-Exp}
&&a^{\mu} = a^+ n^{*\,\mu} + a^- n^{\mu} + a^\mu_\perp,
\nonumber\\
&&a^{\mu,+}\stackrel{{\rm def}}{=} a^+ n^{*\,\mu},
\quad a^{\mu,-}\stackrel{{\rm def}}{=} a^- n^{\mu}.
\end{eqnarray}
In what follows we will not be so precise about writing the covariant and contravariant vectors in any kinds of
summations over the four-dimensional vectors, except the cases where this trick may lead to misunderstanding.

\subsection{Drell-Yan hadron tensor: Derivation of Wilson lines}

The polarized DY process is very convenient process to study the role of twist three by exploring
of different kinds of SSAs.
For example, one can study the left-right asymmetry which means the transverse momenta
of the leptons are correlated with the direction
$\textbf{S}\times \textbf{e}_z$ where $S_\mu$ implies the
transverse polarization vector of the nucleon and $\textbf{e}_z$ is a beam direction \cite{Barone}.

Generally speaking, any single spin asymmetries can be presented in the symbolical form as
\begin{eqnarray}
\label{SSA-A}
{\cal A} \sim d\sigma^{(\uparrow)} - d\sigma^{(\downarrow)}
\sim {\cal L}_{\mu\nu}\, {\cal W}_{\mu\nu}\, ,
\end{eqnarray}
where ${\cal L}_{\mu\nu}$ is an unpolarized leptonic tensor and
${\cal W}_{\mu\nu}$ stands for the hadronic tensor. At the moment,
we do not specify the phase space in Eqn.~(\ref{SSA-A}) because the exact expression for SSA is irrelevant
for our discussion. Instead, we mainly pay our attention on the hadron tensor which can be presented as
\begin{eqnarray}
\label{W-over-g}
{\cal W}_{\mu\nu}&&\hspace{-.3cm}={\cal W}^{(0)}_{\mu\nu} + {\cal W}^{(1)}_{\mu\nu}(g|A) +
{\cal W}^{(2)}_{\mu\nu}(g|A) + (g^n\text{-terms}|n\geq 2)
\nonumber\\
&&\hspace{-.3cm}=
\overline{{\cal W}}^{(0)}_{\mu\nu}(A^{\pm}) +{\cal W}^{(1)}_{\mu\nu}(g|A^\perp) +
{\cal W}^{(2)}_{\mu\nu}(g|A^\perp)+ \cdots,
\end{eqnarray}
where $g$ denotes the strong interaction coupling constant and
\begin{eqnarray}
\label{barW}
\overline{{\cal W}}^{(0)}_{\mu\nu}(A^{\pm})=
{\cal W}^{(0)}_{\mu\nu} + {\cal W}^{(1)}_{\mu\nu}(g|A^+) +
{\cal W}^{(2)}_{\mu\nu}(g|A^-) + \cdots\,.
\end{eqnarray}
The hadron tensor representations can be found below.
In our case, the single transverse spin asymmetry is only generated by the hadron tensors
${\cal W}^{(1)}_{\mu\nu}(g|A^\perp)$ and ${\cal W}^{(2)}_{\mu\nu}(g|A^\perp)$ where
the twist three contributions related to $\langle\bar\psi\gamma^+A^\perp\psi\rangle$
have been extracted.
As shown below, the $\langle\bar\psi\gamma^+A^\pm\psi\rangle$-correlators
in the hadron tensors ${\cal W}^{(1,2)}_{\mu\nu}(g|A)$
participate in forming of the corresponding Wilson lines which appear in the
quark-antiquark correlators of the hadron tensor $\overline{{\cal W}}^{(0)}_{\mu\nu}(A^{\pm})$.
In the frame of usual axial gauge ($A^+=0$), this kind of contributions can be
discarded.
However, we work in the contour gauge which is, first, a non-local generalization of the well-know axial gauge.
Second, the contour gauge contains the important and unique additional information (needed to fix
the prescription in the gluon poles) which is invisible in the case of usual (local) axial gauge.
From this point of view, before we discard the terms with $A^+$, we have to determine
the relevant fixed path in the restored Wilson line with $A^+$
which eventually leads to the certain prescriptions in the gluon poles.

\begin{figure*}[ht]
\centerline{\includegraphics[width=0.45\textwidth]{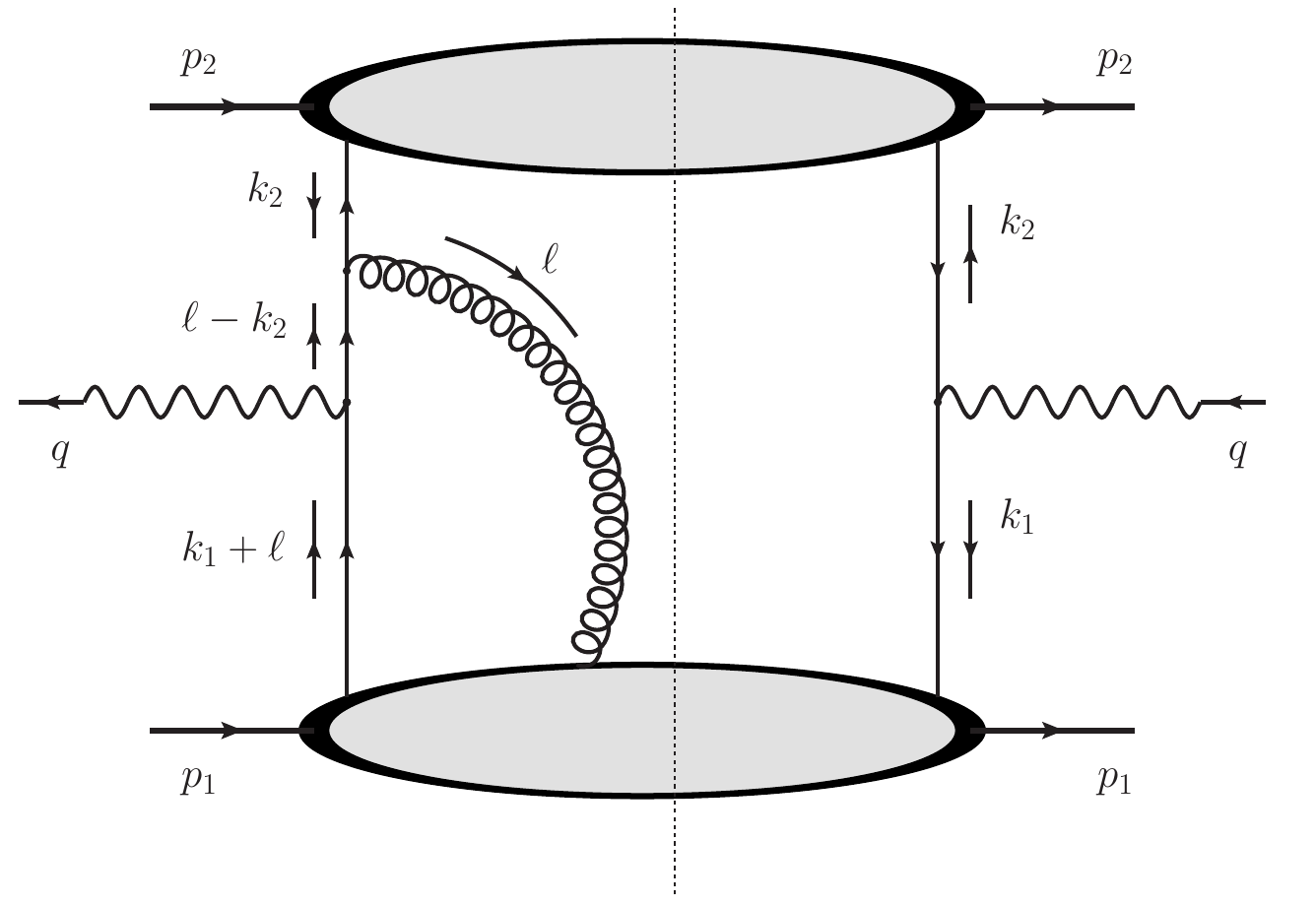}
\hspace{0.2cm}
\includegraphics[width=0.45\textwidth]{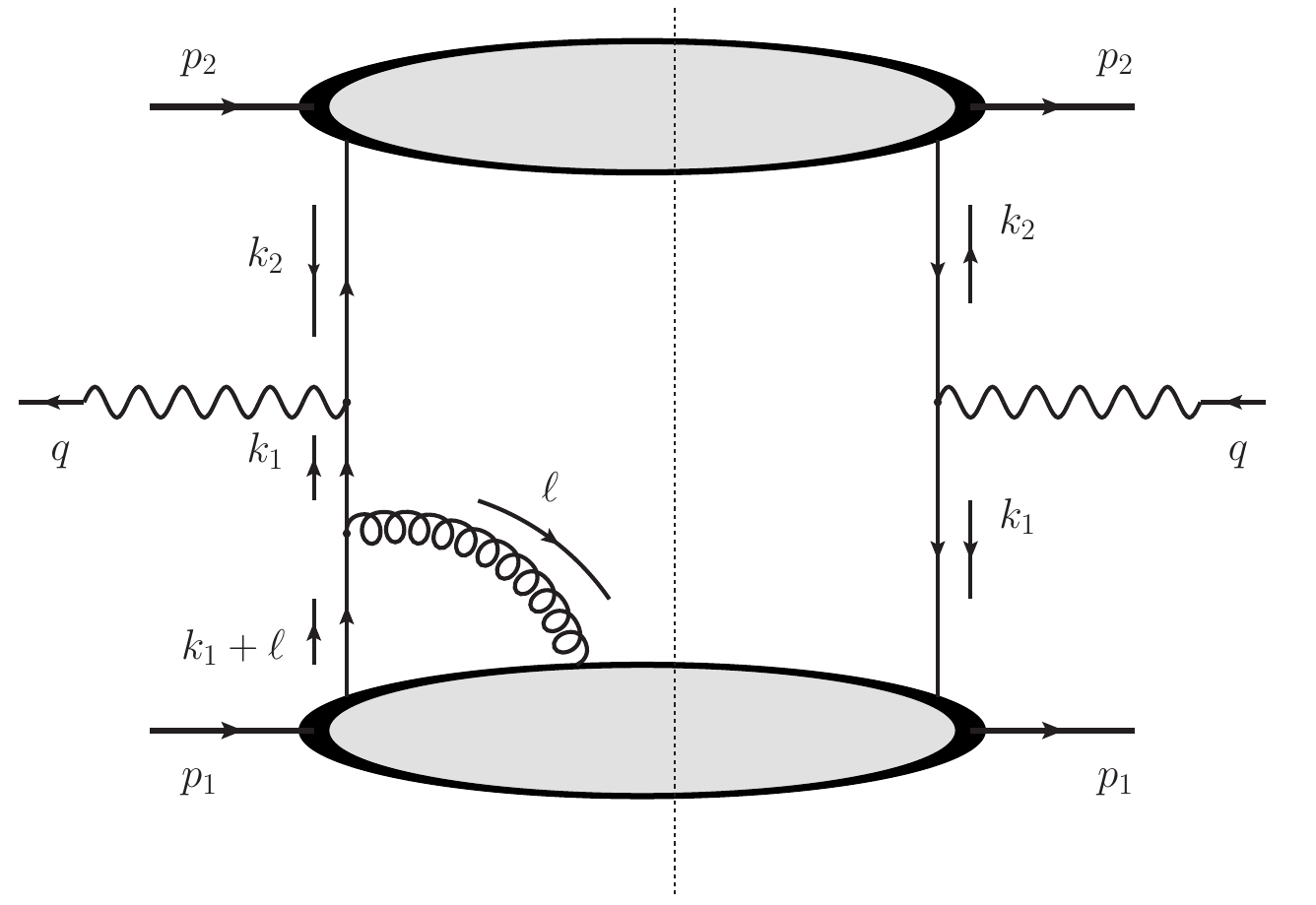}}
  \caption{The Feynman diagrams which contribute to the polarized Drell-Yan hadron tensor:
  the standard (the left panel) and non-standard diagrams (the right panel).}
\label{Fig-DY}
\end{figure*}

In this section, we analyse the part of the DY hadron tensor which is generated by the diagram
in Fig.~\ref{Fig-DY}, the left panel. This is the standard hadron tensor which can be written
in non-factorized form as
\begin{eqnarray}
\label{HadTen1-2}
&&{\cal W}^{(1)}_{\mu\nu}(g|A)=\int d^4 k_1\, d^4 k_2 \, \delta^{(4)}(k_1+k_2-q)
\, \bar\Phi^{[\gamma^-]} (k_2) \times
\nonumber\\
&&\int d^4 \ell \,
\Phi^{(A)\,[\gamma^+]}_\alpha (k_1,\ell) \,
\text{tr}\Big[
\gamma_\mu  \gamma^- \gamma_\nu \gamma^+ \gamma_\alpha\times
\nonumber\\
&&
\frac{(\ell^+-k_2^+)\gamma^- + (\ell^- - k_2^-)\gamma^+ -
(\vec{\ell}_\perp - \vec{k}_{2\,\perp})\vec{\gamma}_\perp}
{(\ell - k_2)^2 + i\epsilon}
\Big] \, ,
\end{eqnarray}
where
\begin{eqnarray}
\label{PhiF-1}
&&\hspace{-0.8cm}\Phi^{(A)\,[\gamma^+]}_\alpha (k_1,\ell)
\stackrel{{\cal F}_2}{=}
\langle p_1, S^T | \bar\psi(\eta_1)\gamma^+  gA_{\alpha}(z)  \psi(0) | S^T, p_1\rangle ,
\\
\label{PhiF-2}
&&\hspace{-0.8cm}\bar\Phi^{[\gamma^-]}(k_2)\stackrel{{\cal F}_1}{=}
\langle p_2 | \bar\psi(\eta_2)\gamma^- \psi(0)| p_2\rangle .
\end{eqnarray}
In Eqns.~(\ref{PhiF-1}) and (\ref{PhiF-2}), ${\cal F}_1$ and  ${\cal F}_2$
denote the Fourier transformation with the measures defined as
\begin{eqnarray}
d^4\eta_2\, e^{ik_2\cdot\eta_2}\,\,\, \text{and} \,\,\,
d^4\eta_1\, d^4 z\, e^{-ik_1\cdot\eta_1-i\ell\cdot z} ,
\end{eqnarray}
respectively. For the sake of shortness, we will omit $S^T$ in
the hadron states that indicates the transverse polarization of hadron.
%
We now analyze the tensor structure of the trace in Eqn.~(\ref{HadTen1-2}).
We can see that the first term of the quark propagator, $\ell^+-k_2^+)\gamma^-$, singles out
only the transverse components of gluon field in the quark-gluon correlator, see Eqn.~(\ref{PhiF-1}).
At the same time, the second term of the quark propagator, $(\ell^- - k_2^-)\gamma^+$, separates out
only the longitudinal component $A^+$ in the quark-gluon correlator. This second term is very important for
derivation of the corresponding Wilson line which defines in our approach the contour gauge.
And, the third term of the quark propagator give us the quark-gluon correlator with both indices $\alpha=(+, \perp)$.

The collinear factorization procedure for the process under consideration can be introduced
by the following steps (for the details see, e.g.,
Refs. \cite{Efremov:1984ip, Anikin:2009bf}):

\noindent (a) the decomposition of loop integration momenta around the corresponding dominant direction:
$$k_i = x_i p + (k_i\cdot p)n + k_T$$
within the certain light cone basis formed by the vectors $p$ and $n$ (in our case, $n^*$ and $n$);

\noindent (b) the replacement:
$$d^4 k_i \Longrightarrow d^4 k_i \,dx_i \delta(x_i-k_i\cdot n)$$
that introduces the fractions with the appropriated spectral properties;

\noindent (c) the decomposition of the corresponding propagator products, which will finally form the hard part,
around the dominant direction. It is necessary to notice that
in the DY process case the corresponding $\delta$-functions appeared in the hadron tensor
and expressed the momentum conservation law should be also referred to the hard parts.
This statement was argued in \cite{An} in the context of the so-called factorization links;

\noindent (d) the use of the collinear Ward identity if it is necessary within the given of accuracy level;

\noindent (e) performing of the Fierz decomposition for $\psi_\alpha (z) \, \bar\psi_\beta(0)$ in
the corresponding space up to the needed projections.

Let us first dwell on the second term, $(\ell^- - k_2^-)\gamma^+$, contribution.
This term is responsible for forming of the Wilson line in the gauge-invariant quark-antiquark string operator.
Indeed, making used the collinear factorization ($\ell^-\approx 0, (\ell-k_2)^2\approx -2\ell^+k_2^-$),
the above-mentioned term contributes in the hadron tensor as
\begin{eqnarray}
\label{HadTen1-3}
&&{\cal W}^{(1)\, [k_2^-]}_{\mu\nu}(g|A^+)=\int d\mu (k_i;x_1,y) \,
\, \bar\Phi^{[\gamma^-]} (k_2)\frac{1}{2} \int dz^-
\times
\nonumber\\
&&
\text{tr}\Big[
\gamma_\mu  \gamma^- \gamma_\nu \gamma^+ \gamma^- \gamma^+
\Big]
 \int d\ell^+\, \frac{e^{ - i\ell^+ z^-}}{\ell^+ - i\epsilon}
\int d^4\eta_1  \, e^{-ik_1\cdot\eta_1}\times
\nonumber\\
&&
\langle p_1 | \bar\psi(\eta_1)\, \gamma^+ \, gA^+(0^+,z^-,\vec{{\bf 0}}_\perp) \, \psi(0)
|p_1\rangle \, ,
\end{eqnarray}
where the integration measure reads
\begin{eqnarray}
&&\hspace{-0.3cm}d\mu (k_i;x_1,y)= dx_1 d^4k_1 \delta\Big(x_1-\frac{k_1^+}{p_1^+}\Big)
\,dy d^4k_2 \delta\Big(y-\frac{k_2^-}{p_2^-}\Big)\times
\nonumber\\
&&\hspace{-0.3cm}\big[ \delta^{(4)}(x_1p_1+yp_2 - q) \big].
\end{eqnarray}
The prescription $-i\epsilon$ in the denominator of (\ref{HadTen1-3})
directly follows from the standard causal
prescription for the massless quark propagator in (\ref{HadTen1-2}) (cf. \cite{Braun}).

Integration over $\ell^+$ in (\ref{HadTen1-3}), using the well-known integral representation
\begin{eqnarray}
\label{theta-func}
\theta(\pm x) =\frac{\pm i}{2\pi} \int\limits_{-\infty}^{+\infty} dk\, \frac{e^{-ikx}}{k\pm i\epsilon},
\end{eqnarray}
leads to the following expression:
\begin{eqnarray}
\label{HadTen1-4}
&&\hspace{-0.4cm}{\cal W}^{(1)\,  [k_2^-]}_{\mu\nu}(g|A^+)= \int d\mu (k_i;x_1,y) \times
\\
&&\hspace{-0.4cm}\text{tr}\Big[
\gamma_\mu  \gamma^- \gamma_\nu \gamma^+
\Big] \, \bar\Phi^{[\gamma^-]} (k_2)\int d^4\eta_1 \, e^{-ik_1\cdot\eta_1} \times
\nonumber\\
&&\hspace{-0.4cm}
\langle p_1| \bar\psi(\eta_1)\, \gamma^+
ig \int\limits_{-\infty^-}^{0^-} dz^- A^+(0^+,z^-,\vec{{\bf 0}}_\perp)
 \psi(0) | p_1\rangle ,
\nonumber
\end{eqnarray}
where we use
\begin{eqnarray}
\frac{1}{2} \gamma^+ \gamma^- \gamma^+ = \gamma^+.
\end{eqnarray}
It is important to stress that the leading order hadron tensor ${\cal W}^{(0)}_{\mu\nu}(g^0)$ differs from the
hadron tensor (\ref{HadTen1-2}) by overall sign: the leading hadron tensor has a pre-factor $i^2$ due to
two photon vertices, while the hadron tensor (\ref{HadTen1-2}) is accompanying by a pre-factor $i^4$
thanks for two photon and one gluon vertices together with the pre-factor from
the massless quark propagator $(-1)/i$ (we use the convention as in \cite{BogoShir}).

Thus, if we include all gluon emissions from the antiquark going from the upper blob
in Fig.~\ref{Fig-DY}, the left panel, (the so-called initial state interactions),
we are able to get the corresponding $P$-exponential in
$\Phi^{(A)\,[\gamma^+]}_\alpha (k_1,\ell)$. The latter is now represented by the
following matrix element:
\begin{eqnarray}
\label{me-Pexp}
\int d^4\eta_1 \, e^{-ik_1\cdot\eta_1}
\langle p_1 | \bar\psi(\eta_1)\, \gamma^+ \, [-\infty^- ;\, 0^-]_{A^+}
\, \psi(0) | p_1\rangle \, ,
\end{eqnarray}
where
\begin{eqnarray}
\label{Pexp-1}
&&[-\infty^- ;\, 0^-]_{A^+}\equiv [0^+, -\infty^-, \vec{{\bf 0}}_\perp ;\, 0^+, 0^-, \vec{{\bf 0}}_\perp]_{A^+}=
\nonumber\\
&&\mathbb{P}{\rm exp}\Big\{ i g \int\limits^{-\infty^-}_{0^-}
 dz^- \, A^+(0^+,z^-,\vec{{\bf 0}}_\perp) \Big\}\, .
\end{eqnarray}
The collinear twist ($t=d-s_a$) of $A^+$ is equal to zero, therefore
the Wilson line which is summing up all these components does not affect the
twist expansion within the collinear factorization.

If now we include in our consideration the gluon emission from
the incoming antiquark (the mirror contributions), we will obtain the Wilson
line $[\eta_1^-,-\infty^-]$ which
will ultimately give us, together with (\ref{Pexp-1}), the Wilson line connecting
the points $0$ and $\eta_1$ in (\ref{me-Pexp}) contributing to
$\overline{{\cal W}}^{(0)}_{\mu\nu}$.
This is exactly what happens, say, in the spin-averaged DY process \cite{Efremov:1978xm}.
However, for the SSA, these
two diagrams should be considered individually.
Indeed, their contributions to SSAs, contrary to spin-averaged case,
differ in sign and the dependence on the boundary point at $-\infty^-$
does not cancel.

For the pedagogical reason, we want to show the exponentiation of the transverse gluon field
(here, we mainly follow to \cite{Belitsky:2002sm}),
although we are restricted by the twist three case and the inclusion of all degrees of the transverse gluon field
exceeds our accuracy.
Let us consider the third term, $(\vec{\ell}_\perp - \vec{k}_{2\,\perp})\vec{\gamma}_\perp$, contribution
which helps us to demonstrate the exponentiation of the transverse gluon fields.
The corresponding hadron tensor part takes the following form
\begin{eqnarray}
\label{HadTen1-4-1}
&&{\cal W}^{(1)\, [\vec{\ell}_\perp]}_{\mu\nu}(g|A^\perp)=
\\
&&\int d\mu (k_i;x_1,y) \,
\, \bar\Phi^{[\gamma^-]} (k_2)
\text{tr}\Big[
\gamma_\mu  \gamma^- \gamma_\nu \gamma^+ \gamma^\perp_\alpha \vec{\gamma}_i^\perp
\Big]\times
\nonumber\\
&& \int d^4\ell \, \frac{(\vec{\ell}^\perp - \vec{k}_2^\perp)_i}{2\ell^+k_2^- + \vec{\ell}_\perp^{\,2} - i\epsilon}
\Phi_\alpha^{(A^\perp)[\gamma^+]}(k_1,\ell)
\equiv
\nonumber\\
&& \int d\mu (k_i;x_1,y) \,
\, \bar\Phi^{[\gamma^-]} (k_2)
\text{tr}\Big[
\gamma_\mu  \gamma^- \gamma_\nu \gamma^+ \gamma^\perp_\alpha \vec{\gamma}_i^\perp
\Big] \mathfrak{L}_{i,\alpha}
\,,
\nonumber
\end{eqnarray}
where we assume that $\vec{k}_{2\,\perp}\approx 0$. In Eqn.~(\ref{HadTen1-4}) let us focus
on the $\ell$-integration, we have
\begin{eqnarray}
\label{HadTen-ell}
&&\mathfrak{L}_{i,\alpha}= \int d\ell^+d\ell^- d^2\vec{\ell}_\perp \,
\frac{\vec{\ell}^\perp_i}{2\ell^+k_2^- + \vec{\ell}_\perp^{\,2} - i\epsilon}\times
\\
&&\int d^4\eta_1\,d^4z\, e^{-ik_1\eta_1 - i \ell z}
\langle p_1| \bar\psi(\eta_1)\gamma^+  gA^\perp_{\alpha}(z) \psi(0) | p_1\rangle.
\nonumber
\end{eqnarray}
We now use the $\alpha$-representation for the denominator that stems from the
quark propagator:
\begin{eqnarray}
\label{alpha-reps}
\frac{1}{2\ell^+k_2^- + \vec{\ell}_\perp^{\,2} - i\epsilon}=
i\int\limits_0^\infty d\alpha\, e^{-i\alpha[2\ell^+k_2^- + \vec{\ell}_\perp^{\,2} - i\epsilon]}.
\end{eqnarray}
Next, in Eqn.~(\ref{HadTen-ell}) we perform the integrations over $d\ell^-$ and $d\ell^+$
which give $\delta(z^+)$ and $\delta(z^- + 2\alpha k_2^-)$, respectively.
We remind that the variables $\alpha$ in (\ref{alpha-reps}) are dimensionful and
$\text{dim}_M[\alpha]=-2$.

Therefore, the integral  $\mathfrak{L}$ takes the following form (cf. \cite{Belitsky:2002sm})
\begin{eqnarray}
\label{HadTen-ell-2}
&&\mathfrak{L}_{i,\alpha}= i \int d^2\vec{\ell}_\perp \,\vec{\ell}^\perp_i
\int\limits_0^\infty d\alpha\, e^{-i\alpha[\vec{\ell}_\perp^{\,2} - i\epsilon]}
\int d^4\eta_1 \, d^2 \vec{z}_\perp
\times
\\
&&e^{-ik_1\eta_1 + i \vec{\ell}_\perp \vec{z}_\perp}
\langle p_1| \bar\psi(\eta_1)\gamma^+  gA^\perp_{\alpha}(0^+, -\infty^-, \vec{z}_\perp) \psi(0) | p_1\rangle.
\nonumber
\end{eqnarray}
In Eqn.~(\ref{HadTen-ell-2}) the transverse gluon field operator can be presented as
\begin{eqnarray}
\label{At}
A^\perp_{\alpha}(0^+, -\infty^-, \vec{z}_\perp) = \frac{\partial}{\partial z^\perp_\alpha}
\int\limits_{\mathds{C}}^{z^\perp} d\omega^\perp_\beta A^\perp_\beta(0^+, -\infty^-, \vec{\omega}^\perp),
\end{eqnarray}
where we fix the arbitrary constant $\mathds{C}$ to be $-\vec{\infty}_\perp$.
By making use of the representation (\ref{At}), after integration over $\alpha$ we arrive at
\begin{eqnarray}
\label{HadTen-ell-3}
&&\hspace{-0.8cm}
\mathfrak{L}_{i,\alpha}= i \int d^2\vec{\ell}_\perp \,
\frac{\vec{\ell}^\perp_i \ell^\perp_\alpha}{\vec{\ell}_\perp^{\,2} - i\epsilon}
\int d^4\eta_1 \, d^2 \vec{z}_\perp
e^{-ik_1\eta_1 + i \vec{\ell}_\perp \vec{z}_\perp}\times
\nonumber\\
&&\hspace{-0.8cm}
\langle p_1| \bar\psi(\eta_1)\gamma^+
g \int\limits_{-\infty^\perp}^{z^\perp} d\omega^\perp_\beta A^\perp_\beta(0^+, -\infty^-, \vec{\omega}^\perp)
\psi(0) | p_1\rangle.
\end{eqnarray}
We insert the obtained expression for $\mathfrak{L}_{i,\alpha}$, see Eqn.~(\ref{HadTen-ell-3}),
into the expression for hadron tensor (\ref{HadTen1-4}).
After integration over $d^2\vec{\ell}_\perp$ and, then, after integration over  $d^2\vec{z}_\perp$
we get the following expression for the $\vec{\ell}_\perp$-term of the hadron tensor:
\begin{eqnarray}
\label{HadTen1-5}
&&\hspace{-0.3cm}{\cal W}^{(1)\, [\vec{\ell}_\perp]}_{\mu\nu}(g|A^\perp)=\int d\mu (k_i;x_1,y) \,
\, \bar\Phi^{[\gamma^-]} (k_2)
\text{tr}\Big[
\gamma_\mu  \gamma^- \gamma_\nu \gamma^+ \Big]\times
\nonumber\\
&&\hspace{-0.3cm}\int d^4\eta_1 \,
e^{-ik_1\eta_1}\times
\\
&&\hspace{-0.3cm}
\langle p_1| \bar\psi(\eta_1)\gamma^+
ig \int\limits_{-\infty^\perp}^{0^\perp} d\omega^\perp_\beta A^\perp_\beta(0^+, -\infty^-, \vec{\omega}^\perp)
\psi(0) | p_1\rangle \,.
\nonumber
\end{eqnarray}
As well as for the case of longitudinal gluons, if we now include all gluon emissions
from the antiquark going from the upper blob in Fig.~\ref{Fig-DY}, left panel,
we reproduce the corresponding $P$-exponential with the transverse gluons in
$\Phi^{(A)\,[\gamma^+]}_\alpha (k_1,\ell)$. Together with the result obtained above for the $A^+$-fields, we finally
have
\begin{eqnarray}
\label{me-Pexp-2}
&&\int d^4\eta_1 \, e^{-ik_1\cdot\eta_1}
\langle p_1 | \bar\psi(0^+,\eta_1^-, \vec{\bf 0}_\perp)\, \gamma^+ \times
\\
&&[0^+,-\infty^-,\vec{\bf 0}_\perp \,; 0^+,0^-,\vec{\bf 0}_\perp]_{A^+}
\times
\nonumber\\
&&[0^+,-\infty^-,-\vec{\bf\infty}_\perp \,; 0^+,-\infty^-,\vec{\bf 0}_\perp]_{A^\perp}
\psi(0) | p_1\rangle \, ,
\nonumber
\end{eqnarray}
where
\begin{eqnarray}
\label{Pexp-1-1}
&&[0^+,-\infty^-,-\vec{\bf\infty}_\perp \,; 0^+,-\infty^-,\vec{\bf 0}_\perp]_{A^\perp} =
\nonumber\\
&&\mathbb{P}{\rm exp}\Big\{ i g \int\limits^{-\infty^\perp}_{0^\perp}
 d\omega_\beta^\perp A_\beta^\perp(0^+,-\infty^-,\vec{{\bf\omega}}_\perp) \Big\}\, .
\end{eqnarray}
The transverse components of gluon fields, $A^\perp$, have the collinear twist
which equals to $1$. Therefore, the Wilson line in Eqn.~(\ref{Pexp-1-1}) represents
the infinite amount of the sub-dominant contributions. Within our frame,
it is enough to be limited by the collinear twist three contributions only.
In other words, we leave only the terms which include the first order of $A^\perp$.


The next step of our consideration is the contribution of the non-standard
diagram, depicted in Fig.~\ref{Fig-DY}, the right panel. The DY hadron tensor
receives the contribution from the non-standard diagram as (before factorization)
\begin{eqnarray}
\label{HadTen-NS-1}
&&{\cal W}^{(2)}_{\mu\nu}(g|A)=
\int d^4 k_1\, d^4 k_2 \, \delta^{(4)}(k_1+k_2-q)\times
\\
&&\text{tr}\big[
\gamma_\mu  {\cal F}(k_1) \gamma_\nu \bar\Phi(k_2)
\big]\, ,
\nonumber
\end{eqnarray}
where the function ${\cal F}(k_1)$ reads
\begin{eqnarray}
\label{PhiF2}
&&{\cal F}(k_1)= S(k_1) \gamma_\alpha \int d^4\eta_1\, e^{-ik_1\cdot\eta_1}\times
\\
&&\langle p_1| \bar\psi(\eta_1) \, gA_{\alpha}(0) \, \psi(0) | p_1\rangle \, .
\nonumber
\end{eqnarray}
Performing the above-described factorization procedure, the non-standard hadron tensor
takes the following form:
\begin{eqnarray}
\label{HadTen-NS-2}
&&{\cal W}^{(2)}_{\mu\nu}(g|A)=  \int dx_1 \, dy \,
\big[\delta(x_1-x_B) \delta(y-y_B)\big] \, \bar q(y) \times
\nonumber\\
&&\text{tr}\biggl[
\gamma_\mu \biggl( \int d^4 k_1\,
\delta(x_1p_1^+ - k_1^+) {\cal F}(k_1)\biggr) \gamma_\nu \hat p_2 \biggr]\equiv
\nonumber\\
&&\int dx_1 \, dy \,
\big[\delta(x_1-x_B) \delta(y-y_B)\big] \, \bar q(y)\,p_2^- \mathfrak{N}^+_{\mu\nu}(x_1)
 \, .
\end{eqnarray}
We now consider the integral over $k_1$ in (\ref{HadTen-NS-2}), we write
\begin{eqnarray}
\label{FacF2}
&&\mathfrak{N}^+_{\mu\nu}=\int d^4 k_1\, \delta(x_1p_1^+ - k_1^+)\times
\\
&&
\text{tr}\Big[ \gamma_\mu \frac{k_1^+\gamma^- + k_1^-\gamma^+ - \vec{k}_{1\perp}\vec{\gamma}_\perp}
{2k_1^+k_1^- - \vec{k}_{1\perp}^2 +i\epsilon} \gamma_\alpha \gamma^- \gamma_\nu \gamma^+
\Big]\times
\nonumber\\
&&
\int d^4\eta_1\, e^{-ik_1\cdot\eta_1}
\langle p_1 | \bar\psi(\eta_1) \gamma^+  gA_{\alpha}(0) \psi(0) | p_1\rangle.
\nonumber
\end{eqnarray}
Technically, derivation of the longitudinal Wilson line for this case differs from the
derivation we implemented for the standard hadron tensor.
We notice that for the non-standard hadron tensor the quark propagator has been included
in the soft part.

Let us consider
the first term, $k_1^+\gamma^-$, in the quark propagator, see Eqn.~(\ref{FacF2}).
Thanks for the $\gamma$-structure, this term singles out the $A^-$-field in the corresponding
correlator. Moreover, the Fourier image of the quark-gluon correlator can be presented
in the equivalent form as
\begin{eqnarray}
\label{FI-1}
&&\int d^4\eta_1\, e^{-ik_1\cdot\eta_1-ik_1 z}
\langle p_1 | \bar\psi(\eta_1) \gamma^+ \times
\\
&&g
\frac{\partial}{\partial z^+} \int\limits_{-\infty^+}^{z^+} d\omega^+ A^-(\omega^+,0^-,\vec{\bf 0}_\perp)
\Bigg|_{z=0}
 \psi(0) | p_1\rangle,
\nonumber
\end{eqnarray}
where the derivative with respect to $z^+$ can be shifted to the exponential function
$e^{-ik_1^- z^+}$. As a result, we have
\begin{eqnarray}
\label{FI-2}
&&i k_1^- \int d^4\eta_1\, e^{-ik_1\cdot\eta_1}\times
\\
&&\langle p_1 | \bar\psi(\eta_1) \gamma^+
g
\int\limits_{-\infty^+}^{0^+} d\omega^+ A^-(\omega^+,0^-,\vec{\bf 0}_\perp)
 \psi(0) | p_1\rangle.
\nonumber
\end{eqnarray}
Using Eqn.~(\ref{FI-2}), the tensor $\mathfrak{N}_{\mu\nu}$ takes the form of
($\vec{k}_{1\perp}^{\,2}\approx 0$)
\begin{eqnarray}
\label{FacF3}
&&\mathfrak{N}^+_{\mu\nu}=\int d^4 k_1\, \delta(x_1p_1^+ - k_1^+)
\,\text{tr}\big[ \gamma_\mu \gamma^- \gamma_\nu \gamma^+
\big]\times
\nonumber\\
&&
\int d^4\eta_1\, e^{-ik_1\cdot\eta_1}\times
\\
&&
\langle p_1 | \bar\psi(\eta_1) \gamma^+
ig\int\limits_{-\infty^+}^{0^+} d\omega^+ A^-(\omega^+,0^-,\vec{\bf 0}_\perp) \psi(0) | p_1\rangle.
\nonumber
\end{eqnarray}
Thus, the first term finally contributes to the non-standard part of the hadron tensor as
\begin{eqnarray}
\label{HadTen-NS-3}
&&\hspace{-.8cm}\overline{\cal W}^{(0)}_{\mu\nu}(A^-)=  \int dx_1 \, dy \,
\big[\delta(x_1-x_B) \delta(y-y_B)\big] \, \bar q(y) \times
\nonumber\\
&&\hspace{-.8cm}\int d^4 k_1\, \delta(x_1p_1^+ - k_1^+)
\,\text{tr}\big[ \gamma_\mu \gamma^- \gamma_\nu \gamma^+
\big]
\int d^4\eta_1\, e^{-ik_1\cdot\eta_1}\times
\\
&&\hspace{-.8cm}
\langle p_1 | \bar\psi(\eta_1) \gamma^+
[-\infty^+,0^-,\vec{\bf 0}_\perp \,; 0^+,0^-,\vec{\bf 0}_\perp]_{A^-}
\psi(0) | p_1\rangle\,.
\nonumber
\end{eqnarray}
The exponentiation of $A^-$ has been presented in Appendix~\ref{Exp-Aminus:App:A}.

Despite the minus component, $A^-$, has formally the collinear twist $2$
(the so-called sub-sub-dominant component),
the Wilson line with $A^-$ in Eqn.~(\ref{HadTen-NS-3})
will play the substantial role for the residual gauge fixing, see discussion in the next section.

\begin{figure}[t]
\centerline{\includegraphics[width=0.45\textwidth]{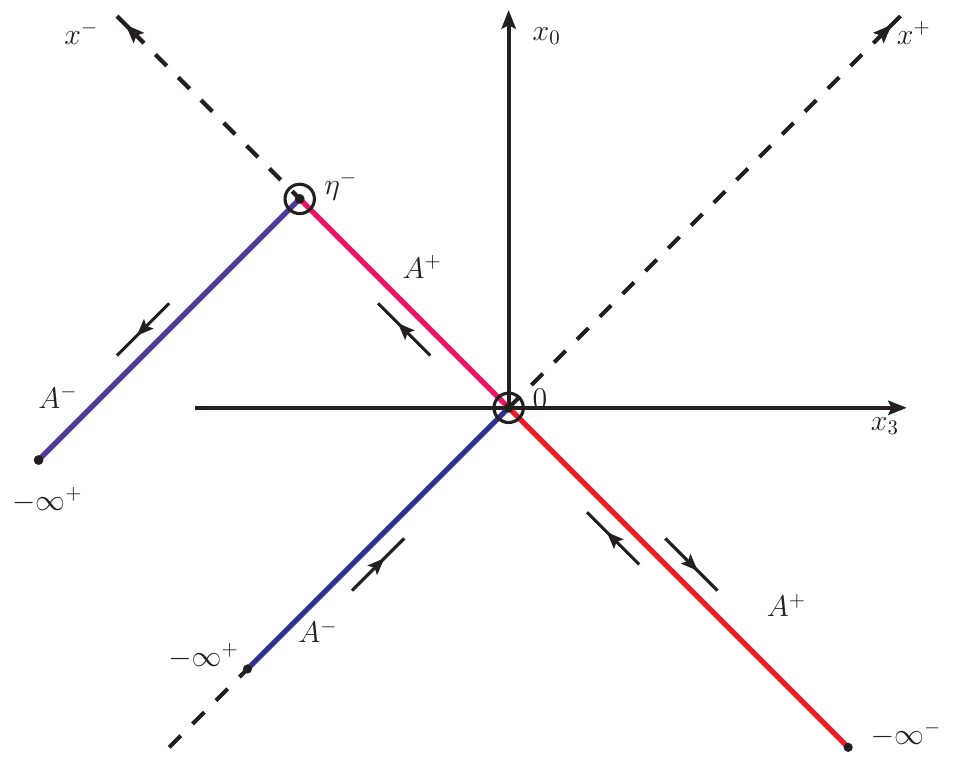}}
  \caption{The longitudinal Wilson lines related to the standard (red lines, the
  exponentials with $A^+$) and
  non-standard (blue lines, the exponentials with $A^-$) Drell-Yan hadron tensor.
  The circles single out the interception points which the continuity conditions
  are defined for.}
\label{Fig-WL}
\end{figure}

To conclude the section, we restore all the longitudinal Wilson lines which emanate from
both the standard and non-standard hadron tensors, see Fig.~\ref{Fig-WL}.

\subsection{Contour Gauge: Elimination of longitudinal Wilson lines}

The axial gauge $A^+=0$ (as well as the Fock-Schwinger gauges)
is in fact a particular case of the most general non-local contour gauge
determined by a Wilson line with a fixed path. Indeed, the straightforward line
in the Wilson exponential which
connects $\pm\infty$ with $x$ gives us the axial gauge, while
the straightforward line connecting $x_0$ with $x$ leads to the Fock-Schwinger gauge.
Notice that two different contour gauges can correspond to the same
local axial gauge. Meanwhile, to distinguish different contour gauge
is very crucial to fix the prescriptions in the gluon poles.

In the past, the contour gauge was very popular subject of intense studies
(see, for example, \cite{ContourG1, ContourG2}).
One of the advantages of using the contour gauge
is that the quantum gauge theory becomes free from the Gribov ambiguities.
On the other hand, the contour gauge gives the simplest way to fix the
gauge including the residual gauge freedom.
In contrast to the usual axial gauge, in the contour gauge we first  fix an arbitrary
point $(x_0, \textbf{g}(x_0))$
in the fiber. Then, we define two directions: one of them in the base, the other in
the fiber. The direction in the base $\mathbb{R}^4$ is nothing else than the tangent vector of a curve which
goes through the given point $x_0$. The fiber direction  can be uniquely determined as the
tangent subspace related to the parallel transport. Finally, we are able to
define uniquely the point in the fiber bundle.

We continue to work with the Drell-Yan hadron tensor.
As shown the standard (direct and mirror) diagrams lead to the following
Wilson lines in the quark-antiquark nonlocal operator which forms the hadron tensor, see Fig.~\ref{Fig-WL}:
\begin{eqnarray}
\label{WL1-mir}
&&[0^+,\eta^-,\vec{\bf 0}_\perp; \, 0^+,-\infty^-,\vec{\bf 0}_\perp]_{A^+},\quad \text{and}
\\
\label{WL1-dir}
&&[0^+,-\infty^-,\vec{\bf 0}_\perp; \, 0^+,0^-,\vec{\bf 0}_\perp]_{A^+},
\end{eqnarray}
{\it i.e.} the gauge invariant quark string operator takes the form of
\begin{eqnarray}
\label{OP1}
&&\bar\psi(0^+,\eta^-,\vec{\bf 0}_\perp) [0^+,\eta^-,\vec{\bf 0}_\perp; \, 0^+,-\infty^-,\vec{\bf 0}_\perp]_{A^+}
\Gamma\times
\nonumber\\
&&[0^+,-\infty^-,\vec{\bf 0}_\perp; \, 0^+,0^-,\vec{\bf 0}_\perp]_{A^+} \psi(0^+,0^-,\vec{\bf 0}_\perp).
\end{eqnarray}
Here $\Gamma$ implies a relevant combination of $\gamma$-matrices.
The Wilson line (\ref{WL1-mir}) is a result of summation in the mirror diagram and
the Wilson line (\ref{WL1-dir}) appears in the direct diagram.

The sum of direct and mirror diagram contributions takes place
if we study the spin-average DY hadron tensor. While, for the single transverse
spin asymmetry, we deal individually with only the direct (or mirror) diagram contribution
because the direct and mirror diagrams differ in sign to construct the corresponding SSA.
For our further considerations in the context of contour gauge, it is not so
crucial what kind of hadron tensors we work with.

The non-standard (direct and mirror) diagrams give us the contributions with
the Wilson lines
\begin{eqnarray}
\label{WL2}
&&[-\infty^+,\eta^-,\vec{\bf 0}_\perp; \, 0^+,\eta^-,\vec{\bf 0}_\perp]_{A^-},\quad \text{and}
\nonumber\\
&&[0^+,0^-,\vec{\bf 0}_\perp; \, -\infty^+,0^-,\vec{\bf 0}_\perp]_{A^-},
\end{eqnarray}
and, therefore, we have the string operator
\begin{eqnarray}
\label{OP2}
&&\bar\psi(0^+,\eta^-,\vec{\bf 0}_\perp) [-\infty^+,\eta^-,\vec{\bf 0}_\perp; \, 0^+,\eta^-,\vec{\bf 0}_\perp]_{A^-}
\Gamma\times
\nonumber\\
&&[0^+,0^-,\vec{\bf 0}_\perp; \, -\infty^+,0^-,\vec{\bf 0}_\perp]_{A^-} \psi(0^+,0^-,\vec{\bf 0}_\perp).
\end{eqnarray}

According to the contour gauge conception,
we eliminate all the Wilson lines with the longitudinal (unphysical) gluon fields $A^+$ and $A^-$.
We note that the ideologically similar approach can be found in \cite{Belitsky:2002sm}.

We begin with the Wilson lines shown in Eqns.~(\ref{WL1-mir})-(\ref{WL1-dir}),
we write the following gauge fixing conditions:
\begin{eqnarray}
\label{CG-1w1}
&&[0^+,\eta^-,\vec{\bf 0}_\perp; \, 0^+,-\infty^-,\vec{\bf 0}_\perp]_{A^+}=\mathds{1},
\nonumber\\
&&[0^+,-\infty^-,\vec{\bf 0}_\perp; \, 0^+,0^-,\vec{\bf 0}_\perp]_{A^+}=\mathds{1}
\end{eqnarray}
explicit solutions of which read
\begin{eqnarray}
\label{Aplus-1}
&&A^+(0^+,\mathds{L}_{0^-,-\infty^-},\vec{\bf 0}_\perp)=0,
\\
\label{Aplus-2}
&&A^+(0^+,\mathds{L}_{-\infty^-,\eta^-},\vec{\bf 0}_\perp) = 0.
\end{eqnarray}
Here $\mathds{L}_{x,y}$ denotes the straightforward line in the Minkowski space connecting point $x$ with point $y$.
In the contour gauge (\ref{CG-1w1})-(\ref{Aplus-2}), the remaining gluon field components can be represented as
(with $\mu=-,\perp$)
\begin{eqnarray}
\label{InRep-1}
&&A^{\mu}_G(0^+,x^-,\vec{\bf 0}_\perp)\Big|_{c.g.(\ref{CG-1})-(\ref{Aplus-2})}
= \int\limits_{-\infty^-}^{x^-} dz_\alpha \frac{\partial z_\beta}{\partial x_\mu}
G^{\alpha\beta}(z|A^{\mu}_{c.g.})
\nonumber\\
&&
=\tilde{n}^-\int\limits_{0}^{\infty} ds e^{-\epsilon s} G^{+\mu}(x^-- \tilde{n}^-s|A^{\mu}_{c.g.})
\end{eqnarray}
with the boundary condition
\begin{eqnarray}
\label{bc}
A^{\mu}_{b.c.}(0^+,x^--\tilde{n}^-\infty,\vec{\bf 0}_\perp)\Big|_{c.g.(\ref{CG-1})-(\ref{Aplus-2})}=0.
\end{eqnarray}
In Eqn.~(\ref{InRep-1}), we use the parametrization of $\mathds{L}_{-\infty^-,x^-}$ as
\begin{eqnarray}
\label{ParL}
&&z(s)=\big(0^+,x^--\tilde{n}^-\lim\limits_{\epsilon\to 0}\frac{1-e^{-\epsilon s}}{\epsilon},\vec{\bf 0}_\perp\big),
\\
&&dz_{\alpha}\Big|^{-\infty}_{x}=\tilde{n}_\alpha ds e^{-\epsilon s} \Big|_{0}^{\infty}.
\nonumber
\end{eqnarray}

We now dwell on the gauge conditions for $A^-$ gluon component.
We put the Wilson lines (\ref{WL2}) to be equal to $1$ too, {\it i.e.}
\begin{eqnarray}
\label{CG-2}
&&[-\infty^+,\eta^-,\vec{\bf 0}_\perp; \, 0^+,\eta^-,\vec{\bf 0}_\perp]_{A^-}=\mathds{1},
\nonumber\\
&&[0^+,0^-,\vec{\bf 0}_\perp; \, -\infty^+,0^-,\vec{\bf 0}_\perp]_{A^-}=\mathds{1}.
\end{eqnarray}
These conditions yield
\begin{eqnarray}
\label{Aminus-1}
&&A^-(\mathds{L}_{0^+,-\infty^+},\eta^-,\vec{\bf 0}_\perp)=0,
\\
\label{Aminus-2}
&&A^-(\mathds{L}_{-\infty^+,0^+},0^-,\vec{\bf 0}_\perp) = 0.
\end{eqnarray}
As above, in the contour gauge (\ref{CG-2})-(\ref{Aminus-2}), the remaining
gluon fields have the integral representations which read
(here $\mu=+,\perp$)
\begin{eqnarray}
\label{InRep-2}
&&A^{\mu}_G(x^+,\eta^-,\vec{\bf 0}_\perp)\Big|_{c.g.(\ref{CG-2})-(\ref{Aminus-2})}
= \int\limits^{-\infty^+}_{x^+} dz_\alpha \frac{\partial z_\beta}{\partial x_\mu}
G^{\alpha\beta}(z|A^{\mu}_{c.g.})
\nonumber\\
&&=-\breve n^+ \int\limits_{0}^{\infty} dt e^{-\epsilon t}
G^{-\mu}\big(x^+ - \breve n^+ t|A^{\mu}_{c.g.}\big)
\end{eqnarray}
with the boundary condition
\begin{eqnarray}
\label{bc-1}
A^{\mu}_{b.c.}(x^+ -\breve n^+ \infty,\eta^-, \vec{\bf 0}_\perp)
\Big|_{c.g.(\ref{CG-2})-(\ref{Aminus-2})}=0.
\end{eqnarray}
In Eqn.~(\ref{InRep-2}) the path parametrization of $\mathds{L}_{x,-\infty}$ is given by
\begin{eqnarray}
\label{ParL-1}
&&z(s)=\big(x^+ -\breve n^+\lim\limits_{\epsilon\to 0}\frac{1-e^{-\epsilon t}}{\epsilon},\eta^-,\vec{\bf 0}_\perp\big),
\\
&&dz_{\alpha}\Big|^{-\infty}_{x}=-\breve n^+_{\alpha} dt e^{-\epsilon t} \Big|_{0}^{\infty}.
\nonumber
\end{eqnarray}

Further, the gluon field $A^{-}_G$ of Eqn.~(\ref{InRep-1}) has to be
compatible with the gluon field $A^-$ of Eqn.~(\ref{Aminus-1}).
Also, the same inference has to be valid for the gluon fields $A^{+}_G$ of Eqn.~(\ref{InRep-2})
and $A^+$ of Eqn.~(\ref{Aplus-2}).
We thus require the analytical continuity for these gluon fields at the interception points,
see Fig.~\ref{Fig-WL},
and we finally arrive at the following conditions (here we omit the subscript $G$)
\begin{eqnarray}
\label{Gfix-1}
&&A^+(0^+,x^-=\eta^-,\vec{\bf 0}_\perp) = A^+(x^+=0^+,\eta^-,\vec{\bf 0}_\perp) =0,
\nonumber\\
&&A^-(x^+=0^+,\eta^-,\vec{\bf 0}_\perp) = A^-(0^+,x^-=\eta^-,\vec{\bf 0}_\perp) =0,
\nonumber\\
\end{eqnarray}
respectively. Having used these conditions, we stay with the
physical gluon fields $A^\perp$ only.

\subsection{Gluon Propagator}

We now go over to consideration of the gluon propagator.
In the case of local axial gauge $A^+=0$, the gluon propagator
is still not a well-defined object because of the spurious singularity
related to the residual gauge transformations.
In other words, the axial gauge cannot fix completely the unique element of each
orbit defined on the gauge group. In Appendix~\ref{G-Rsymmetry:App:B},
we present the handbook material regarding the gauge and residual gauge fixing.
It is clear that if, in the local axial gauge $A^+=0$, we fix the residual gauge
by requiring $\theta^a_0(k^-,\vec{\bf k}_\perp)=0$ (see, Eqns.~(\ref{ReAx-con-p-1})-(\ref{ReAx-con-reg}))
we immediately get that $A^-=0$ as well. The same inference can be reached by the simplest way
if we use the contour gauge conception (see, Eqn.~(\ref{Gfix-1})).
Notice that the maximal gauge fixing which is based on the contour gauge conception
does not relate technically to the problem of finding the inverse kinematical operator
(see, Eqns.~(\ref{Kop})-(\ref{Cont-Eqn})). The contour gauge approach is, therefore,
an alternative method of gauge fixing compared to the ``classical'' approaches based on the
corresponding effective Lagrangian (see, for example, \cite{Chirilli:2015fza}).

So, we perform our calculation in the contour gauge defined by Eqns.~(\ref{CG-1}) and/or (\ref{CG-2})
together with the conditions of Eqn.~(\ref{Gfix-1}) where the only physical gluons are presented.
In the framework of collinear factorization under our consideration,
the gluon momentum has the plus dominant components.

Having used the Wilson lines from the standard and non-standard diagrams,
we calculate the gluon propagator which reads
\begin{eqnarray}
\label{GP-1}
\langle 0| T A^{\mu}_\perp(0^+,x^-, \vec{\bf 0}_\perp)
A^{\nu}_\perp(0^+,0^-, \vec{\bf 0}_\perp)|0\rangle = D^{\mu\nu}_\perp(x^-).
\end{eqnarray}
Using the integral representation (\ref{InRep-1}), the gluon propagator takes the form of
\begin{eqnarray}
\label{GP-2}
&&D^{\mu\nu}_\perp(x^-)=
n_\alpha n_\beta \int\limits_0^\infty ds ds^\prime e^{-\epsilon s - \epsilon s^\prime}
\times
\nonumber\\
&&
\langle 0| T G^{\mu\alpha}(x^--\tilde n^-s) G^{\nu\beta}(0^--\tilde n^-s^\prime)|0\rangle=
\nonumber\\
&&\int(d^4 \ell) e^{-i\ell^+x^-} \frac{1}{\ell^2+i0}
\frac{(\ell^+)^2 d^{\mu\nu}_\perp(\ell)}{(\ell^++i\epsilon)(\ell^+-i\epsilon)}.
\end{eqnarray}
In Eqn.~(\ref{GP-2}), we have explicitly performed the integration over $ds(ds^\prime)$:
\begin{eqnarray}
\label{s-int}
\int\limits_0^\infty ds e^{\pm is(\ell^+ \pm i\epsilon)} = \frac{\pm i}{\ell^+ \pm i\epsilon}
\end{eqnarray}
which emanates from the path parametrization.
It is worth to emphasize the gluon pole prescription can be traced from this
kind of integrations.
The transverse tensor $d^{\mu\nu}_\perp$ has been constructed as
\begin{eqnarray}
\label{d-tensor}
d^{\mu\nu}_\perp(\ell)= g^{\mu\nu}- \frac{\ell^{\mu,+} n^\nu + \ell^{\nu,+} n^\mu}{[\ell^+]_{\rm reg}}
\end{eqnarray}
where the spurious singularity $[\ell^+]_{\rm reg}$ has to be regularized.

We consider the combination
\begin{eqnarray}
\label{com}
\frac{(\ell^+)^2}{(\ell^++i\epsilon)(\ell^+-i\epsilon)}
d^{\mu\nu}_\perp(\ell).
\end{eqnarray}
The first term of Eqn.~(\ref{com}) includes the combination
\begin{eqnarray}
g^{\mu\nu}\ell^+ \frac{\ell^+}{(\ell^++i\epsilon)(\ell^+-i\epsilon)}
\end{eqnarray}
which has to be treated only as
\begin{eqnarray}
\label{g-met}
g^{\mu\nu}\frac{\ell^+}{2}\Big( \frac{1}{\ell^++i\epsilon} + \frac{1}{\ell^+-i\epsilon} \Big)
=g^{\mu\nu}\ell^+\frac{\mathcal{P}}{\ell^+}=g^{\mu\nu}.
\end{eqnarray}

On the other hand, for $x^- >0$ (see, the momentum integral (\ref{GP-2})), the integration contour
has to be closed in the lower semi-plane, $\Im\text{m} \ell^+ < 0$.
Hence, for the $g_{\mu\nu}$-term, we obtain the integrand
\begin{eqnarray}
\label{int-g}
g^{\mu\nu} \frac{\ell^+}{\ell^++i\epsilon}
\end{eqnarray}
where the denominator $\ell^+-i\epsilon$ has been cancelled by one of $\ell^+$ in the numerator.
It is clear that the remaining combination in Eqn.~(\ref{int-g}) yields $g_{\mu\nu}$ (cf. Eqn.~(\ref{g-met})).

Regarding the second term of Eqn.~(\ref{com}), we propose two ways of reasoning.

\noindent
{\it The first way:} We don't specify explicitly the tensor structure of this term.
The second term of Eqn.~(\ref{com}) can be written in the following form
(here the momentum flux direction is not fixed):
\begin{eqnarray}
\label{sec-term-1}
&&\frac{(\ell^+)^2}{(\ell^++i\epsilon)(\ell^+-i\epsilon)} \frac{L^{\mu\nu}(\ell,n)}{[\ell^+]_{\rm reg}}=
\nonumber\\
&&\ell^+\frac{\mathcal{P}}{\ell^+} \frac{L^{\mu\nu}(\ell,n)}{[\ell^+]_{\rm reg}},
\end{eqnarray}
where we use
\begin{eqnarray}
\frac{\mathcal{P}}{\ell^+}=\frac{\ell^+}{(\ell^+  + i\epsilon)(\ell^+ - i\epsilon)}.
\end{eqnarray}

To well-define the product of two generalized functions the pole $1/[\ell^+]_{\rm reg}$ must be treated only as
\begin{eqnarray}
\label{Reg-pv}
\frac{1}{[\ell^+]_{\rm reg}}=\frac{\mathcal{P}}{\ell^+}.
\end{eqnarray}
Indeed, we have
\begin{eqnarray}
\frac{\mathcal{P}}{\ell^+} \ell^+ \frac{\mathcal{P}}{\ell^+} = \frac{\mathcal{P}}{\ell^+}.
\end{eqnarray}
On the other hand, if we let $1/[\ell^+]_{\rm reg}$ be equal to $1/(\ell^+\pm i\epsilon)$,
we will face on the wrong-defined product of two generalized functions \cite{Vladimirov}:
\begin{eqnarray}
&&\frac{\mathcal{P}}{\ell^+} \ell^+ \frac{1}{\ell^+\pm i\epsilon} =
\frac{\mathcal{P}}{\ell^+} \ell^+ \Big( \frac{\mathcal{P}}{\ell^+} \mp i\pi \delta(\ell^+)\Big)
\nonumber\\
&&\Longrightarrow \frac{\mathcal{P}}{\ell^+} \ell^+ \delta(\ell^+)\quad \text{-- wrong-defined product}.
\end{eqnarray}

\noindent
{\it The second way:} We take into account that the tensor structure includes
the plus component of the gluon momentum. Hence, the second term of of Eqn.~(\ref{com})
reads
\begin{eqnarray}
\label{sec-term-2}
\frac{(\ell^+)^2}{(\ell^++i\epsilon)(\ell^+-i\epsilon)} \frac{\ell^{\mu,+} n^\nu + \ell^{\nu,+} n^\mu}{[\ell^+]_{\rm reg}}.
\end{eqnarray}
Here, as shown above, for the first factor, we can again use that
\begin{eqnarray}
\label{sec-term-2-1}
\frac{(\ell^+)^2}{(\ell^++i\epsilon)(\ell^+-i\epsilon)}=\ell^+\frac{\mathcal{P}}{\ell^+}=1
\end{eqnarray}
and, for the second factor, we have
\begin{eqnarray}
\label{sec-term-2-2}
\frac{\ell^{\mu,+} n^\nu + \ell^{\nu,+} n^\mu}{[\ell^+]_{\rm reg}} =
\frac{\ell^+}{[\ell^+]_{\rm reg}} \big( n^{*\,\mu} n^\nu + n^{*\,\nu} n^\mu\big).
\end{eqnarray}
Based on this expression, it is clear that the only possibility
is to define $1/[\ell^+]_{\rm reg}$ through the principle value, see Eqn.~(\ref{Reg-pv}).

Thus, in the contour gauge generated by both the standard and non-standard diagrams,
the gluon propagator reads
\begin{eqnarray}
\label{GP-f1}
&&\hspace{-0.5cm}D^{\mu\nu}_\perp(x^-)=
\\
&&\hspace{-0.5cm}\int(d^4 \ell) e^{-i\ell^+x^-} \frac{1}{\ell^2+i0}
\Big\{ g^{\mu\nu} - \frac{\mathcal{P}}{\ell^+}\big(\ell^{\mu,+} n^\nu + \ell^{\nu,+} n^\mu\big) \Big\}
\nonumber
\end{eqnarray}
or, using Eqn.~(\ref{sec-term-2-1}), we obtain
\begin{eqnarray}
\label{GP-f1-2}
D^{\mu\nu}_\perp(x^-)=
\int(d^4 \ell) e^{-i\ell^+x^-} \frac{g^{\mu\nu}_\perp}{\ell^2+i0},
\end{eqnarray}
where $g^{\mu\nu}_\perp = g^{\mu\nu} -  n^{*\,\mu} n^\nu - n^{*\,\nu} n^\mu$.

We notice that the gluon propagator presented in Eqn.~(\ref{GP-f1-2}) takes place for the
very specific case of the polarized DY hadron tensor under our consideration.
In the case of deep-inelastic scattering process, where the corresponding Wilson lines are different,
the gluon propagator derived in the contour gauge frame has the form similar to Eqn.~(\ref{GP-f3}), see below.
We also stress that, in Eqns.~(\ref{GP-f1}) and (\ref{GP-f1-2}), the gluon momentum flux is not important and is not specified.

We now consider a particular case wherein only the standard diagram exists.
For example, this can be achieved if we neglect the higher twist correlators
$\langle \bar\psi A^- \psi\rangle$
which appear in the non-standard diagram.
Moreover, the gluon field co-ordinates are not
necessarily on the minus direction and
the gluon momentum flux is fixed
in the positive direction from the $\nu$-vertex to $\mu$-vertex.
In this case, the gluon propagator reads
\begin{eqnarray}
\label{GP-f2}
&&D^{\mu\nu}(x)\Big|^{\text{stand. dia.}}_{\text{fixed flux}}=
\int(d^4 \ell) e^{-i\ell x} \frac{1}{\ell^2+i0}\times
\\
&&\Big\{ g^{\mu\nu} - \frac{\mathcal{P}}{\ell^+}\bigg(\ell^{\mu} n^\nu\theta(\ell^+)
+ \ell^{\nu} n^\mu \theta(-\ell^+)\bigg) \Big\}
\nonumber
\end{eqnarray}
where the corresponding $\theta$-functions specify the momentum flux.
Using the Cauchy theorem in Eqn.~(\ref{GP-f2}), we finally arrive at
\begin{eqnarray}
\label{GP-f3}
&&D^{\mu\nu}(x)\Big|^{\text{stand. dia.}}_{\text{fixed flux}}=\int(d^4 \ell)  \frac{e^{-i\ell x}}{\ell^2+i0}\times
\\
&&\Big\{ g^{\mu\nu} - \frac{\ell^{\mu} n^\nu}{\ell^+- i\epsilon} -
\frac{\ell^{\nu} n^\mu}{\ell^+ + i\epsilon} \Big\}
\nonumber
\end{eqnarray}
which coincides with the results in \cite{Belitsky:2002sm}, \cite{Chirilli:2015fza}.
This expression is sensitive to the definition of the positive (negative)
flux direction (see, Eqn.~(\ref{GP-f2})).
Hence, the symmetry over $\mu\leftrightarrow\nu$ takes place only together with
the simultaneous replacement $\ell^+ \leftrightarrow -\ell^+$ in the second and third terms of
Eqn.~(\ref{GP-f3}).

\section{Contour gauge and the decomposition theorem for gluons}
\label{CG-Dec-Th}

In this section, we present the other practical example where the usage of contour gauge
plays a significant role.
%
%
We now concentrate on a reexamination of the gluon decomposition given by
\begin{eqnarray}
\label{Decom-1}
A_\mu(x)= A_\mu^{\text{pure}}(x) + A_\mu^{\text{phys.}}(x).
\end{eqnarray}
We want to make a clarification of the conditions
which provide the decomposition validity.
We intend to consider Eqn.~(\ref{Decom-1}) as a statement
which must be proven at least within the gauge condition that is more suitable for
a demonstration of Eqn.~(\ref{Decom-1}).
To this end, we adhere the contour gauge conception.

At the beginning, we remind that, within the Hamiltonian formulation of gauge theory \cite{Faddeev:1980be},
the extended functional integration measure over
the generalized momenta, $E_i$, and coordinates, $A_i$, includes
two kinds of the functional delta-functions. The first kind of delta-functions reflects the
primary (secondary etc) constraints on $E_i$ and $A_i$,
while the second kind of delta-functions refers to the so-called additional constraints (or gauge conditions)
the exact forms of which have been dictated by the gauge freedom.
If the primary (secondary etc) constraints are needed to exclude the unphysical gauge field components,
the gauge conditions would allow, in the most ideal case, to fix the corresponding Lagrange factor related to the gauge orbit.
Focusing on the Lagrangian formulation, since the infinite volume of gauge orbit
is factorized out in the functional measure over the gauge field components, the gauge conditions work for
the elimination of unphysical gluon components.

In this connection, the contour gauge implies that in order to fix completely the gauge function (orbit representative) or
to eliminate the unphysical gluons, one can demand the Wilson path functional
between the starting point $x_0$ and the final destination point $x$, $P(x_0,x)$, to be equal to unity, {\it i.e.}
${\bf g}(x | A)=[x\, ; \, x_0 ]_A =\mathbb{I}$, see Eqn.~(\ref{CG-1}).
We remind that
the path $P(x_0,x)$ is now fixed and $x_0$ is a very special starting point that might depend on
the destination point $x$, see also \cite{Weigert:1992my}.

In fact, the well-known axial gauge, like $A^+=0$, is a particular case of the most general non-local contour gauge
determined by the condition of Eqn.~(\ref{CG-1}) if the fixed path is the straightforward line
connecting $\pm\infty$ with $x$.

We remind that, by construction, the contour gauge does not suffer from the residual gauge freedom.
It gives, from the technical point of view, the simplest way to fix totally the gauge in the finite space.
We can thus uniquely define the point in the fiber bundle, ${\cal P}({\cal X}, \pi \,|\, G)$, which has the unique
horizontal vector corresponding to the given tangent vector at $x\in {\cal X}$.
The tangent vector at the point $x$ is uniquely determined by the
given path passing through $x$.
That is, within the Hamilton formalism based on the geometry of gluons
the condition of Eqn.~(\ref{CG-1}) corresponds to the determining of the surface
on ${\cal P}({\cal X}, \pi \,|\, G)$. This surface is parallel to the base plane with the path and
singles out the identity element, ${\bf g}=1$, in every fiber of ${\cal P}({\cal X}, \pi \,|\, G)$ \cite{An-Sym}.

As above-mentioned, the contour gauge
naturally generalises the familiar local axial-type of gauges.
In contrast to the local axial gauge,
the contour gauge does not possess the residual gauge freedom in the finite region of a space.
However, as shown below, the boundary gluon configurations can generate the special class of the
residual gauges.

We are going over to the discussion of the contour gauge defined by the condition of Eqn.~(\ref{CG-1}).
Having used the path-dependent gauge transformations for gluons (see Eqn.~(\ref{A-trans-CG})), and
having calculated the derivation of the Wilson line \cite{Durand:1979sw}, we readily derive that
in the gauge $[x\, ; \, - \infty ]_A =\mathbb{I} $ the gluon field can be presented as the following decomposition
\begin{eqnarray}
\label{A-cg}
A^{\text{c.g.}}_\mu(x) = \int_{-\infty}^{ x} d\omega_\alpha G_{\alpha\mu}(\omega | A^{\text{c.g.}}) +
A^{\text{c.g.}}_\mu( x - n\, \infty ),
\end{eqnarray}
where $G_{\mu\nu}$ is the gluon strength tensor;
the starting point is now equal to $-\infty$ and the path parametrization is given by
\begin{eqnarray}
\label{Path-par}
\omega \Big|_{x}^{ - \infty} = x - n \lim_{\epsilon\to 0} \frac{1-e^{-s\epsilon}}{\epsilon} \Big|_{s=0}^{s=\infty}.
\end{eqnarray}
This path parametrization includes the vector $n$ defined a given fixed direction.
As usual, the vector $n$ becomes a minus light-cone basis vector, $n=(0^+, n^-, {\bf 0}_\perp)$, within the approaches where
the light-cone quantization formalism has been applied.

Notice that the decomposition of Eqn.~(\ref{A-cg}) differs substantially from \cite{Hatta:2011zs}
by the absence of $\epsilon$-function. Indeed, the given contour gauge chooses either one $\theta$-function
or the other, see \cite{Anikin:2015xka} for details.

From Eqn.~(\ref{A-cg}), we can see that the contour gauge allows the gluon field to be naturally separated on
the $G$-dependent and $G$-independent components. That is, instead of Eqn.~(\ref{A-cg})
it is instructive to write the separation as (cf. \cite{Bashinsky:1998if})
\begin{eqnarray}
\label{A-sep}
A^{\text{c.g.}}_\mu(x) = A_\mu(x | G) + A^{\text{b.c}}_\mu(-\infty),
\end{eqnarray}
where $A_\mu(x | G)$  is nothing but the
first term of Eqn.~(\ref{A-cg}) and
the boundary gluon configuration defined as
$A^{\text{b.c}}_\mu(-\infty)\equiv A^{\text{c.g.}}_\mu(x - n\, \infty)$.
It is worth to notice that (a) the $G$-dependent configuration  $A_\mu(x | G)$ stems from the nontrivial deformation of a path
\cite{Durand:1979sw}; (b) the gluon separation presented by Eqn.~(\ref{A-sep})
resembles the equation of  \cite{Bashinsky:1998if} but differs slightly by meaning.

In the contour gauge, see Eqn.~(\ref{CG-1}), the boundary gluon configurations have to fulfil the condition as
\begin{eqnarray}
\label{bc-cond}
\mathbb{P}\text{exp} \Big\{ ig A^{\text{b.c}}_\mu(-\infty)\int_{-\infty}^x d\omega_\mu \Big\}=\mathbb{I}.
\end{eqnarray}
Therefore, since the integral over $d\omega_\mu$ in Eqn.~(\ref{bc-cond}) is divergent as $1/\epsilon$ at $\epsilon$ goes to zero,
the combination $n_\mu A^{\text{b.c}}_\mu(-\infty)$  should behave as $\epsilon^2$.
Indeed, the exponential function of Eqn.~(\ref{bc-cond}) reads (here, we deal with the space where the dimension is $D=4$)
\begin{eqnarray}
\label{bc-cond-2}
&&A^{\text{b.c}}_\mu(-\infty)\int_{-\infty}^x d\omega_\mu \equiv
A^{\text{b.c}}_\mu(x -\infty \,n)\int_{-\infty}^x d\omega_\mu=
\nonumber\\
&&\Big\{ \lim_{\epsilon\to 0} \Big(\frac{1}{\epsilon} \Big)^{-1}\Big\}
A^{\text{b.c}}_\mu(n) \, n_\mu  \Big\{ \lim_{\delta\to 0} \frac{1}{\delta}\Big\} = 0.
\end{eqnarray}
Hence, the boundary gluon configurations obey the transversity condition as
\begin{eqnarray}
\label{bc-cond-3}
n_\mu(\theta_i,\, \varphi) \, A^{\text{b.c}}_\mu\big (n(\theta_i,\, \varphi) \big)  = 0.
\end{eqnarray}
Here, since the vector $n$ defines the fixed direction it is more convenient to use the spherical co-ordinates in the Euclidean space
(or the pseudo-spherical system in the Minkowski space) where the vector $n$ depends on the angle co-ordinates
$(\theta_i, \varphi)$ only, see below.
If the space dimension is $D>4$, the transversity condition of Eqn.~(\ref{bc-cond-3}) is not necessary to fulfil the contour gauge
condition.

We are in position to show that in the contour gauge the boundary gluon configurations have the only form of the pure gauge
configurations.
First of all, the starting point $x_0$ plays the special role in the
considered formalism because all the paths originate from this point and the base ${\cal X}$ touches the
principle fiber bundle ${\cal P}$ only at this point in the general path-dependent gauge by construction.

\begin{figure}[t]
\centerline{\includegraphics[width=0.4\textwidth]{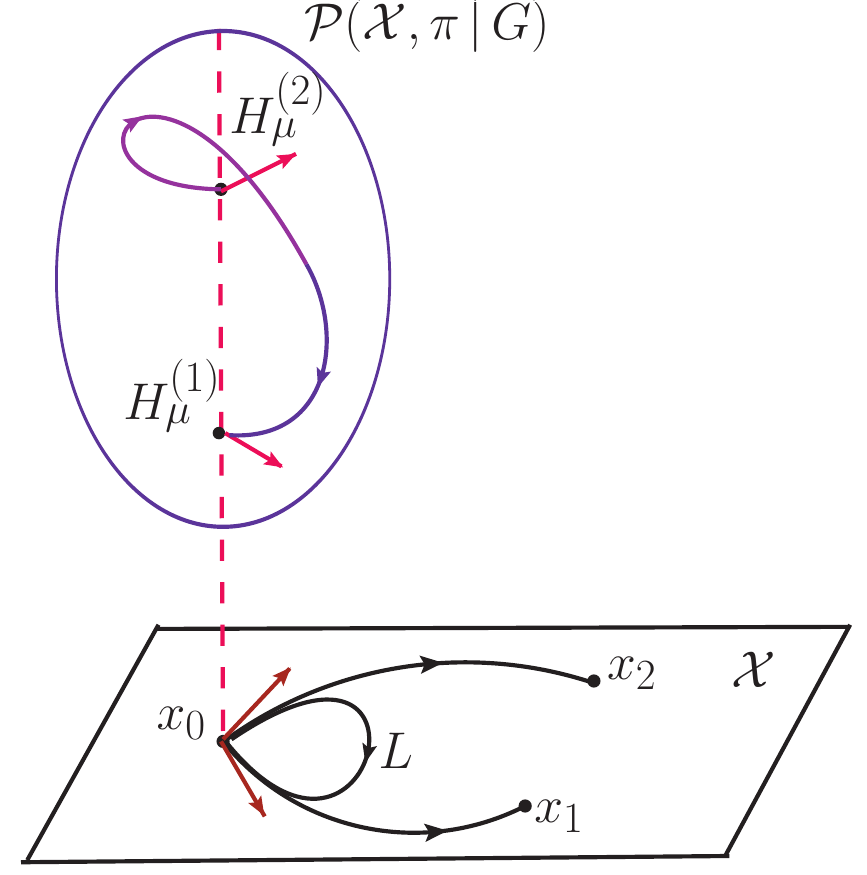}}
\caption{The holonomy: $H^{(i)}_\mu$ with $i=1,2$ denote the corresponding horizontal vectors defined on the
given fiber of ${\cal P}$. }
\label{Fig-1}
\end{figure}

Let us consider the point $x_0$ where, say, two different paths are started, see Fig.~(\ref{Fig-1}).
This starting point has two tangent vectors associated with $P(x_0, x_1)$ and $P(x_0, x_2)$.
In its turn, every of tangent vectors has the unique horizontal vector $H_\mu^{(i)}$ defined in the fiber.
Then, making use of Eqn.~(\ref{CG-1}) we can obtain that
\begin{eqnarray}
\label{bc-cond-4_1}
&&\mathbb{P}\text{exp} \Big\{ ig \int_{L(x_0)} d\omega_\mu A_\mu(\omega)\Big\}=\mathbb{I}
\quad \text{and} \quad
\\
&&
\label{bc-cond-4_2}
\mathbb{P}\text{exp} \Big\{ ig \int_{\Omega} d\omega_\mu \wedge d\omega_\nu G_{\mu\nu}(\omega)\Big\}
=\mathbb{I},
\end{eqnarray}
where $L(x_0)$ implies the loop with a basepoint $x_0=-\infty$ and $\Omega$ is the corresponding surface related to the
loop $L(x_0)$, see Fig.~\ref{Fig-1}.
%
%
%
Hence, we directly get that
\begin{eqnarray}
\label{bc-cond-5}
A_\mu(\omega) = \frac{i}{g}\,U(\omega) \partial_\mu U^{-1}(\omega)
\end{eqnarray}
from Eqn.~(\ref{bc-cond-4_1}), and  $G_{\mu\nu}(\omega)=0$ from  Eqn.~(\ref{bc-cond-4_2}) after the Stocks theorem has been used.

In the path group theory it states that any loop as a element of the loop subgroup can homotopically be transformed to the
"null element" which is, in our case, the basepoint $x_0=-\infty$.
As a result, the pure gauge representation of Eqn.~(\ref{bc-cond-5})
is valid for the boundary configurations  as well, {\it i.e.} we have
\begin{eqnarray}
\label{bc-cond-6}
A^{\text{b.c.}}_\mu(x_0) = \frac{i}{g}\,U(x_0) \partial_\mu U^{-1}(x_0).
\end{eqnarray}

Finally, combining Eqns.~(\ref{A-sep}) and (\ref{bc-cond-6}), one can prove that
the gluon field can be really decomposed only within the contour gauge approach: 
\begin{eqnarray}
\label{conc-1}
A^{\text{c.g.}}_\mu(x) = A_\mu (x | G)  + \frac{i}{g}\,U(x_0) \partial_\mu U^{-1}(x_0)\Big|_{x_0=-\infty},
\end{eqnarray}
where both terms are perpendicular to the chosen direction vector $n_\mu$.

\begin{figure}[t]
\centerline{\includegraphics[width=0.4\textwidth]{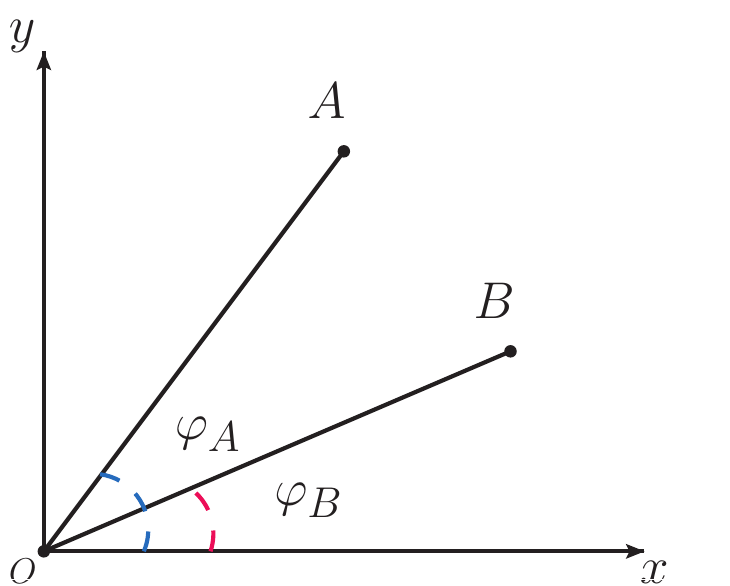}}
\caption{The angle independence of the starting point $O$: $A=(R_A, \varphi_A)$ and
$B=(R_B, \varphi_B)$; $\lim_{R_{A}\to 0} A=\lim_{R_{B}\to 0} B=O$  where
$O=(0\cdot\cos\varphi_i, 0\cdot\sin\varphi_i)$ with $i=A, B$.}
\label{Fig-2}
\end{figure}

Eqn.~(\ref{conc-1}) demonstrates the fact that the residual gauge of contour gauge is entirely located at the boundary.
To understand the nature of the residual gauge related to the boundary gluon configurations, 
we consider the simplest example of $\mathbb{R}^2$ where $A$ and $B$ have the only starting point $O$,
see Fig.~\ref{Fig-2}. In terms of the spherical system, we have that
$A(R_A, \varphi_A)\equiv (R_A \cos\varphi_A, R_A\sin\varphi_A)$ etc.
If the radius vectors of both $A$ and $B$ are not equal to zero (even in an infinitesimal sense), we are able to 
distinguish these two points.
But, if $R_A=R_B=0$, the starting point $O$ does not possesses any information on the vectors $A$ and $B$
(the vectors $A$ and $B$ merely disappear). 
It is true because of
$O=(0\cdot\cos\varphi_A, 0\cdot \sin\varphi_A)=(0\cdot \cos\varphi_B, 0\cdot \sin\varphi_B)$. 
In this case, the angles can be arbitrarily chosen.
Therefore, we may say that the starting point $O$ is the angle independent point.

If $x_0=\lim_{R\to 0} X(R, \theta_1, \theta_2, \varphi)$, we obtain
\begin{eqnarray}
\label{res-g-1}
A^{\text{b.c.}}_\mu\big(  \bar\epsilon n(\theta_i, \varphi) \big) =\frac{i}{g}\,
U\big( \bar\epsilon n(\theta_i, \varphi) \big)
\partial_\mu U^{-1}\big( \bar\epsilon n(\theta_i, \varphi)  \big),
\end{eqnarray}
where $\bar\epsilon\to -\infty$ and
$\theta_i$ and $\varphi$ are unfixed.  It ensures that 
the residual gauge freedom exists in the similar way as in Fig.~\ref{Fig-2}.
%
%
%

The convenient way how to fix the residual gauge freedom is as follows.  We assume
\cite{Anikin:2010wz, Anikin:2015xka}
\begin{eqnarray}
\label{res-g-2}
A^{\text{b.c.}}_\mu\big( \bar\epsilon n(\theta_i, \varphi) \big)\equiv
A^{\text{pure}}_\mu\big( \bar\epsilon n(\theta_i, \varphi) \big)=0.
\end{eqnarray}
Given that, the decomposition of Eqn.~(\ref{Decom-1}) becomes a trivial one.

We have, thus, demonstrated that the use of contour gauge approach provides the natural decomposition
of gluon fields. This natural decomposition involves the $G$-dependent gluon component (this is the physical one)
and the unphysical gluon component associated with the configuration of pure gauge.
Besides, in the finite region of space, the contour gauge does not suffer from the residual gauge.
At the same time, the remaining possible residual gauge has entirely been isolated on the infinite boundary
of space.

\subsection{Non-zero boundary gluon configurations}

Let us study the influence of non-zero boundary gluon configurations on the definitions of different parton distributions.
We first emphasize that our decomposition of Eqns.~(\ref{A-cg}) and (\ref{A-sep}) relates in the meaning to the
decomposition of \cite{Bashinsky:1998if}. Indeed, we are able to rewrite Eqn.~(\ref{A-sep}) as
(here, we use the limit of $\bar\epsilon\to - \infty$)
\begin{eqnarray}
\label{A-dec-BJ}
\tilde A^{\text{l.c.}}_\mu(k^+; \tilde x) = G_\mu (k^+; \tilde x)  +
\delta(k^+) A^{\text{b.c.}}_\mu\big( \bar\epsilon n^-(\pi/4, 0, 0) ; \tilde x \big),
\end{eqnarray}
where
the light-cone gluon field $\tilde A^{\text{l.c.}}_\mu$ is the Fourier image of $A^{\text{l.c.}}_\mu$ with respect to $x^-$ only,
{\it i.e.}
\begin{eqnarray}
\label{F-im}
A^{\text{l.c.}}_\mu (x^-; \tilde x)\stackrel{{\cal F}}{=} \tilde A^{\text{l.c.}}_\mu(k^+; \tilde x),
\end{eqnarray}
and, therefore, we have
\begin{eqnarray}
\label{M-BJ}
&&G^\mu (k^+; \tilde x) \stackrel{{\cal F}}{=} \int_{-\infty^-}^{ x^-} d\omega^- G^{ + \mu}(\omega^-, \tilde x | A^{\text{c.g.}})
\nonumber\\
&&\delta(k^+)\,A^{\text{b.c.}}_\mu \big( \bar\epsilon n^- ; \tilde x \big)\stackrel{{\cal F}}{=}
A^{\text{b.c.}}_\mu\big( \bar\epsilon n^- ; \tilde x \big).
\end{eqnarray}
Here, we underline that Eqns.~(\ref{A-dec-BJ}), as well as  Eqn.~(\ref{A-sep}), has been derived by direct solution of
the contour gauge requirement, see Eqn.~(\ref{CG-1}).
As mentioned, the important finding of the present paper is that despite the contour gauge fixes the whole gauge freedom in the finite domain
of space, it is still possible to deal with the residual gauge which is, however, located at the boundary field configurations only.
The non-trivial topological effects due to the boundary field configurations are forthcoming in the further our studies.

In \cite{Bashinsky:1998if}, the representation that is similar to our Eqn.~(\ref{A-dec-BJ}) has rather been guessed in the local axial gauge,
$A^+=0$, where the corresponding residual gauge freedom is incorporated into the inhomogeneous term with $\delta(k^+)$.
In turn, the gauge $A^+=0$ with the fixed residual gauge
freedom in the finite domain of space is actually identical to the unique contour gauge \cite{Anikin:2015xka}.

In the frame of the path group formalism, we have the following path-dependent transformation,
which  generates the usual translation transformation,
\begin{eqnarray}
\label{PG-tran-1}
\big[ \mathbb{U}^{P(x, x+y)}\psi \big](x) = [x+y\, ; \, x ]^{-1}_A \, \psi(x+y),
\end{eqnarray}
where $\psi(x)$ belongs to the spinor fundamental representation and has defined on
the Minkowski space $M=P/L$ ($P$ denotes the corresponding path group,
$L$ stands for the loop subgroup of $P$) as a invariant function of the conjugacy classes,
{\it i.e.} $\psi(x)={\bf g}(p)\Psi(p)$ with $p=(x, {\bf g}) \in{\cal P}$ \cite{Mensky:2012iy}.
Besides, in Eqn.~(\ref{PG-tran-1}) the operator $\mathbb{U}$ which acts on the spinor manifold has
the form of
\begin{eqnarray}
\label{U-op-1}
\mathbb{U}^{P(x, x+y)} = \mathbb{P}\text{exp} \Big\{- ig\int_{x}^{x+y} d\omega_\mu {\cal D}_\mu \Big\}.
\end{eqnarray}

In the contour gauge where the Wilson line of Eqn.~(\ref{PG-tran-1}) is fixed to be equal to unity,  the
transport operator $\mathbb{U}_q (y)$ takes the trivial form of
\begin{eqnarray}
\label{U-op-2}
\mathbb{U}^{P(x, x+y)} \Big|_{\text{c.g.}}= \mathbb{P}\text{exp} \Big\{- ig\int_{x}^{x+y} d\omega_\mu \partial_\mu \Big\}.
\end{eqnarray}
This operator does not include any information on the boundary configurations even if, say, $y\to\pm \infty$
because the boundary field configurations obey Eqn.~(\ref{CG-1}) too.
Moreover, in our case, the Wilson line of Eqn.~(\ref{PG-tran-1})
is set to unity due to the nullified integrand, $A^+=A^-=0$, and the nullified integral over $A_\perp$.

Hence, if we introduce the quark-gluon operators, forming the spin and orbital AM,
as the residual-gauge invariant operators, we have to use the covariant derivative as
$i\,{\cal D}^{\text{b.c.}}_\mu = i\,\partial_\mu  + g A^{\text{b.c.}}_\mu(-\infty)$.
In this sense, our results and the results of \cite{Bashinsky:1998if} are not much at variance .
For example, we readily obtain that
\begin{eqnarray}
\label{BJ-1}
&&f_{L_q}(x) = {\cal N}\, \int_{-\infty}^{+\infty} dz^- e^{\,i x\,P^+ z^-} \int_{-\infty}^{+\infty} d^2{\bf y}_\perp
\\
&&\times
\langle P |
\bar \psi(y_\perp) \gamma^+ y^{[ i }_\perp \, i {\cal D}^{j ]}_{\text{b.c.}} \psi(y_\perp + z^-)
|P \rangle,
\nonumber
\end{eqnarray}
where the antisymmetric combination $[i\, j]$ has been introduced  with $i, j= 1, 2$ and ${\cal N}$ is the normalization factor defined as in  \cite{Bashinsky:1998if}.

As above-mentioned, $f_{L_q}(x) $ as the physical quantity does not depend on the gauge choice.
At the same time, the axial-type (local or non-local) gauges are correlated with the fixed direction
which is also necessary for the factorization procedure \cite{Anikin:2009bf}.
Therefore, we are able to treat the gauge independency in the meaning of an independency on the chosen direction
as well.
In the frame of the Hamiltonian formalism, we assume that the gauge condition (or an additional condition)
can be completely resolved regarding the gauge function excluding the gauge transforms in the finite region.
In a sense, the physical quark-gluon operators, considered in the contour gauge, are ``gauge invariant'' by
construction because we do not deal with any gauge transforms in the finite region due to the fixed gauge function $\theta_{\text{fix}}$
(as above discussed, due to ${\bf g}=1$ in the fiber for the whole base ${\cal X}$), see \cite{An-Sym} for details.

\section{Conclusions}


In the paper, we have made public the important subtleties based on the mathematical
technique adjusted to the physical language.
We have presented the important explanations and analysis hidden in the preceding publications
which should help to clarify
the main advantages of the use of non-local contour gauges.
To this goal, the combination of the Hamilton and Lagrangian approaches to the
gauge theory has been exploited in our consideration.
Since the contour gauge is mainly backed on the geometrical interpretation of gluon fields as a connection on the
principle fiber bundle, we have provided the illustrative demonstration of geometry of the contour gauge.
In this connection, the Hamiltonian formalism is supposed to be more convenient for understanding
the subtleties of contour gauges. Indeed, the Lagrange factor $\lambda_a$ fixation has a direct treatment
in terms of the orbit group element which has been uniquely chosen by the corresponding plan transecting
the principle fiber bundle, see Fig.~\ref{Fig-S-2}.
While, as shown, the Lagrangian formalism is very well designed for the practical uses to the eliminate
the unphysical gluon degree of freedom from the corresponding amplitudes.

Also, we have reminded the details of that studies where the local axial-type gauge can lead to the certain ambiguities
in the gluon field representation. These ambiguities may finally produce incorrect results.
Meanwhile, as demonstrated in the paper, the non-local contour gauge can fix this kind of problems and, for example,
can provide not only the correct (gauge invariant) final result but also find the new contributions to
the hadron tensors of DY-like processes \cite{Anikin:2010wz,Anikin:2015xka}.
Thus, the use of contour gauge conception gives a possibility
(i) to find a solution of the gauge-invariance problem, discovered in the Drell-Yan hadron tensor, by the correct description
of the gluon pole that is appearing in the corresponding parton distributions;
(ii) to discover the new sizeable contributions to the single-spin asymmetries which are under intensive experimental studies.

In the context of the contour gauge use, the recent progress is mainly related to
the studies of the so-called gluon pole contributions to the
Drell-Yan-like processes \cite{Anikin:2010wz,Anikin:2015xka}. However, the practical profit
of the non-local contour gauge is not limited by the study of gluon poles which manifest
in the different hard processes. With the help of non-local gauges, we plan to adopt the
method based on the geometric quantization \cite{Nair:2016ufy} to the investigation of
different asymptotical (hard) regimes in QFT.


In the contour gauge, from the technical viewpoint, the maximal gauge fixing are not associated with
the problem of finding the inverse kinematical operator. Hence,
the contour gauge approach has to be considered as
the alternative method of gauge fixing in comparison with the ``classical'' approaches based on the
corresponding effective Lagrangians.
It is necessary to stress that the contour gauge contains the important and unique additional information (needed to fix
the prescription in the gluon poles) which is invisible in the case of usual (local) axial gauge.
From this point of view, before we discard the terms with $A^+$, we have to determine
the relevant fixed path in the corresponding Wilson line with $A^+$
which finally leads to the certain prescriptions in the gluon poles.
Moreover, the corresponding Wilson line with $A^-$ in the non-standard diagram, which contributes to the
polarized DY hadron tensor, prompts the way of residual gauge fixing.

We thus advocate the preponderance of the contour gauge use which allows to fix completely
the gauge freedom by the most illustrative and simplest way. We demonstrate that
the non-standard diagram plays the important role in forming of the relevant contour
in the corresponding Wilson line. Hence, from the viewpoint of contour gauge, there is no
way to neglect the additional non-standard diagram.


To conclude, we have expounded the useful correspondence between local and non-local gauges
which is extremely important to avoid the substantially wrong conclusions appeared in the literature.

We have proposed the proof of the following statement which is valid in the non-Abelian theory:
in the contour gauge the gluon field can be presented in the form
of decomposition
on the gluon configuration $A_\mu(x | G)$ being the physical degree of freedom and
the pure gauge gluon configuration $A_\mu^{\text{pure}}(x_0)$ that is totally isolated on the boundary
and includes the special type of residual gauge freedom.
We have demonstrated that the contour gauge condition cannot finally eliminate this, new-found, special residual gauge
the nature of which has been illustrated in detail.

In the case of the trivial boundary conditions, {\it i.e.}
$A^{\text{b.c.}}_\mu=0$, in the contour gauge
the decomposition of Eqn.~(\ref{Decom-1})  does not make a sense in the non-Abelian theory
because only the boundary gluon configurations can be presented as the pure gauge gluon configurations.
Moreover, if the boundary configurations have been nullified, there is no the gauge freedom at all
and, therefore, we deal with the gauge invariant operators by construction modulo the global gauge transformations that are
not essential for the bilinear forms.

As a last point, we want to mention that the gluon decomposition of \cite{Chen:2008ag}, which is formally similar to Eqn.~(\ref{Decom-1}),
has a status of the ansatz rather then a strong inference formulated and proven in our studies.
Moreover, it has a distinguished feature that
the gluon fields are separated into the physical and pure gauge gluon configurations
before the gauge condition has been fixed. Hence, in this case, in order to formulate
the ansatz they should demand to impose
an addition requirement to extract $A^{\text{pure}}_\mu(x)$ which is finally defined by $G^{\text{pure}}_{\mu\nu}(x) = 0$.
In its turn, this requirement appears naturally working within contour gauge conception, see Eqn.~(\ref{conc-1}).
Eqn.~(\ref{BJ-1}), is formally not at odds with \cite{Chen:2008ag,Bashinsky:1998if} but,
in a sense,  we are in contradiction with \cite{Ji:1996ek,Wakamatsu:2010cb}.

\section*{Acknowledgements}
We thank Xurong~Chen, I.~Cherednikov, Y.~Hatta, C.~Lorce, A.V.~Pimikov, L.~Szymanowski, O.V.~Teryaev, A.S.~Zhevlakov
for useful discussions.  A special thank goes to Prof.~Jian-Hui~Zhang for the illuminating discussions.
Also, the author thanks his colleagues  from 
the Chinese University of Hong Kong for a very warm hospitality. The work has been supported in part by PIFI Program.

\section*{Conflicts of interest}
The author claims no conflict of interests. The essential and determinative contributions of the author to the papers
which have been used for the present review are undisputed.

\appendix
\renewcommand{\theequation}{\Alph{section}.\arabic{equation}}
\section*{Appendix}

\section{Exponentiation of component $A^-$}
\label{Exp-Aminus:App:A}

In this Appendix, we demonstrate the method of the $A^-$ component exponentiation.
In fact, there are several methods how to exponentiate the gluon fields,
see e.g. \cite{Gross:1971wn, Balitsky:1987bk, Radyushkin:1983mj}. Here, we present an alternative
frame-independent and most efficient method mainly based on the approach described in \cite{BogoShir}, see \S 46.

\subsection*{Some conventional notations}

Before going further, we remind several conventions regarding how the gauge transformations match
the Wilson lines. Taking, for the sake of simplicity, the Abelian gauge theory
(in the case of interest the distinction between Abelian and non-Abelian groups is irrelevant),
let us assume that the fermion and gauge fields are transformed as
\begin{eqnarray}
\label{Wl-GT-1}
&&\psi^{\omega}(x)=e^{\pm i\theta(x)} \psi(x),
\\
\label{Wl-GT-1-2}
&&A_{\mu}^\omega(x)=A_\mu(x) \pm \partial_\mu\theta(x)
\end{eqnarray}
where  $\omega$ stands here for the gauge transformation.
Generally speaking, the signs at the gauge function $\theta$ in Eqns.~(\ref{Wl-GT-1}) and (\ref{Wl-GT-1-2}) are
conventional. If we fix the transformations as in Eqns.~(\ref{Wl-GT-1}) and (\ref{Wl-GT-1-2}), {\it i.e.}
the same sings in both expressions,
then we can readily see that the covariant derivative and
the {\it gauge-invariant} fermion string operator become
\begin{eqnarray}
\label{Wl-GT-2}
&&i{\cal D}_{\mu} = i\partial_\mu + g A_\mu(x),
\nonumber\\
&&\mathbb{O}^{\text{g.-inv.}}(x,y)=\bar\psi(y) [y ;\, x]_{A} \psi(x),
\end{eqnarray}
where the Wilson line is given by
\begin{eqnarray}
\label{Def-WL}
&&[y \,;\, x]_{A} \stackrel{{\rm def}}{=}
\mathbb{P}{\rm exp}\Big\{+ i g \int\limits^{y}_{x}
 dz_\mu \, A^\mu(z) \Big\}=
\\
&&\lim_{N\to\infty} [y \,;\, x_N]_{A}\,[x_N \,;\, x_{N-1}]_{A}...[x_1 \,;\, x]_{A}=
\nonumber\\
&&\lim_{N\to\infty} \big[1+igA(x_N)\cdot (y-x_N)\big]...\big[1+igA(x)\cdot (x_1-x)\big].
\nonumber
\end{eqnarray}
In Eqn.~(\ref{Def-WL}), the starting point $x$ and final point $y$ are connected by the certain path $\mathbb{P}\in \mathds{R}^4$
which allows the arrangement by pounding $\big\{ x_N \big\}^{y}_{x}$.

However, if the signs in both fermion and gauge boson transformations differ from each other, {\it i.e.}
\begin{eqnarray}
\label{Wl-GT-1-d}
&&\psi^{\omega}(x)=e^{\pm i\theta(x)} \psi(x),
\\
\label{Wl-GT-1-2-d}
&&A_{\mu}^\omega(x)=A_\mu(x) \mp \partial_\mu\theta(x),
\end{eqnarray}
the covariant derivative takes the form of
\begin{eqnarray}
\label{Wl-GT-2-d}
i{\cal D}_{\mu} = i\partial_\mu - g A_\mu(x),
\end{eqnarray}
while the {\it gauge-invariant} fermion string operator, in this case, reads
(see, for example, \cite{Belitsky:2002sm})
\begin{eqnarray}
\label{Wl-GT-2-2-d}
\mathbb{O}^{\text{g.-inv.}}(x,y)=\bar\psi(y) [x \,;\, y]_{A} \psi(x)
\end{eqnarray}
or
\begin{eqnarray}
\label{Wl-GT-2-2-d-2}
\mathbb{O}^{\text{g.-inv.}}(x,y)=\bar\psi(y) [y \,;\, x]^{-1}_{A} \psi(x)
\end{eqnarray}
with the Wilson line defined as in Eqn.~(\ref{Def-WL}).

In our paper, we adhere the conventions as in Eqn.~(\ref{Wl-GT-2}).

\subsection*{Description of the method}

We begin with the most illustrative subject which is the Green function in the external field.
The gluon radiation from the proper spinor line as shown in Fig.~\ref{Fig-DY}, the right panel,
is actually relevant to the Green function in the external field.

Consider the differential equation for the Green function
\begin{eqnarray}
\label{GF-1}
\big[i\widehat{\partial} + g \widehat{A}(x) \big] {\mathrm S}(x,y) = -\delta^{(4)}(x-y),
\end{eqnarray}
where the wide hat denotes the convolution with $\gamma$-matrices as $\widehat{A}=\gamma\cdot A$ etc.

We emphasize that the Green function defined by Eqn.~(\ref{GF-1}) is {\it not} gauge-invariant subject
(see, for example, \cite{Gross:1971wn, Balitsky:1987bk}).
As one can see below, namely the {\it gauge-noninvariant} Green function ensures the appearance of
the {\it gauge-invariant} fermion string operator in the corresponding hadron matrix element.

For the sake of simplicity and without the loss of generality, we assume that
$\partial^\mu=(0^+,\partial^-, \vec{\bf 0}_\perp)$, $A^\mu=(0^+,A^-, \vec{\bf 0}_\perp)$
and we, therefore, study the tensor combination as
${\mathrm S}^{[\gamma^+]}(x,y)\stackrel{{\rm def}}{=}\gamma^+{\mathrm S}(x,y)$. That is,
instead Eqn.~(\ref{GF-1}) we deal with the following differential equation
\begin{eqnarray}
\label{GF-2}
\big[i\partial^- + g A^-(x) \big] {\mathrm S}^{[\gamma^+]}(x,y) = -\delta^{(4)}(x-y).
\end{eqnarray}
Hence, in the operator forms, the Green function takes the form of
\begin{eqnarray}
\label{GF-3}
{\mathrm S}^{[\gamma^+]}(x,y) = -\frac{1}{\big[i\hat\partial^- + g \hat A^-(x) \big]}\delta^{(4)}(x-y)
\end{eqnarray}
where the small hat now denotes the corresponding operators. From the mathematical point of view,
the inverse operator is defined via the integral representation as
\begin{eqnarray}
\label{GF-4}
\frac{i}{\big[i\hat\partial^- + g \hat A^-(x) \big]} =\lim_{\epsilon\to 0}
\int\limits_{0}^{\infty}d\nu  e^{i\nu\big[i\hat\partial^- + g \hat A^-(x) + i\epsilon\big]}.
\end{eqnarray}
Hence, we can write the Green function as
\begin{eqnarray}
\label{GF-5}
{\mathrm S}^{[\gamma^+]}(x,y)&=&i \int\limits_{0}^{\infty}d\nu  e^{i\nu\big[i\hat\partial^- + g \hat A^-(x) + i\epsilon\big]} \delta^{(4)}(x-y)
\nonumber\\
&\equiv& i \int\limits_{0}^{\infty}d\nu {\cal U}(\nu).
\end{eqnarray}
Here and in what follows the limit symbol has been omitted.
In the momentum representation,  ${\cal U}(\nu)$ takes the form of
\begin{eqnarray}
\label{GF-6}
{\cal U}(\nu)=\int(d^4p) e^{-ip(x-y)+i\nu \hat p + i {\cal K}(x,\nu)-\epsilon \nu}
\end{eqnarray}
where the integration measure $(d^4p)$ includes all needed normalization constants and  we use
\begin{eqnarray}
e^{-\nu \hat\partial^-} e^{-ip(x-y)}= e^{-ip(x-y)}  e^{i \nu \hat p^-}
\end{eqnarray}
which defines how the operator acts.
In Eqn.~(\ref{GF-6}), the function ${\cal K}(x,\nu)$ is an unknown function
which we have to derive.

Since the function ${\cal U}(\nu)$ obeys (we can check that by straightforward calculations)
\begin{eqnarray}
\label{GF-7}
-i \frac{\partial {\cal U}(\nu)}{\partial\nu} = \big[i\hat\partial^- + g \hat A^-(x) + i\epsilon\big] {\cal U}(\nu),
\end{eqnarray}
the function ${\cal K}(x,\nu)$ has to satisfy the following differential equation
\begin{eqnarray}
\label{GF-8}
\frac{\partial {\cal K}(x,\nu)}{\partial\nu} = - \partial^- {\cal K}(x,\nu) + g A^-(x)
\end{eqnarray}
provided ${\cal K}(x,\nu=0)=0$.
A solution of Eqn.~(\ref{GF-8}) can be easily found (see, \cite{BogoShir}), it reads
\begin{eqnarray}
\label{GF-9}
{\cal K}(x,\nu)&&= g \int\limits_{0}^{\nu} ds \int(d^4k) e^{-ik(x-s\breve{n}^+)} A^-(k)
\nonumber\\
&&=g\int\limits_{0}^{\nu} ds A^-(x-s \breve{n}^+).
\end{eqnarray}
Using Eqn.~(\ref{GF-9}), the corresponding Green function takes the form of
\begin{eqnarray}
\label{GF-10}
{\mathrm S}^{[\gamma^+]}(x,y) =&&
i \int\limits_{0}^{\infty}d\nu e^{-\epsilon \nu}
\delta^{(4)}(x-y -\nu \breve{n}^+) \times
\nonumber\\
&&{\rm exp}\Big\{-ig\int\limits_{x}^{y} dz^+ A^-(z^+) \Big\}
\end{eqnarray}
where the standard integral representation for $\delta$-function has been used,
\begin{eqnarray}
\delta^{(4)}(x-y -\nu \breve{n}^+)=\int(d^4p) e^{-ip(x-y)+i\nu \breve{n}^+ p},
\end{eqnarray}
and we trade $x-\nu \breve{n}^+$ (see, the upper integral limit in integration over $dz^+$) for $y$
thanks for the argument of $\delta$-function.

The final stage is to write the integration of $\delta$-function as
\begin{eqnarray}
&&i\int\limits_{0}^{\infty} d\nu e^{-\epsilon\nu} \delta^{(4)}(x-y -\nu \breve{n}^+)=
\nonumber\\
&&i\int\limits_{0}^{\infty} d\nu e^{-\nu \hat \partial^- -\epsilon\nu} \delta^{(4)}(x-y)=
\nonumber\\
&&=-\frac{1}{[i\hat\partial^- + i\epsilon]} \delta^{(4)}(x-y)\equiv S^{c\,[\gamma^+]}(x-y).
\end{eqnarray}
Thus, we derive that
\begin{eqnarray}
\label{f-GF-1}
\hspace{-0.5cm}{\mathrm S}^{[\gamma^+]}(x,y)=S^{c\,[\gamma^+]}(x-y)\,[x \,;\, y]_{A^-}
\end{eqnarray}
where $S^c(x-y)$ is defined through $\langle 0| T \psi(x) \bar\psi(y)|0\rangle$ and we use
the obvious property $ [x \,;\, y]_{A}= [y \,;\, x]^{-1}_{A}$.

The extension to the non-Abelian gauge group is straightforward.

From Eqn.~(\ref{f-GF-1}), we can conclude that the fermion field operator in the external field reads
\begin{eqnarray}
\label{f-GF-2}
\hspace{-0.5cm}\Psi(x^+ | A)=\psi(x^+) {\rm exp}\Big\{ig\int\limits_{\mathds{C}}^{x^+} dz^+ A^-(z^+) \Big\},
\end{eqnarray}
where $\mathds{C}$ is, in principle, an arbitrary point which however we choose to be equal to $-\infty^+$.

We stress that the fermion in the external field differs from the gauge-invariant fermion field which appears
in the string operator, see Eqn.~(\ref{Wl-GT-2}).
Indeed, as well-known (see, for example, \cite{Balitsky:1987bk, Radyushkin:1983mj}) in order to get the
gauge-invariant string operator it is necessary to include the gauge boson (gluon) radiations from the fermions
after the interaction of them with photons (or other gauge bosons) as shown in Fig.~\ref{Fig-DY}, the left panel.
Otherwise, we deal with the fermions in the external fields which are {\it not} gauge-invariant
(see, Fig.~\ref{Fig-DY}, the right panel).

To illustrate the last statement, let us consider the simplest case of Compton-like amplitude (see also \cite{Radyushkin:1983mj}).
We have
\begin{eqnarray}
\label{CA-1}
T^{\mu\nu}=\int (d^4x) e^{-iq\cdot x} \langle p | T J^{\mu}(x) J^{\nu}(0)|p \rangle.
\end{eqnarray}
On the handbag diagram level, we have
\begin{eqnarray}
\label{CA-2}
&&T^{\mu\nu}=\int (d^4x) e^{-iq\cdot x}\times
\\
&&\langle p |
: \bar\psi(x)\gamma^{\mu}
\contraction{}{\psi}{(x)}{\bar\psi} \psi(x) \bar\psi(0)
\gamma^{\nu}\psi(0) : |p \rangle.
\nonumber
\end{eqnarray}
In order to include all gauge boson radiations from the fermion propagator given
by the fermion contraction, we merely make a substitution
(modulo the conventional normalizations which are now irrelevant)
\begin{eqnarray}
\label{SubProp-1}
\contraction{}{\psi}{(x)}{\bar\psi} \psi(x) \bar\psi(0) = S^c(x)\Longrightarrow
\mathrm{S}(x,0)
\end{eqnarray}
with $\mathrm{S}(x,0)$ being {\it gauge-noninvariant} Green function, see Eqn.~(\ref{f-GF-1}).
Using the relation which is similar to Eqn.~(\ref{f-GF-1}), we can obtain that
\begin{eqnarray}
\label{CA-3}
&&T^{\mu\nu}=\int (d^4x) e^{-iq\cdot x}\times
\\
&&\langle p |
: \bar\psi(x)\gamma^{\mu}
S^{c}(x)\,[x \,;\, 0]_{A}
\gamma^{\nu}\psi(0) : |p \rangle.
\nonumber
\end{eqnarray}
After the factorization procedure, the matrix combination
$\gamma^{\mu}\,S^{c}\,\gamma^{\nu}$ refers to the so-called hard part,
while the non-perturbative hadron matrix element involves the {\it gauge-invariant}
string operator defined as
\begin{eqnarray}
\langle p |
: \bar\psi(x)\,[x \,;\, 0]_{A} \psi(0) : |p \rangle.
\end{eqnarray}

\section{Local and nonlocal gauge transform convention}
\label{L-NL-g-conv:App:B}

Before going further, it is important to remind the convention system regarding the gauge transformations which match
the corresponding Wilson path functional (see \cite{Anikin:2016bor} for more details).
In what follows, for the sake of shortness, we say simply the Wilson line independently on the form of a path unless it
leads to misunderstanding.
For the non-Abelian gauge theory, let us now assume that the fermion and gauge fields are transformed as
\begin{eqnarray}
\label{Wl-GT-1}
&&\psi^{\theta}(x)=e^{+i\theta(x)} \psi(x)\equiv U(x) \psi(x) ,
\\
\label{Wl-GT-1-2}
&&A_{\mu}^\theta(x)= U(x)A_\mu(x) U^{-1}(x) + \frac{i}{g} U(x)\partial_\mu U^{-1}(x),
\end{eqnarray}
where $\theta=\theta^a T^a$ with $T^a$ being the generators of corresponding representations.
With the local transformations fixed as in Eqns.~(\ref{Wl-GT-1}) and (\ref{Wl-GT-1-2}),
we can readily see that the covariant derivative and
the gauge-invariant fermion string operator take the following forms:
\begin{eqnarray}
\label{Wl-GT-2}
i\, {\cal D}_{\mu} = i\, \partial_\mu + g A_\mu(x),
\quad
\mathbb{O}^{\text{g.-inv.}}(x,y)=\bar\psi(y) [y \,;\, x]_{A} \psi(x)
\end{eqnarray}
with the Wilson line defined as
\begin{eqnarray}
\label{WL-def-1}
[x\, ; \, x_0 ]_A =
\mathbb{P}\text{exp} \Big\{ ig\int_{P(x_0,x)} d\omega_\mu A_\mu(\omega)\Big\}
= {\bf g}(x | A)\equiv {\bf g}(P),
\end{eqnarray}
where $P(x_0,x)$ stands for a path connecting the starting $x_0$
and destination $x$ points in the Minkowski space.

Inserting the point $x_0$ in the Wilson line of the gauge-invariant string operator, see Eqn.~(\ref{Wl-GT-2}), we get that
\begin{eqnarray}
\label{WI-GT-3}
\mathbb{O}^{\text{g.-inv.}}(x,y)=\bar\psi(y) [y \,;\, x_0]_{A} [x_0 \,;\, x]_{A} \psi(x).
\end{eqnarray}
Eqn.~(\ref{WI-GT-3}) hints that the path-dependent non-local gauge transformation of fermions can be introduced in the form of
\begin{eqnarray}
\label{Wl-GT-4}
\psi^{{\bf g}}(x)  =  {\bf g}^{-1}(x | A) \psi(x),
\end{eqnarray}
where $\psi^{{\bf g}}(x)$ is nothing but the Mandelstam gauge-invariant
fermion field $\Psi(x | A)$, modulo the global gauge transforms \cite{Mandelstam:1962mi,DeWitt:1962mg}.
This transformation leads, in turn, to (cf. Eqn.~(\ref{Wl-GT-1-2}))
\begin{eqnarray}
\label{A-trans-CG}
A_{\mu}^{{\bf g}}(x)= {\bf g}^{-1}(x | A)A_\mu(x) {\bf g}(x | A) + \frac{i}{g} {\bf g}^{-1}(x | A)\partial_\mu {\bf g}(x | A).
\end{eqnarray}
Therefore, we have the following correspondence between local and non-local gauge transformation
\begin{eqnarray}
\label{Corr-loc-nonloc}
 U(x) \Leftrightarrow {\bf g}^{-1}(x | A)
\end{eqnarray}
which is extremely important for the further discussions because the wrong correspondence results in the
substantially wrong conclusions (see for example \cite{Lorce:2012rr}).

\section{Gauge and residual gauge symmetries}
\label{G-Rsymmetry:App:B}

In this Appendix, we remind some subtleties related to the residual gauge transformations
in different gauge theories.

\subsection*{Classical $U(1)$-gauge theory (Abelian theory)}

The $U(1)$-gauge theory where the gauge transformation
\begin{eqnarray}
\label{U1}
A_\mu^\Lambda(x)=A_\mu(x) + \partial_\mu \Lambda(x),
\end{eqnarray}
defines an orbit on the $U(1)$-group.
In the Abelian case, the strength  tensor $F_{\mu\nu}$ is gauge-invariant and, therefore,
only the longitudinal (unphysical) components of field, $A^L_\mu$, can be gauge-transformed.
Indeed, in the classical gauge theory for both $k^2=0$ and $k^2\neq 0$,
the solution of the Maxwell equation in vacuum, $\partial_\mu F_{\mu\nu}=0$, reads (modulo the complex conjugated terms),
(see, e.g., \cite{RubakovBook})
\begin{eqnarray}
\label{SolMeq}
&&A_{\mu}(x)=A^L_{\mu}(x) + A^\perp_{\mu}(x)=
\\
&&\int(d^4k)e^{ikx} k_\mu a_L(k) +
\int(d^4k)e^{ikx} \delta(k^2) e^{\perp\,(\alpha)}_\mu a_\perp^{(\alpha)}(k),
\nonumber
\end{eqnarray}
where $(d^4 k)$ stands for the corresponding integration measure with an appropriate normalization,
$\alpha=(1,2)$ and $k\cdot e^{\perp\,(\alpha)}=0$. With this expression, we can easily derive the gauge transformations in
$p$-space ($=$ the momentum representation)
\begin{eqnarray}
\label{GTr-pspace-1}
k_\mu a_{L}^{\Lambda}(k) + e^{\perp}_\mu a_{\perp}^{\Lambda}(k) =
k_\mu a_{L}(k) + e^{\perp}_\mu a_{\perp}(k) + k_\mu\tilde\Lambda(k),
\end{eqnarray}
where the imaginary factor $i$ is absorbed in the definition of $\tilde\Lambda$.
In what follows summation over $\alpha$ and dimensionful normalizations are not shown explicitly
unless it leads to misunderstanding.

Since $k\cdot e^{\perp\,(\alpha)}=0$, we conclude that
\begin{eqnarray}
\label{GTr-pspace-2}
a_{L}^{\Lambda}(k)= a_{L}(k) + \tilde\Lambda(k), \quad a_{\perp}^{\Lambda}(k) = a_{\perp}(k),
\end{eqnarray}
or, equivalently,
\begin{eqnarray}
\label{GTr-pspace-3}
&&A^{L,\,\Lambda}_\mu(k)= A^{L}_\mu(k) + k_\mu \tilde\Lambda(k),
\nonumber\\
&&A^{\perp,\,\Lambda}_\mu(k) = A^{\perp}_\mu(k).
\end{eqnarray}
Moreover, it is easy to demonstrate that
\begin{eqnarray}
\label{AL}
A^{L}_\mu(x) = -i \partial_\mu\alpha(x),
\end{eqnarray}
where $\alpha(x)$ is a scalar function which is related to $a_{L}(k)$ via the Fourier transformation,
$\alpha(x)\stackrel{\text{F}}{=}a_{L}(k)$, and
$a_{L}(k)=\xi(k)/k^2$ with $\xi(k)\stackrel{\text{def}}{=}k\cdot A^{L}(k)\neq 0$
for $k^2\neq 0$. Notice that if $k^2=0$, the Maxwell equation takes the simplest form, $k\cdot A(k)=0$,
in the $p$-space and, therefore, $k\cdot A^{L}(k)= k^2 a_{L}(k)=0$ or, in other words,
$a_{L}(k)=\xi(k)/k^2\sim 0/0$.

As well-known, to fix the certain  representative on the group orbit we have to impose a
gauge condition $F(A^\Lambda)=0$ on the gauge-transformed fields in order to find a solution with respect to
the gauge parameter $\Lambda$. Here, we do not discuss the appearance of Gribov's ambiguity.

{\bf The Lorentz gauge.} As the first example, we consider the Lorentz (covariant) condition which states
\begin{eqnarray}
\label{L-con}
\partial_\mu A^\Lambda_\mu(x)= \partial_\mu A_\mu + \partial^2 \Lambda(x)
=0.
\end{eqnarray}
In $p$-space, the condition (\ref{L-con}) takes the following form:
\begin{eqnarray}
\label{L-con-2}
k_\mu A^L_\mu(k) + k^2 \tilde\Lambda(k)=0
\end{eqnarray}
which gives us the relation $a_L(k)=-\tilde\Lambda(k)$ for the case of $k^2\neq 0$.
Notice that if $k^2=0$, then the functions $a_L(k)$ and $\tilde\Lambda(k)$ in the
combination $a_L(k)+\tilde\Lambda(k)$ are free functions and they are independent of each other.

However, the gauge condition (\ref{L-con}) (or (\ref{L-con-2})) can not fix the orbit
representative uniquely. Indeed, there is still the so-called residual gauge freedom
defined by $F(A^\Lambda)=F(A)=0$. For the Lorentz condition,
two simultaneous conditions:
\begin{eqnarray}
\label{ResG-1}
\partial_\mu A^\Lambda_\mu(x)=0\quad\text{and}\quad \partial_\mu A_\mu(x)=0
\end{eqnarray}
lead to
\begin{eqnarray}
\label{ResG-2}
\partial^2 \Lambda_0(x)=0
\end{eqnarray}
where the gauge function (parameter) $\Lambda_0$ defines the residual gauge freedom.
That is, the residual gauge transformation with the function $\Lambda_0$ keeps the gauge
condition, $F(A)=0$, gauge-invariant.  Hence, the gauge freedom fixing
means that one fixes all gauge freedom including the residual gauge.
In other words, if there is no residual gauge transformation, the
given gauge condition fixes the gauge freedom completely and we deal with
one representative on a gauge orbit.

Let us consider the second gauge condition in Eqn.~ (\ref{ResG-1}). In $p$-space, it leads to
the following possibilities ($k\cdot a_\perp=0$ by definiton)
\begin{eqnarray}
\label{L-con-p-1}
k^2 a_L(k)=0\,\, \Longrightarrow
\begin{cases}
k^2=0, &a_L(\vec{k})-\,\text{arbitrary }\,\\
k^2\neq 0, &a_L(k)=0.\\
\end{cases}
\end{eqnarray}
Hence, we can see that the gauge condition (\ref{ResG-1}) cannot eliminate the
unphysical field $A^L_\mu$ for the case of $k^2=0$. Working with the equation
(\ref{ResG-2}), in the same manner, we conclude that the gauge function
$\tilde\Lambda_0(\vec{k})$ is not fixed and generates the residual gauge transformation provided
$k^2=0$.

It is instructive to consider the condition (\ref{ResG-2}) in the coordinate representation ($x$-space).
Solutions (\ref{ResG-2}) can be easily found and represented, for instance, the following form:
\begin{eqnarray}
\label{Sol-ReG-x}
\Lambda_0(x)=\begin{cases}
\text{const} &\,\\
1/x^2, &\text{for}\,\,\, x^2\neq 0\,\\
C_0 e^{i(x_0-\vec{x}\vec{N})}&\text{with},\,\,\, |\vec{N}|=1.\\
\end{cases}
\end{eqnarray}
Notice that the scalar function $\alpha(x)$ in Eqn.~(\ref{AL}) which obeys the second condition in Eqn.~ (\ref{ResG-1}),
{\it i.e.} $\partial^2\alpha(x)=0$,  has formally the same form as (\ref{Sol-ReG-x}).

For $k^2\neq0$, the scalar gauge function $\Lambda$ gives also
the longitudinal (unphysical) field $A^L_\mu$, see (\ref{L-con-2}). Therefore,
the first two solutions of (\ref{Sol-ReG-x}) are irrelevant for our study.
In order to get matched with the corresponding condition (\ref{L-con-p-1}) in the momentum representation,
we have to put $C_0$ be equal to zero, $C_0=0$. However, for the case of $k^2=0$,
as above-mentioned, the functions $\alpha(x)$ and $\Lambda_0(x)$ are independent and arbitrary
due to the different free constant pre-factors in the plane wave solution.

We can also consider the Lorentz gauge condition (\ref{L-con}) as an inhomogeneous
differential equation with respect to $\Lambda(x)$, {i.e.}
\begin{eqnarray}
\label{L-con-de}
\partial^2\Lambda(x)=\eta(x)
\end{eqnarray}
where $\eta(x)\stackrel{\text{def}}{=}-\partial_\mu A_\mu(x)$. Solving (\ref{L-con-de}), we obtain that
\begin{eqnarray}
\label{Sol-L-con-de}
\Lambda(x)=\Lambda_0(x)+\int d^4y \,G(x-y) \eta(y),
\end{eqnarray}
where the Green function $G(x)$ is defined as
\begin{eqnarray}
G(x)=\frac{1}{[\partial^2]_{\text{reg}}}\delta^{(4)}(x)
\end{eqnarray}
with the suitable regularization of operator stemmed from
the corresponding boundary conditions, see \cite{BogoShir}.

{\bf The Coulomb gauge.} Using the condition $A_0^\Lambda(x)=0$ to amplify the Lorentz condition (\ref{L-con}),
we can get the Coulomb gauge condition which reads
\begin{eqnarray}
\label{C-con}
\vec{\partial}\vec{A}^\Lambda(x)= \vec{\partial}\vec{A}(x) + \Delta\Lambda(x)=0.
\end{eqnarray}
In $p$-space, the condition (\ref{C-con}) is transformed to (recall that $\vec{\partial}\vec{A}^\perp=0$
by construction)
\begin{eqnarray}
\label{C-con-p}
\vec{k}^2 a_L(k) + \vec{k}^2 \tilde\Lambda(k) = 0.
\end{eqnarray}
Again, let us study the corresponding residual gauge freedom:
\begin{eqnarray}
\label{RegC-con-1}
\vec{\partial}\vec{A}(x)=0\quad\text{and}\quad  \Delta\Lambda(x)=0.
\end{eqnarray}
For the sake of simplicity, we dwell on the case of $k^2=0$ which leads to $\vec{k}^2\neq 0$. With this,
instead of (\ref{C-con-p}), it is enough to stop on the equation
\begin{eqnarray}
\label{C-con-L}
\vec{k}^2 \tilde\Lambda(\vec{k})=0.
\end{eqnarray}
Hence,  the only solution of (\ref{C-con-L}) is $\tilde\Lambda=0$ which means that there is no any
residual freedom at all.

Therefore, in the Coulomb gauge there are no the longitudinal field components and we deal with the
physical gauge field $A^\perp_\mu$ only.

{\bf The Hamilton and axial gauges.} In the similar manner, we can study the residual gauge symmetries in
the Hamilton ($A_0^\Lambda=0$) and axial ($A^{+,\,\Lambda}=0$) gauges. The residual gauge transformations
are given by the corresponding free (unfixed) gauge function $\tilde\Lambda(k)$ provided
$k_0=0$ or $k^+=0$.

\subsection*{Classical $SU(3)$-gauge theory (Non-abelian theory)}

The next subject of our discussion is a non-Abelian gauge theory with
$SU(3)$ gauge group. In this case  case, the gauge transformation is given by
\begin{eqnarray}
\label{SU3}
A^\omega_\mu(x)=\omega(x) A_\mu(x) \omega^{-1}(x) +
\frac{i}{g}\omega(x)\partial_\mu \omega^{-1}(x)
\end{eqnarray}
which gives in the infinitesimal form
\begin{eqnarray}
\label{SU3inf}
A^{a,\,\omega}_\mu(x)= A^a_\mu(x) +
f^{abc} A^b_\mu(x) \theta^c(x) +
\frac{1}{g} \partial_\mu \theta^a(x),
\end{eqnarray}
where $\omega(x)=\exp\big( i \theta^a(x) t^a\big)$.
The decomposition of field components in the longitudinal and transverse
components is similar to the Abelian case, see above.
In contrast to the $U(1)$ gauge group, the strength tensor $G_{\mu\nu}$
is gauge-covariant. It means that all field components may change under
gauge transformations.

{\bf The Lorentz gauge.} We again begin with the Lorentz gauge condition:
\begin{eqnarray}
\label{SU3inf-L-con}
&&\partial_\mu A^{a,\,\omega}_\mu(x)=
\\
&&\partial_\mu A^a_\mu(x) +
f^{abc} \partial_\mu \big( A^b_\mu(x) \theta^c(x)\big) +
\frac{1}{g} \partial^2 \theta^a(x)=0.
\nonumber
\end{eqnarray}
As above-mentioned, the gauge condition is invariant under
the residual gauge transformation:
\begin{eqnarray}
\label{ReL-con-SU-1}
\partial_\mu A^{a,\,\omega}_\mu(x)=\partial_\mu A^a_\mu(x)=0
\end{eqnarray}
or, equivalently,
\begin{eqnarray}
\label{ReL-con-SU-2}
{\cal D}_\mu^{ac}\partial_\mu \theta^c(x)=0,
\end{eqnarray}
where ${\cal D}_\mu^{ac}=\partial_\mu\delta^{ac} + g f^{abc} A^b_\mu(x)$.

In $p$-space, the condition (\ref{ReL-con-SU-2}) takes the form of
\begin{eqnarray}
\label{ReL-con-SU-p-1}
-k^2 \theta^a(k) + i g f^{abc} k_\mu A^{b,\,L}_\mu(k) \theta^c(k)=0.
\end{eqnarray}
If $k^2=0$ and, therefore, $k_\mu A^{b,\,L}_\mu(k)=0$, then the gauge function $\theta(x)$ cannot be fixed
and generates the residual gauge transformation.

{\bf The Hamilton gauge.} The similar situation occurs in the Hamilton gauge,
$A^\omega_0=0$. The residual transformation is induced by the gauge function which obeys
\begin{eqnarray}
\label{ReH-con}
\partial_0 \theta^a(x_0,\vec{x})=0.
\end{eqnarray}
Hence, the solution of this equation is rather trivial: $\theta$-function is the time-independent function, $\theta_0(\vec{x})$.

In the momentum representation, the condition (\ref{ReH-con}) gives us the equation
\begin{eqnarray}
\label{ReH-con-p-1}
\int (d^4k)e^{ikx} k_0 \theta^a(k_0,\vec{k})=0
\end{eqnarray}
which has a solution as
\begin{eqnarray}
\label{Sol-ReH-con-p}
\theta^a_0(k)=\delta(k_0)\theta^a_0(\vec{k}).
\end{eqnarray}
Therefore, we find in the coordinate representation
\begin{eqnarray}
\int (d^4k)e^{ikx} \delta(k_0)\theta^a_0(\vec{k}) = \theta^a_0(\vec{x})
\end{eqnarray}
which coincides with the results of the preceding paragraph.

{\bf The axial gauge.} Working in the axial gauge, $A^{+,\,\omega}=0$, in the similar manner
we are able to find the gauge function that is responsible for the residual gauge symmetry.
We impose the condition
\begin{eqnarray}
\label{RecAx-con-1}
A^{+,\,\omega}(x)=A^{+}(x)=0
\end{eqnarray}
or, in the equivalent form,
 \begin{eqnarray}
\label{RecAx-con-2}
\partial^{+}\theta^a(x^+,x^-,\vec{\bf x}_\perp)=0\quad \text{with}\quad
\partial^+=\partial_-=\frac{\partial}{\partial x^-}.
\end{eqnarray}
The solution of this trivial differential equation is the $x^-$-independent function
$\theta^a_0(x^+,\vec{\bf x}_\perp)$ which has the following form in $p$-space
(cf. (\ref{ReH-con-p-1}) and (\ref{Sol-ReH-con-p})):
\begin{eqnarray}
\label{ReAx-con-p-1}
\theta^a_0(k^+,k^-,\vec{\bf k}_\perp)=\delta(k^+)\theta^a_0(k^-,\vec{\bf k}_\perp)
\end{eqnarray}
where $\theta^a_0(k^-,\vec{\bf k}_\perp)$ is an arbitrary gauge function related to the residual symmetry.

It is instructive to focus on the finite gauge transformations and corresponding gauge condition, namely
\begin{eqnarray}
\label{ReAx-f-con}
&&A^{+,\,\omega}(x)=
\\
&&\omega(x) A^+(x) \omega^{-1}(x) +
\frac{i}{g}\omega(x)\partial^+ \omega^{-1}(x)=0.
\nonumber
\end{eqnarray}
The solution of this equation can easy be found, it reads
\begin{eqnarray}
\label{Sol-Ax-con-f}
\omega_0(x)=\mathbb{P}{\rm exp}\Big\{ ig \int\limits^{x^-}_{\mathds{C}}
dz^- A^+(x^+,z^-,\vec{\bf x}_\perp) \Big\}
\end{eqnarray}
where, generally speaking, $\mathds{C}$ is an arbitrary constant.
We stress that the solution $\omega_0(x)$ is valid for $\forall x \in \mathds{R}^4$.
At the same time, this function can be multiplied by an arbitrary $x^-$-independent
gauge function to produce another solution of equation (\ref{ReAx-f-con}), {\it i.e.}
\begin{eqnarray}
\label{Omega-gen}
W(x^+,x^-,\vec{\bf x}_\perp) = \bar\omega(x^+,\vec{\bf x}_\perp) \omega_0(x^+,x^-,\vec{\bf x}_\perp),
\end{eqnarray}
where $\bar\omega(x^+,\vec{\bf x}_\perp)=\exp\big( i\theta^a(x^+,\vec{\bf x}_\perp)t^a \big)$.
Indeed, one can demonstrate that $A^{+,\,W}(x)=0$.

To study the residual symmetry, we have to demand that $A^+(x)=0$ for any $x$.
Therefore, from (\ref{Omega-gen}), we obtain that the function
\begin{eqnarray}
\label{Omega-gen-res}
W(x^+,x^-,\vec{\bf x}_\perp)\Big|_{A^+=0} = \bar\omega(x^+,\vec{\bf x}_\perp)
\end{eqnarray}
generates the residual transformation we are interested in.

Let us now return to the gauge function presented by (\ref{ReAx-con-p-1}).
The case of $k^+=0$ (which provides us the residual symmetry) leads to the
so-called spurious singularity in the gluon propagator in the axial gauge, see the next subsection.
If we adopt a procedure to regularize this singularity with the help of some well-defined procedure,
$[ k^+]_{reg}\neq 0$, then the existence condition for the residual symmetry, see (\ref{RecAx-con-2}),
has to be given by (in the momentum representation)
\begin{eqnarray}
\label{ReAx-con-reg}
\int (d^4k)e^{ikx} [k^+]_{reg} \delta(k^+)\theta^a_0(k^-,\vec{\bf k}_\perp)=0.
\end{eqnarray}
Hence, the only possibility to satisfy this equation is to demand that $\theta^a_0(k^-,\vec{\bf k}_\perp)=0$ which
means that we fix the remaining residual symmetry.
Thus, we conclude that the spurious singularity is fixed if and only if we do not have the residual gauge symmetry.
On the other hand, we may say that the residual gauge fixing is enough for the elimination of spurious singularity.

\subsection*{Spurious singularity of gluon propagator}

Let us return to the issue of the spurious singularity which appears in the gluon propagator in
the axial gauge $A^+=0$.

The generating functional for gluons (gluonodynamics)
in the most general gauge $F(A^\theta)=0$
\begin{eqnarray}
\label{Z}
&&\mathbb{Z}=N \int {\cal D} A_\mu e^{iS[A]} =
\\
&&\tilde N \int {\cal D} A_\mu \, \Delta_c[A]\, \delta\big(F(A)\big) \, e^{iS[A]},
\nonumber
\end{eqnarray}
where $\tilde N$ involves the infinite gauge group volume, $\int d\theta$, and we use
\begin{eqnarray}
\mathds{1}= \int d\theta \Delta_c[A] \delta\big( F(A^\theta)\big),
\quad \Delta_c[A^\theta] =  \Delta_c[A].
\end{eqnarray}
Instead of solving the gauge condition $F(A^\theta)=0$ with respect to the group function
$\theta$ within the generalized Hamilton formalism, we separate out the infinite group volume, $\int d\theta$,
in the generating functional (the Faddeev-Popov approach).

The next trick is related to the exponentiation of $\delta\big(F(A)\big)$. We introduce the generalized gauge condition as
$F(A)=C$ with $\delta C/\delta A_\mu=0$.  The generalizing functional $\mathbb{Z}$ must be independent on $C$.
Therefore, to get the $C$-independent functional we have to integrate out over this parameter $C$.
Using the integration measure defined as
\begin{eqnarray}
dC \exp\big( - \frac{i}{2\xi} \int d^4 x \,  C^2(x) \big),
\end{eqnarray}
we have
\begin{eqnarray}
\label{Z-2}
&&\hspace{-0.5cm}\mathbb{Z}=\tilde N \int dC e^{\big( - \frac{i}{2\xi} \int d^4 x C^2(x) \big)}
\int {\cal D} A_\mu \, \Delta_c[A]\, \delta\big(F(A)-C\big) \, e^{iS[A]}
\nonumber\\
&&=\tilde N
\int {\cal D} A_\mu \, \Delta_c[A] \, e^{iS[A]- \frac{i}{2\xi} \int d^4 x F^2(A)}.
\end{eqnarray}
In (\ref{Z-2}), the effective action  with the gauge-fixing term,
\begin{eqnarray}
S_{\text{fix}}=- \frac{1}{2\xi} \int d^4 x F^2(A),
\end{eqnarray}
is now not gauge-invariant anymore. As a result of this trick, we don't need to solve
the gauge condition with respect  to the gauge function.

Let the gauge condition $F(A)=0$ be $A^+=0$ with $n^2=0$. In this case, the determinant $\Delta_c[A]$ is independent on $A$
and, therefore, we are able to include this determinant in the normalization of functional.
Thus, the effective Lagrangian reads
\begin{eqnarray}
\label{Leff}
{\cal L}_{\text{eff}}= -\frac{1}{4} G_{\mu\nu}G_{\mu\nu} - \frac{1}{2\xi} \big( n\cdot A \big)^2.
\end{eqnarray}
This Lagrangian yields the effective action which can be written as
\begin{eqnarray}
\label{Seff}
S_{\text{eff}}= \frac{1}{2} \int d^4x\, A_\mu(x)\, K_{\mu\nu}(x)\, A_\nu(x),
\end{eqnarray}
where
\begin{eqnarray}
\label{Kop}
K_{\mu\nu}(x)=g_{\mu\nu}\partial^2 - \partial_\mu \partial_\nu - \frac{1}{\xi} n_\mu n_\nu.
\end{eqnarray}
In $p$-space, the operator $K_{\mu\nu}$ has an inverse operator which, in the limit of $\xi\to 0$, is given by
\begin{eqnarray}
\label{InverseK}
&&K^{-1}_{\mu\nu}(k)=\frac{d_{\mu\nu}(k,n)}{k^2+i0},\quad
\nonumber\\
&&d_{\mu\nu}(k,n)=g_{\mu\nu}-\frac{k_\mu n_\nu + k_\nu n_\mu}{k^+}.
\end{eqnarray}
As we have demonstrated in the preceding subsection, when we fix/regularize
the spurious singularity $[k^+]_{reg}$ it means that we fix the residual gauge
symmetry defined by the gauge function $\theta^a(k^-,\vec{\bf k}_\perp)$ and {\it vise versa}.

We also remind that it is not possible to fix the residual gauge simply by means of adding of
\begin{eqnarray}
\label{Amin-Leff}
\frac{1}{2\xi_2}\big( n^*\cdot A \big)^2
\end{eqnarray}
in Eqn.~(\ref{Leff}). In this case, the inverse kinematical operator (see, Eqn.~(\ref{InverseK})) does not
exist due to the fact that the free (without the coefficients) tensors $n_{\mu}n_{\nu}$ and
$n^*_{\mu}n^*_{\nu}$ present in the corresponding equation to determine the coefficients.
Indeed, introducing the Lorentz parametrization (where the coefficients have to be determined)
\begin{eqnarray}
\label{Lor-par-d}
&&d_{\nu\rho}(k,n,n^*)=g_{\nu\rho} + a_1 k_{\nu}k_{\rho}+
b_2 k_{\nu}n_{\rho}+
b_3 n_{\nu}k_{\rho}+
\nonumber\\
&&b_4 k_{\nu}n^*_{\rho}+
b_5 n^*_{\nu}k_{\rho}+
c_6 n_{\nu}n_{\rho}+
c_7 n^*_{\nu}n^*_{\rho},
\end{eqnarray}
where
\begin{eqnarray}
\text{dim}_M[a_1]=-2,\, \text{dim}_M[b_i]=-1,\, \text{dim}_M[c_j]=0,
\end{eqnarray}
the contraction equation on the coefficients (or, in other words, the Green function equation)
\begin{eqnarray}
\label{Cont-Eqn}
K_{\mu\nu}d_{\nu\rho}=g_{\mu\rho}
\end{eqnarray}
involves the tensors $n_{\mu}n_{\nu}$ and $n^*_{\mu}n^*_{\nu}$ which stay without
coefficients. It means that the inverse operator cannot be derived.


\end{document}